\documentclass[a4paper,11pt]{article}
\pdfoutput=1 % if your are submitting a pdflatex (i.e. if you have
\usepackage{jcappub} % for details on the use of the package, please
\usepackage[T1]{fontenc} % if needed

\usepackage{hyperref}
\usepackage{graphicx}
\usepackage{amsmath,amssymb}
\usepackage{float}
\usepackage[caption = false]{subfig}

\usepackage{multirow}

\newcommand{\be}{\begin{equation}}
\newcommand{\ee}{\end{equation}}
\newcommand{\bea}{\begin{eqnarray}}
\newcommand{\eea}{\end{eqnarray}}

\newcommand{\Msun}{\ensuremath{M_{\odot}}}

\def\kMpc{\, h \, {\rm Mpc}^{-1}}

\begin{document}

\author[a,b]{Andrej Obuljen,}
\author[c,d]{Emanuele Castorina,}
\author[e]{Francisco Villaescusa-Navarro,}
\author[a,b,f]{Matteo Viel}

\affiliation[a]{SISSA- International School for Advanced Studies, Via Bonomea 265, 34136 Trieste, Italy}
\affiliation[b]{INFN -- National Institute for Nuclear Physics, Via Valerio 2, I-34127 Trieste, Italy}
\affiliation[c]{Berkeley Center for Cosmological Physics, University of California, Berkeley, CA 94720, USA}
\affiliation[d]{Lawrence Berkeley National Laboratory, 1 Cyclotron Road, Berkeley, CA 93720, USA}
\affiliation[e]{Center for Computational Astrophysics, 162 5th Ave, New York, NY, 10010, USA}
\affiliation[f]{INAF, Osservatorio Astronomico di Trieste, Via Tiepolo 11, I-34131 Trieste, Italy}

\emailAdd{aobuljen@sissa.it}
\emailAdd{ecastorina@berkeley.edu}
\emailAdd{fvillaescusa@flatironinstitute.org}
\emailAdd{viel@sissa.it}

\title{High-redshift post-reionization cosmology with 21cm intensity mapping} 

\abstract{We investigate the possibility of performing cosmological studies in the redshift range $2.5<z<5$ through suitable extensions of existing and upcoming radio-telescopes like CHIME, HIRAX and FAST. We use the Fisher matrix technique to forecast the bounds that those instruments can place on the growth rate, the BAO distance scale parameters, the sum of the neutrino masses and the number of relativistic degrees of freedom at decoupling, $N_{\rm eff}$. We point out that 
quantities that depend on the amplitude of the 21cm power spectrum, like $f\sigma_8$, are completely degenerate with $\Omega_{\rm HI}$ and $b_{\rm HI}$, and propose several strategies to independently constrain them through cross-correlations with other probes. Assuming $5\%$ priors on $\Omega_{\rm HI}$ and $b_{\rm HI}$, $k_{\rm max}=0.2~h{\rm Mpc}^{-1}$ and the primary beam wedge, we find that a HIRAX extension can constrain, within bins of $\Delta z=0.1$: 1) the value of $f\sigma_8$ at $\simeq4\%$, 2) the value of $D_A$ and $H$ at $\simeq1\%$. In combination with data from Euclid-like galaxy surveys and CMB S4, the sum of the neutrino masses can be constrained with an error equal to $23$ meV ($1\sigma$), while $N_{\rm eff}$ can be constrained within 0.02 ($1\sigma$). We derive similar constraints for the extensions of the other instruments. We study in detail the dependence of our results on the instrument, amplitude of the HI bias, the foreground wedge coverage, the nonlinear scale used in the analysis, uncertainties  in the theoretical modeling and the priors on $b_{\rm HI}$ and $\Omega_{\rm HI}$. We conclude that 21cm intensity mapping surveys operating in this redshift range can provide extremely competitive constraints on key cosmological parameters.}

\maketitle
%\flushbottom

\section{Introduction}

The $\Lambda$CDM cosmological model describes how the small, quantum fluctuations in the primordial density field were amplified by gravity to give rise to the distribution of matter we observe today: galaxies concentrated in massive galaxy clusters, fed by thin filaments surrounded by huge empty voids. It is the most accepted framework that is able to explain a very large variety of cosmological observations with just a handful of parameters. Those parameters represent fundamental physical quantities such as the geometry of the Universe, the amount of dark energy, the sum of the neutrino masses or the properties of the Universe initial conditions. 

The goal of the current and upcoming surveys is to infer the value of those cosmological parameters through cosmological observables such as the anisotropies in the cosmic microwave background (CMB), the spatial distribution of galaxies, weak lensing or the properties of the Ly$\alpha$ forest \cite{Planck2015, BOSS17, DES16}.  The idea behind it being that the large scale structure distribution in the Universe is a faithful tracer of the underlying dark matter density field we want to probe.

Cosmic neutral hydrogen (HI) offers another way to map the matter distribution. It can be detected in emission or in absorption (e.g.\ in the Ly$\alpha$-forest). In emission, one of the most efficient mechanism is the production of photons due to hyperfine transition of the ground state of the hydrogen atom. In this case a photon with a wavelength of 21cm (in the rest frame of the atom) is emitted, and it can be detected on Earth with radio-telescopes.

Intensity mapping (IM) \citep{Bharadwaj_2001A, Bharadwaj_2001B,Battye:2004re,McQuinn_2006,Chang_2008,Loeb_Wyithe_2008, Bull2015, Santos_2015, Villaescusa-Navarro_2015a} is a technique that consists in performing a low angular resolution survey to map the 21cm emission from unresolved galaxies or HI clouds. Intensity mapping for the Ly$\alpha$ emission line has already been successfully applied for large scale clustering studies at high redshift using BOSS data \cite{croft16}.
The advantages of this technique over traditional methods, such as galaxy redshift surveys, are many. Firstly, IM can survey efficiently very large cosmological volumes. Secondly, IM is spectroscopic in nature and thus offers high radial resolution. Thirdly, it can efficiently probe the spatial distribution of neutral hydrogen from the local Universe to the dark ages. On the other hand, its practical implementation is affected by some major challenges. The most import one is the fact that the amplitude of the galactic and extragalactic foregrounds is several orders of magnitude larger than the cosmological signal. 
Many intensity maping surveys (CHIME\footnote{\url{http://chime.phas.ubc.ca/}} \cite{CHIME}, HIRAX\footnote{\url{http://www.acru.ukzn.ac.za/~hirax/}}\cite{HIRAX}, Tianlai\footnote{\url{http://tianlai.bao.ac.cn}}, SKA\footnote{\url{https://www.skatelescope.org/}} \cite{SKA}, FAST\footnote{\url{http://fast.bao.ac.cn/en/}} \cite{FAST}) are currently being built and taking data with the goal of measuring the Baryon Acoustic Oscillations (BAO) scale between $0.8<z<2.5$ with unprecedented precision. On the other hand, the Ooty Wide Field Array (OWFA)\footnote{\url{http://rac.ncra.tifr.res.in}}\cite{OWFA} is intended to measure the 21cm HI power spectrum at redshift 3.35 \citep{2017arXiv170705335R, 2017JApA...38...14S, 2017JApA...38...15C,2017arXiv170903984S}.

At the same time galaxy surveys such as \textsc{DESI} \cite{DESI2016}, \textsc{Euclid} \cite{Euclid}, \textsc{LSST} \cite{LSST}, \textsc{WFIRST} \cite{WFIRST} will detect millions of galaxies up to $z\simeq2$ and will allow us to tightly constrain the value of the cosmological parameters. 

The redshift window $2.5<z<5$ remains however vastly unexplored on cosmological scales. DESI will measure the BAO in the three dimensional clustering of the Ly$\alpha$-forest in the range $2<z<3.5$ remarkably well, with a 1-2\% precision, but still far from the cosmic variance limit, and HETDEX \cite{HETDEX} will observe Ly$\alpha$ emitting galaxies over 400 square degrees between  $1.9<z<3.5$ probing both BAO and redshift space distortions (RSD). 
Indeed, carrying out galaxy redshift surveys at high-redshifts becomes increasingly more difficult, as galaxies become fainter and fainter. IM can survey these redshifts without any limitation,  and therefore, assuming its technical difficulties can be overcome, it offers a natural solution to map the high-redshift Universe.
For these and other reasons, a hypothetical 21cm survey at high-redshift has been considered by the Cosmic Vision program \cite{CosmicVision} as one the five possible experiment for the next generation of cosmological surveys.

Although in most models the contribution of dark energy to the energy content of the Universe decreases with redshift, it is important to carry out observations at redshifts $z\geqslant2$ in order to probe the consistency of the model. Alternatively, extensions of the models as for example early dark energy still provide viable solutions to the cosmological constant problem and could be exquisitely probed with IM of the 21cm line \cite{Doran, Verde, carucci17}. The nature of dark matter can be tested by using IM \cite{carucci15}. Another possibility is to constrain extensions of the vanilla $\Lambda$CDM cosmology, e.g.\ to massive neutrinos and curvature. More generally, measurements at high-redshift could increase our lever arm to constrain any deviation from the standard paradigm.

At high-redshift the modeling of RSD is also more robust, as the dark matter field is more linear on a cosmological scales. 
Unfortunately, a major limitation of all IM measurements is the interpretation of the redshift-space  clustering. This is due to the conversion factor from the measured brightness temperature of the 21cm line to the underlying dark matter distribution, which is related to the mean distribution of neutral hydrogen in the Universe. Exploiting the full potential of 21cm IM, would therefore require external information about this astrophysical parameter.
 
In this work we want to study how much the cosmological constraints of a 21cm survey will be affected by uncertainties in the astrophysics of neutral hydrogen. We will discuss novel ways of tackle this problem, and then focus our attention to the performances of a hypothetical 21cm survey in the redshift range $2.5<z<5$. One of the main goals is to find the cosmological parameters that will mostly benefit from a combination of CMB probes, galaxy survey and 21cm observations.
 
This paper is organized as follows.  In section \ref{sec:method} we present the modelling of the 21cm signal and of the noise along with the Fisher matrix formalism employed for this work. The characteristics of the 21cm IM surveys considered are outlined in section \ref{sec:surveys}. The constraints on the growth rate are shown in section \ref{sec:growth}. In sections \ref{sec:BROADBAND_mnu} and \ref{sec:BROADBAND_Neff} we discuss the constraints on two possible extensions of the baseline cosmological model: neutrino masses and effective number of neutrinos, respectively. We finally outline the main conclusions of this work in section \ref{sec:conclusions}.
%\section{Introduction}

%%%%%%%%%%%%%%%%%% METHOD %%%%%%%%%%%%%%%%%%%%%%%
\section{Method}
\label{sec:method}
Estimates of the performance of a given survey in constraining cosmological parameters are usually carried out using the Fisher Matrix formalism \cite{Tegmark}. Its main ingredients are a model for the signal and the noise one would like to measure. In our case we need to define the 21cm power spectrum, $P_{21}(k,z)$, and the noise level of a given configuration of radio antennas.

The fiducial cosmology we use is the Planck 2015 \cite{Planck2015} best-fit $\Lambda\mathrm{CDM}$:
$$\Omega_\mathrm{M}=0.3075,\, h=0.6774, \, \Omega_\mathrm{K}=0,\,n_{\rm s}=0.9667,\,A_{\rm s}=2.142\times10^{-9},\,\sigma_{8,0}=0.828,\,\Omega_\nu=0.$$
For massive neutrinos we assume a fiducial value of $\Sigma m_\nu=0.06$ eV, and for the number of relativistic degrees of freedom $N_\mathrm{eff}=3.046$. In all the calculations with non-zero $\Sigma m_\nu$ we keep the total matter fraction $\Omega_\mathrm{M}=\Omega_{\rm cdm}+\Omega_{\rm b}+\Omega_\nu$ fixed. We do this by including the neutrinos as the part of CDM fraction and decreasing the fiducial CDM fraction $\Omega_\mathrm{cdm}$ by $\Omega_\nu=\Sigma m_\nu/(94.07\,\mathrm{eV})/h^2$. Throughout the whole analysis we keep the baryon fraction $\Omega_\mathrm{b}$ fixed. Power spectra have been computed using CAMB \cite{CAMB}.

%%%%%%%%%%%%% 21cm model %%%%%%%%%%%%%%%%%
\subsection{21cm signal model}
\label{sec:21cmsignal}
After reionization ends we can assume most of the neutral hydrogen lives in halos and galaxies, where it remains self-shielded from the background UV ionizing radiation \citep{FVN2014}. This means that on large enough scales, a good model for the 21cm power spectrum, $P_{21}(k,z,\mu)$, is given by \cite{Kaiser}
\be
\label{eq:P21}
P_\mathrm{21}(k,z,\mu) = \overline{T}_b^2(z) (b_\mathrm{HI}(z) + f(z)\mu^2)^2 P(k,z),
\ee
where $b_\mathrm{HI}$ is the linear bias of the HI field, $f$ is the growth rate we assume to follow $f=\Omega_{\rm M}(z)^{0.545}$, $\mu=k_\parallel/k$, $P(k,z)$ is the total matter power spectrum and $\bar{T}_b$ is the mean brightness temperature given by
\be
\bar{T}_b(z)=180\frac{H_0(1+z)^2}{H(z)}\Omega_{\rm HI}(z)h~{\rm mK},
\ee
where $\Omega_{\rm HI}(z)=\rho_{\rm HI}(z)/\rho_c^0$ and $\rho_c^0$ is the critical density of the Universe at $z=0$. In principle, due to the non-linear effects that tend to damp the BAO peak, eq. \eqref{eq:P21} should contain an additional damping term: $\exp{(-k^2_\perp\Sigma^2_\perp/2 - k^2_\parallel\Sigma^2_\parallel/2)}$, where $\Sigma_\perp$ and $\Sigma_\parallel$ are the non-linear damping scales in the transverse and the radial direction, respectively \cite{SeoEisenstein2007}. These damping scales decrease with redshift roughly as the linear growth factor. Since we will be considering high-redshift window, the effects of non-linearities are pushed to small scales. Furthermore, a way to ameliorate this effects is to perform a BAO reconstruction method on the observed non-linear galaxy distribution \cite{Reconstruction} or the HI intensity maps at low-redshifts \cite{SeoHirata,BAOpixels}. Applying these techniques can effectively half the damping scales, pushing the effects of non-linearities to even smaller scales. In fact, we have explicitly checked that including these effects in the redshift range we consider does not make a significant difference in our results. For simplicity, we will not include the damping term in the model for the 21cm power spectrum.

One immediately sees in eq. \eqref{eq:P21} the main difference with respect to a cosmological analysis of a galaxy power spectrum: any constraint on the amplitude of the fluctuations, $\sigma_8$,  of RSD parameters, e.g.\ $f$, or a combination of the two, is completely degenerate with the amount of HI in the Universe. This quantity is very uncertain in the theory modeling, so any information on it must come from independent datasets. If no external priors are available then RSD measurements with the 21cm line will not be competitive with galaxy surveys. On the other hand, constraints on the BAO scales are still safe, as they are mostly independent of the broadband shape and normalization of the power spectrum.

The good news is that several measurements of the HI mean density are available nowadays \cite{Crighton,Noterdaeme2012,Zafar2013,Bird2016}. They mostly come from the detection in quasars (QSO) spectra of Damped Ly$\alpha$ (DLA) systems. DLAs are objects with high column density, $N_{\rm HI} > 10^{20.3}\,\text{cm}^{-2}$,  and contain more than the 90\% of the total HI in the Universe. The abundance of DLAs gives a direct estimate of  $\Omega_{\rm HI}(z)$, with current errorbars around 5\% at $z<3.5$ and around 30\% at $z>3.5$ \citep[see e.g.][]{Crighton}. In both cases there is a lot of margin for improvement, as existing low redshift measurements are limited by the noise level in the spectra, whereas at high-redshift by the low numbers of QSO spectra publicly available. In the near future surveys like DESI and WEAVE \cite{WEAVE} will collect hundreds of thousands of QSO spectra, 10 times more than current surveys, yielding much better estimates of the abundance of HI.
Motivated by this, in the remaining of this work we will assume three different priors on $\Omega_{\rm HI}$: 2\%, 5\%, 10\%. We take the above priors to be redshift independent, although one could also imagine a prior degrading with increasing redshift.

The other astrophysical parameter which is quite degenerate with cosmological parameters is the HI linear bias $b_\mathrm{HI}$. In particular, any measurement that can break the $\Omega_\mathrm{HI}-b_\mathrm{HI}$ would be of great value for parameter estimation with the 21cm power spectrum. Naive cross-correlation with other kinds of tracers will not help, and weak lensing is not a particularly viable option since it probes the $k_{||}\rightarrow0$ limit of the density field which for 21cm surveys is the most affected by foreground contaminants \cite{Shaw2014,Shaw2015}. 

What we propose, based on the arguments in \cite{EmaPaco}, is to measure the clustering of the same objects one uses to measure $\Omega_\mathrm{HI}$, e.g.\  Lyman-limit systems (LLS) and DLAs, but weighting each of them by the value of its column density. This weighted density field can then be cross-correlated with the Lyman-$\alpha$ forest or the 21cm field itself to partially break the $\Omega_\mathrm{HI}-b_\mathrm{HI}$ degeneracy. 
We stress that such a measurement can already be done with existing datasets, yielding the first ever measurement of the bias of the HI field. 
As indeed shown in \cite{EmaPaco} the bias of the DLAs and of the HI can be written as
\begin{eqnarray}
b_{\rm HI}(z)&=&\frac{c}{H_0}\frac{\int_0^\infty b(M,z)n(M,z)dM\int_{0}^{\infty} d\sigma N_{\rm HI}}{\int_0^\infty f(z,N_{\rm HI})N_{\rm HI}~dN_{\rm HI}}
\label{eq:bHIDLAa}\\
b_{\rm DLA}(z)&=&\frac{c}{H_0}\frac{\int_0^\infty b(M,z)n(M,z) dM\int_{\sigma(10^{20.3})}^\infty d\sigma}{\int_{10^{20.3}}^\infty f(z,N_{\rm HI})~dN_{\rm HI}}
\label{eq:bHIbDLA}
\end{eqnarray}
where $b(M,z)$ is the halo bias, $n(M,z)$ is the halo mass function, $\mathrm{d} \sigma\equiv \mathrm{d}N_{\rm HI}\frac{\mathrm{d}\sigma}{\mathrm{d} N_{\rm HI}}(N_{\rm HI}|M,z)$ is the cross-section of systems with column density $N_{\rm HI}$, i.e.\ the probability of observing an object with $N_{\rm HI}$ in a  halo, and $f(z,N_{\rm HI})$ is the column density distribution function, i.e.\ the abundance of such systems. We see that the difference between the two is just an extra factor $N_{\rm HI}$. This means that if we weight DLAs by the value of their column density and then cross-correlate with another tracer would yield an estimate of the HI bias. The lower integration limit in eq. \eqref{eq:bHIbDLA} is not important in this discussion, since most of the neutral hydrogen, 95\% or more, lives in objects with $N_{\rm HI}\simeq10^{21} {\rm cm}^{-2}$ \cite{Noterdaeme2012,EmaPaco}, and the weighing by $N_{\rm HI}$ will make the integrals above completely dominated by the tail of the column density distribution function. 
In the limiting case the bias of DLAs does not depend on this lower threshold, i.e.\ the DLA cross-section does not depend on halo mass, then $b_{\rm DLA}=b_{\rm HI}$. Recent measurements in \cite{Rafols} show a slight statistical preference for the latter situation.

\begin{figure}[t!]
\centering
\includegraphics[scale = 0.4]{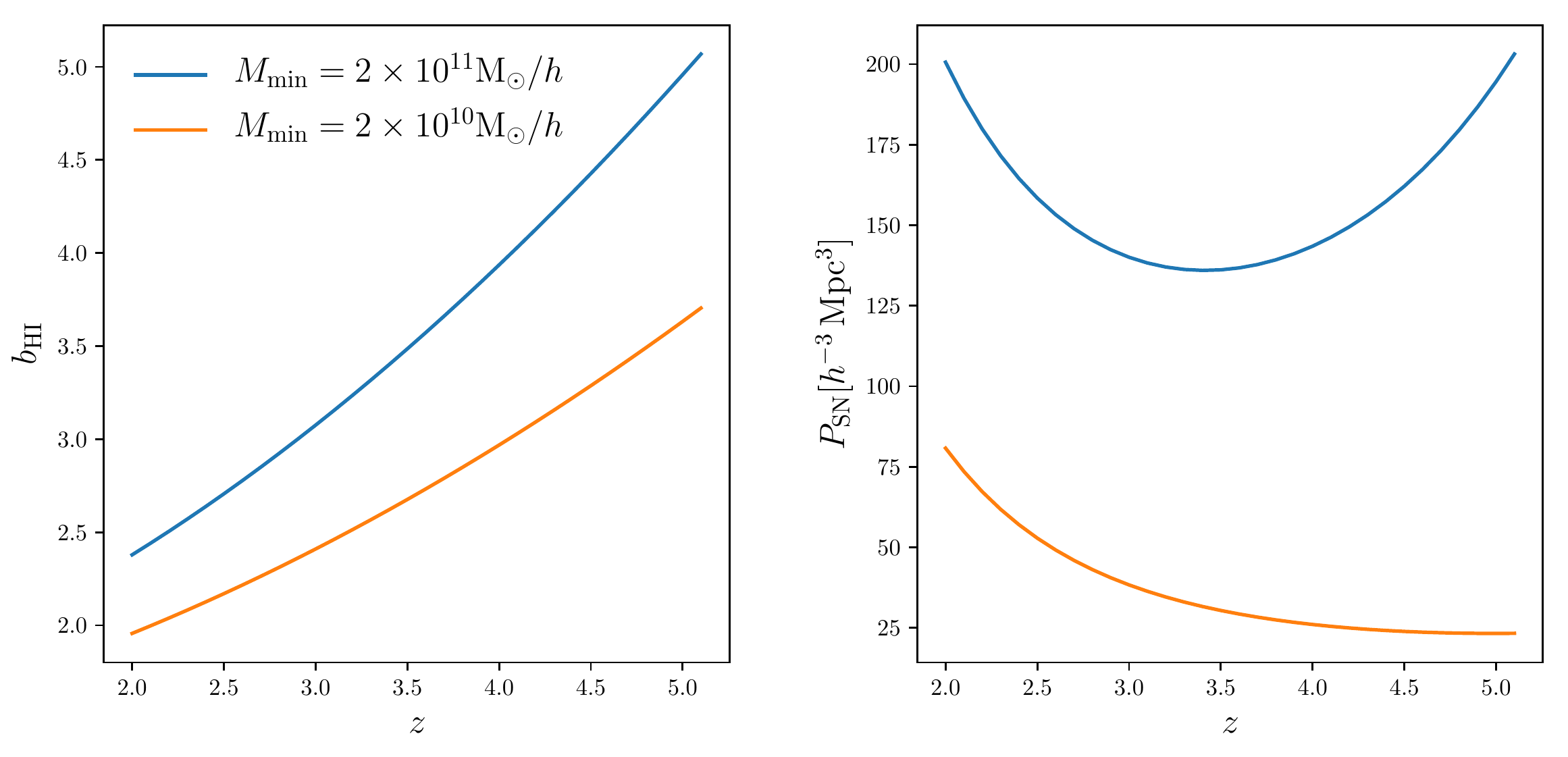}
\caption{The figure shows the dependence on redshift of the HI bias (left) and the HI shot-noise (right) for two  different models of the HI distribution (see text).}
\label{fig:Bias_Shot_noise_nP02SN}
\end{figure}

The cross-correlation of the weighted DLA field, $\delta_{\rm DLA}^w$, with the Lyman-$\alpha$ forest flux, $\delta_F$, will therefore return in the linear regime
\be
\langle\delta_{\rm DLA}^w  \delta_F\rangle = b_{\rm HI}b_F P(k)(1+f \mu^2/b_{\rm HI})(1+b_v f\mu^2/ b_F)
\ee
with $b_F$ and $b_v$ the density and velocity bias of the forest. Projected correlation function will get rid of most of the RSD contribution allowing a clear determination of $b_{\rm HI}$. Recent measurements of the cross-correlation between the forest and DLAs presented in \cite{Font-Ribera2012,Rafols} provide an estimate of $b_{\rm DLA}\simeq2$ with a 5-10\% error depending on different data cuts, which in the limit the DLA cross-section does not depend on halo mass exactly yields a measurement of HI bias, see eq. \eqref{eq:bHIbDLA}.
Another option would be to cross-correlate $\delta_{\rm DLA}^w$ with the 21cm field itself, 
\be
\langle \delta_{\rm DLA}^w  \delta_{21} \rangle=\bar{T}_b b_{\rm HI}^2 P(k)(1+f\mu^2/b_{\rm HI})^2,
\ee
and again we see that the degeneracy between $\Omega_{\rm HI}$ and the bias can be partially broken by combining different probes.

We expect next generation of galaxy surveys to drastically improve on these numbers. We will therefore assume, as we did above for the cosmic neutral fraction, three different redshift independent priors on $b_{\rm HI}$: 2\%, 5\%, 10\%

In a Fisher analysis we also need to specify the fiducial value of both the cosmological and astrophysical parameters. For the latter we need a model for the abundance and spatial distribution of HI, that we briefly describe here and refer the reader to \cite{EmaPaco} for further details. We assume that all HI is bound within dark matter halos and model the $M_{\rm HI}(M,z)$ function as 
\be
M_{\rm HI}(M,z) = A(z)e^{-M_{\rm min}/M}M^\alpha,
\ee
where $A(z)$, $M_{\rm min}$ and $\alpha$ are free parameters. In this work we take $\alpha=1$ and consider two values of $M_{\rm min}=2\times10^{10} h^{-1}M_\odot$ and $2\times10^{11} h^{-1}M_\odot$. The overall normalization, $A(z)$, is chosen such as $\Omega_{\rm HI}(z)=\frac{1}{\rho_c^0}\int_0^\infty n(M,z)M_{\rm HI}(M,z)dM=4\times10^{-4}(1+ z)^{0.6}$ \cite{Crighton}, where $n(M,z)$ is the halo mass function at redshift $z$. The HI bias is then fully determined by the model and is given by
\be
b_{\rm HI}(z)=\frac{\int_0^\infty b(M,z)n(M,z)M_{\rm HI}(M,z)dM}{\int_0^\infty n(M,z)M_{\rm HI}(M,z)dM},
\ee
where $b(M,z)$ represents the bias of halos of mass $M$ at redshift $z$. For both the theoretical halo mass function and the halo bias we use the fitting functions calibrated from N-body simulations presented in \cite{Tinker2010}. The values of the free parameters of our model are motivated by the poorly understood redshift evolution of the HI fraction and the HI bias, both in simulations and data \cite{FVN2014, 2016MNRAS.460.4310S,FVN2016,Penin2017,padma17a,Crighton}. In Figure \ref{fig:Bias_Shot_noise_nP02SN} we show the fiducial value for the bias and the effective HI shot noise -- $P_{\rm SN}$, for the two different choices of minimum mass $M_{\rm min}=2\times10^{11}~M_\odot/h$ (blue) and $M_{\rm min}=2\times10^{10}~M_\odot/h$ (orange). Similar halo models for the HI are also presented in \cite{padma17a,padma17}.

%%%%%%%%%%%% SURVEYS %%%%%%%%%%%%%
\subsection{Noise model and 21cm IM surveys}
\label{sec:surveys}
Current and planned 21cm IM surveys focus either on the reionization epoch at $z>6$ (e.g. HERA, PAPER, SKA) or on the low-redshift Universe $z<2.5$, targeting the BAO feature in the 21cm power spectrum (e.g.\ CHIME, Tianlai, HIRAX, FAST, SKA1-MID). Some surveys (e.g.\ SKA1-LOW) will cover both redshift ranges but at the price of low angular resolution \cite{PacoBull}, while OWFA will focus on a single redshift \cite{Sarkar}. Our analysis focuses on the range $2.5<z<5$, and therefore we first have to make a few choices on instrument specifications. We consider two approaches: interferometer instruments (similiar to CHIME and HIRAX) and single-dish telescopes (similar to FAST). We modify the configuration of both types of instruments such that they can probe higher redshifts. The main requirement is to keep angular resolution good enough for the scales we want to probe, as a fixed physical distance subtend a smaller angular scale at higher redshift. This usually means we have to increase the size of the longest baseline in an interferometer or the size of the dishes in auto-correlation mode. We dub the interferometers as Ext-CHIME and Ext-HIRAX, while we dub the single-dish with highzFAST.  Table \ref{table:Characteristics} shows the characteristics of Ext-CHIME, Ext-HIRAX and highzFAST.

\begin{table}[ht!]
\centering
\begin{tabular}{lccc}
\hline \hline
& Ext-CHIME & Ext-HIRAX & highzFAST\\
\hline	
\hline
$T_\mathrm{a} \mathrm{(K)}$ & 10 & 10 & 10\\
$L_\mathrm{cyl} \mathrm{(m)}$  & 200 & - & -\\
$W_\mathrm{cyl}  \mathrm{(m)}$ & 20 & - & -\\
$N_\mathrm{cyl}$ & 10 & - & -\\
$N_\mathrm{f}$ & 256 & - & -\\
$D_\mathrm{dish} \mathrm{(m)}$ & - & 10 & 500\\
$N_\mathrm{dish}$ & - & $64\times64$ & 1\\ 
$\eta$ & 0.7 & 1.0 & -\\
$D_\mathrm{max}\mathrm{(m)}$ & 269 & 800 & -\\
$D_\mathrm{min}\mathrm{(m)}$ & $L_\mathrm{cyl}/N_\mathrm{f}$ & 10 & -\\
$N_\mathrm{b}$ & 2560 & - & -\\
$S_{21}$ $(\deg^2)$ & 25000 & 25000 & 5000\\
$n_\mathrm{pol}$  & 2 & 2 & 2\\ 
$\Delta\nu \mathrm{(kHz)}$ & - & - & 50\\
\hline
\end{tabular}
\caption{Main characteristics of the considered surveys.}
\label{table:Characteristics}
\end{table}

\subsubsection{Thermal noise power spectrum - Interferometers}
We model the thermal noise power spectrum of an interferometer as in \cite{Bull2015},
\be
P^\mathrm{th}_\mathrm{N}(z) = \frac{T^2_\mathrm{sys}(z)X^2(z)Y(z) \lambda^4(z) S_{21}}{A^2_\mathrm{eff}\mathrm{FOV}(z)t_0 n_\mathrm{pol}n(\mathbf{u},z)},
\ee
which we use for Ext-CHIME and Ext-HIRAX.
The system temperature is the sum of antenna and the background sky temperature $T_\mathrm{sys}(z)=T_\mathrm{a}+T_\mathrm{sky}(z)$. We take the antenna temperature $T_\mathrm{a}=10\,\mathrm{K}$ and we model the sky temperature as $T_\mathrm{sky}(z)=60\,\mathrm{K}\times(\nu_\mathrm{21}(z)/300\mathrm{MHz})^{-2.55}$\cite{Bull2015}. While the background sky temperature is usually below antenna temperature at low-redshifts, going to higher redshifts it increases significantly and represents the major noise contribution.

The terms $X(z)$ and $Y(z)$ are used to convert from angular and frequency space to physical space. $X(z)$ is the comoving distance and $Y(z)=\frac{c(1+z)^2}{\nu_{21}H(z)}$, where $c$ is the speed of light, $H(z)$ is the Hubble parameter, and $\nu_{21}$ is the frequency of the 21cm line (1420 MHz). The wavelength of the 21cm line at a given redshift is $\lambda(z)=\lambda_{21}(1+z)$.
We assume a survey area $S_{21}=25000 \deg^2$, the observing time of $t_\mathrm{obs}=10000\,\mathrm{hours}$ and the number of polarisation states $n_\mathrm{pol}=2$ for both Ext-CHIME and Ext-HIRAX.

The last term in the denominator is the number density of baselines in visibility space. We will assume a uniform distribution of $n(u;z)=N_\mathrm{b}(N_\mathrm{b}-1)/2\pi/(u^2_\mathrm{max}(z)-u^2_\mathrm{min}(z))$ between $u_\mathrm{min}=D_\mathrm{min}/\lambda(z)$ and $u_\mathrm{max}=D_\mathrm{max}/\lambda(z)$. We will use an approximation of this expression $n(u;z)\simeq N_\mathrm{b}^2/2\pi/u^2_\mathrm{max}(z)$. The total number of beams $N_\mathrm{b}$ is the total number of antenna feeds $N_\mathrm{f}\times N_\mathrm{cyl}$ in the case of Ext-CHIME, while for Ext-HIRAX it is the number of dishes $N_\mathrm{dish}$.

\subparagraph{Ext-CHIME}
The effective area per feed is computed as $A_\mathrm{eff}=\eta L_\mathrm{cyl} W_\mathrm{cyl}/N_\mathrm{f}$, where $\eta$ is the efficiency factor that we take $\eta=0.7$; $L_\mathrm{cyl}$ and $W_\mathrm{cyl}$ are the length and the width of the cylinders, respectively, and $N_\mathrm{f}$ is the number of feeds per cylinder. While the primary beam of each element in this set-up is anisotropic, we will assume an isotropic field of view (FOV) for each element and we will use $\mathrm{FOV}\approx 90^\circ\times1.22 \lambda(z) / W_\mathrm{cyl}$ \cite{Bull2015}.

\subparagraph{Ext-HIRAX}
The effective area per beam is $A_\mathrm{eff}=\pi(D_\mathrm{dish}/2)^2$ where $D_\mathrm{dish}$ is the diameter of a single dish. The primary beam of each dish is isotropic and defined as $\mathrm{FOV}=\Big(1.22\frac{\lambda(z)}{D_\mathrm{dish}}\Big)^2$ \citep{Santos_2015}.

\subsubsection{Window functions for interferometers}
In the case we consider -- uniform number density of baselines -- the noise power spectrum is scale-independent. Nevertheless, this will not be the real case and there will be a range of modes that are going to be (un)observable. To account for this we include sharp cut offs in the $k$-range probed by the survey, both in $k_\parallel$ and $k_\perp$.
We take the limits for $k_\perp$ according to \cite{Bull2015}:
\bea
k_\perp^{min}(z)=k_{D_{min}}(z) &=& \frac{2\pi D_{min}}{D(z)\lambda(z)},\\
k_\perp^{max}(z)=k_{D_{max}}(z)&=&\frac{2\pi D_{max}}{D(z)\lambda(z)},
\eea
where $D(z)$ is the comoving distance. The values for $D_\mathrm{min}$ and $D_\mathrm{max}$ depend on the configuration we consider and are given in Table \ref{table:Characteristics}.
For $k_\parallel$(z) we take the upper limit $k_\mathrm{max} (z)$ to be $0.2\,h\mathrm{Mpc}^{-1}$ or the non-linear scale $k_\mathrm{nl}(z)=k_\mathrm{nl,0}(1+z)^{2/3}$, where $k_\mathrm{nl,0}=0.2\,h\mathrm{Mpc}^{-1}$ \cite{Smith2003}. 
We take the lower limit $k_\parallel^\mathrm{min}\mathrm{(z)}=2\pi/V_\mathrm{sur}^{1/3}(z)$. Based on these, we can now define a window function that defines the range of scales available to a given configuration.
$$W(z,k,\mu)=\Theta(k_{D_{max}}-k_\perp)\Theta(k_\perp-k_{D_{min}})\Theta(k_\parallel-k_\parallel^\mathrm{min})\Theta(k_\mathrm{nl}-k\mu)\Theta(k_\mathrm{nl}-k),$$
where $\Theta$ is the Heaviside step function. We show the effect of this window function on the signal power spectrum in Figure \ref{fig:Veff_V_S_N} in the case of Ext-HIRAX and in the bottom left panel of Figure \ref{fig:PHI_noises} in the case of Ext-HIRAX and Ext-CHIME.

\subsubsection{The foreground wedge}
\label{sec:wedge}
The 21cm cosmological signal is several orders of magnitude weaker than the astrophysical foreground emissions. Isolating the cosmological signal will be the major challenge in the data analysis, and it mostly relies on the assumption that foregrounds are spectrally smooth. In principle this means that only the smallest radial wave-numbers $k_\parallel$ are affected by the foregrounds, allowing us to clean the data without much of the cosmological information being lost. However, the chromaticity of the interferometers response function will cause foregrounds to leak into the high-$k_\perp$ modes. This will make the foreground cleaning very difficult. This effect is limited to a certain region in $k_\parallel-k_\perp$ space, acting on low-$k_\parallel$ and high-$k_\perp$, and it is known in the literature as the foreground \textit{wedge} \cite{Datta,Morales2012,Liua,Liub,Parsons2012,Pober2015,SeoHirata}:
\be
k_\parallel < \sin(\theta_\mathrm{FoV})\frac{D(z)H(z)}{c(1+z)}k_\perp,
\ee
where $\theta_\mathrm{FoV}$ is the field-of-view (or the primary beam width) of an interferometer element. We can also express the wedge region in terms of all the modes having $\mu<\mu_\mathrm{min}$ with:
\be
\mu_\mathrm{min}(z)= \frac{k_\parallel}{\sqrt{k_\parallel^2+k_\perp^2}}=\frac{\sin(\theta_\mathrm{FoV})D(z)H(z)/(c(1+z))}{\sqrt{1+\left[\sin(\theta_\mathrm{FoV})D(z)H(z)/(c(1+z))\right]^2}}.
\ee
It is important to stress that the foreground wedge is not an intrinsic limitation of a 21cm interferometric survey, but rather a manifestation of our incomplete knowledge of the calibration of the antennas. Accurate calibration is a difficult task of any data analysis, in IM in particular because of the   $10^5$ difference in signal strength between the foregrounds and the signal one is trying to measure.

Unfortunately, this contaminated region grows at higher redshifts, as both the comoving distance and the Hubble parameter increase faster than $(1+z)$. For instance, at $z=2$ the foreground wedge contaminates all wave modes with $\mu\leq\mu_\mathrm{min}(2)=0.77$, and the situation gets much worse at $z=4$ with $\mu\leq0.9$. The simplest approach is to ignore the modes inside the wedge in data analysis (or in Fisher forecasts). This would mean that we discard 90\% of the available modes at $z=4$. Another possibility is to develop techniques that can recover as much as possible information inside the wedge \cite{Shaw2014,Shaw2015}. While these methods strongly depend on the details of the instrumental set up and of the data analysis procedure, in this paper, for simplicity we will consider few idealised cases: 
\begin{enumerate}
\item Wedge -- we discard all the modes inside the wedge assuming a horizon limit, i.e.\ $\sin(\theta_\mathrm{FoV})=1$;
\item Mid-wedge -- (i) in the case of Ext-HIRAX we use $\mu=\mu_{\rm min}$; (ii) in the case of Ext-CHIME where the primary beam is anisotropic we assume modes with $\mu=\mu_\mathrm{min}/1.5$ are unavailable, i.e.\ one third of the contaminated region is cleaned;
\item No wedge case -- the foreground cleaning works perfectly for all the modes.
\end{enumerate}

\begin{figure}[t!]
\centering
\subfloat{\includegraphics[scale = 0.5]{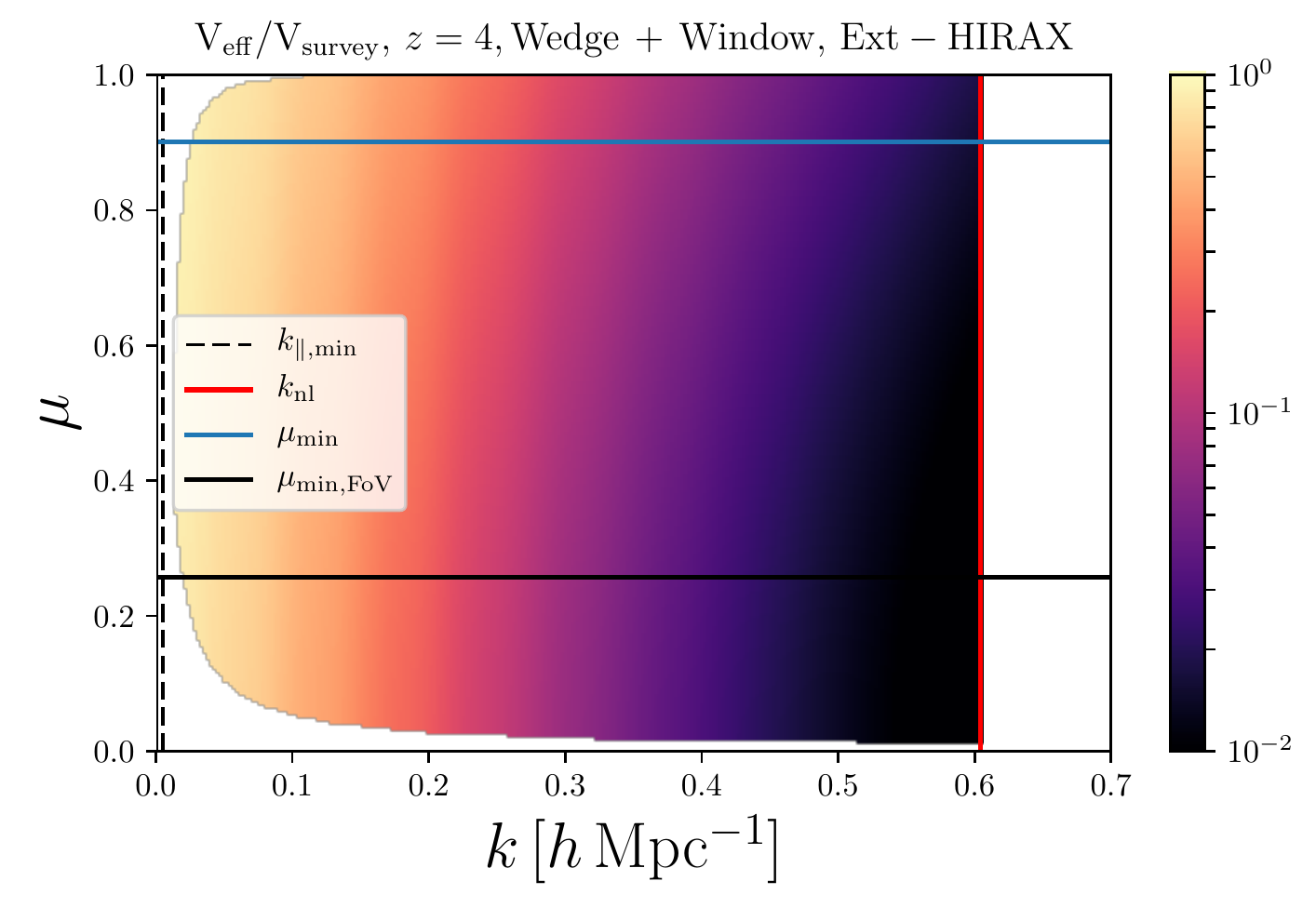}}
\subfloat{\includegraphics[scale = 0.5]{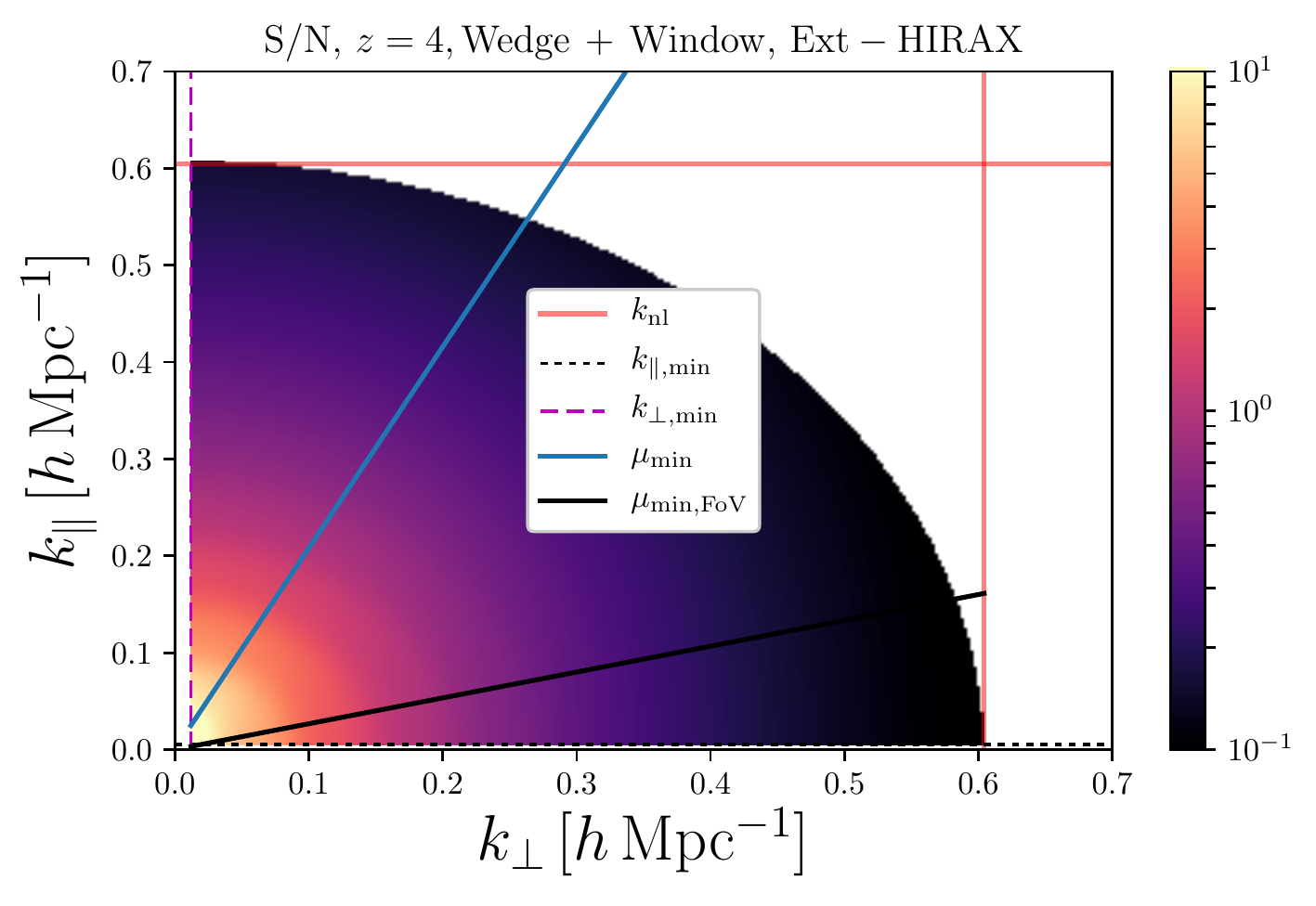}}
\caption{This plot shows the effect of the window function, wedge and $k_{\rm nl}$ on the number of modes available to extract cosmological information for Ext-HIRAX. The left panel displays the ratio between the effective volume to the survey volume as a function of $k$ and $\mu$. The right panel represents the signal-to-noise ratio as a function of $k_\perp$ and $k_\parallel$. For both panels we have considered a redshift bin $\Delta z=0.1$ centered at $z=4$ and assumed $k_\mathrm{max}=k_\mathrm{nl}(z)$. Unavailable modes are shown as white regions. The effect of the wedge can be seen from the solid black and blue lines. If no wedge is present all modes shown in the panels contribute, while only the modes above the black/blue line contribute in the case of full/mid wedge.} 
\label{fig:Veff_V_S_N}
\end{figure}

In Figure \ref{fig:Veff_V_S_N} we show the effect of the wedge and the window function on the available modes in the case of Ext-HIRAX. In the left panel we show the effective to survey volume ratio as a function of $k$ and $\mu$. Available modes in the three cases we consider are above the solid lines of constant $\mu_\mathrm{min}$. We show the $\mu_\mathrm{min}$ in the wedge and mid-wedge case in solid blue and solid line, respectively. Similarly, in the right panel, we show the signal-to-noise ratio as a function of $k_\perp$ and $k_\parallel$, where the effect of the wedge is more clearly seen. Again, solid blue and black line represent the wedge cases we consider and the available modes are on restricted on the left side of these lines. In both of the panels we use $k_\mathrm{max}=k_\mathrm{nl}$.

\subsubsection{Thermal noise power spectrum - highzFAST}

We model the noise power spectrum of a single dish as \cite{Pourtsidou}
\be
P^\mathrm{th}_\mathrm{N}(k,\mu,z) = \sigma^2_\mathrm{pix}(z) V_\mathrm{pix}(z) W^{-2}(k_\perp,z).
\ee
The first term is the pixel thermal noise given by
\be
\sigma^2_\mathrm{pix} = \frac{T^2_\mathrm{sys}}{\Delta\nu t_\mathrm{obs} (\Omega_\mathrm{pix}/S_{21}) N_\mathrm{dish} N_\mathrm{beam}},
\ee
where $T_\mathrm{sys}$ is the system temperature described above, $\Delta\nu$ is the frequency resolution, $t_\mathrm{obs}$ is the total observing time, $\Omega_\mathrm{pix}$ is the pixel area, $S_{21}$ is the survey sky area, while $N_\mathrm{dish}$ and $N_\mathrm{beam}$ are the number of dishes and number of beams per dish, respectively. 
The pixel area is calculated assuming a Gaussian beam $\Omega_\mathrm{pix}=1.13\theta^2_\mathrm{FWHM}$, where $\theta_\mathrm{FWHM}=\lambda(z)/D_\mathrm{dish}$.
The volume of a pixel $V_\mathrm{pix}$ is computed by integrating the comoving volume element of the pixel area along the line-of-sight in the redshift range corresponding to the frequency resolution $\Delta\nu$:
\be
V_\mathrm{pix} = \Omega_\mathrm{pix}\int_{z-\Delta z/2}^{z+\Delta z/2} dz' \frac{c D^2(z')}{H(z')}.
\ee

The last term in the noise power spectrum is the response function describing the angular resolution of a single-dish telescope. Since the frequency resolution we assume here is very high, we will not take into account the radial response and will only consider the angular one. We use
\be
W^2(k_\perp,z)=\exp\left[-k^2_\perp D^2_\mathrm{c}(z)\left(\frac{\theta_\mathrm{FWHM}}{\sqrt{8\ln2}}\right)^2\right].
\ee
For an instrument like the highzFAST the scales which will be accessible are
\bea
k_\perp^{min}(z) &=& \frac{2\pi}{\sqrt{D(z)^2 S_{21}}},\\
k_\perp^{max}(z) &=& \frac{2\pi D_\mathrm{dish}}{D(z)\lambda(z)}.
\eea

We show the noise power spectrum of highzFAST at $z=4$ in the bottom right panel of Figure \ref{fig:PHI_noises} in dashed line. The dashed gray line is the signal power spectrum, while the blue solid line includes the effect of the accessible modes. 

\subsection{Total noise power spectrum}
\begin{figure}[t!]
\centering
\subfloat{\includegraphics[scale = 0.5]{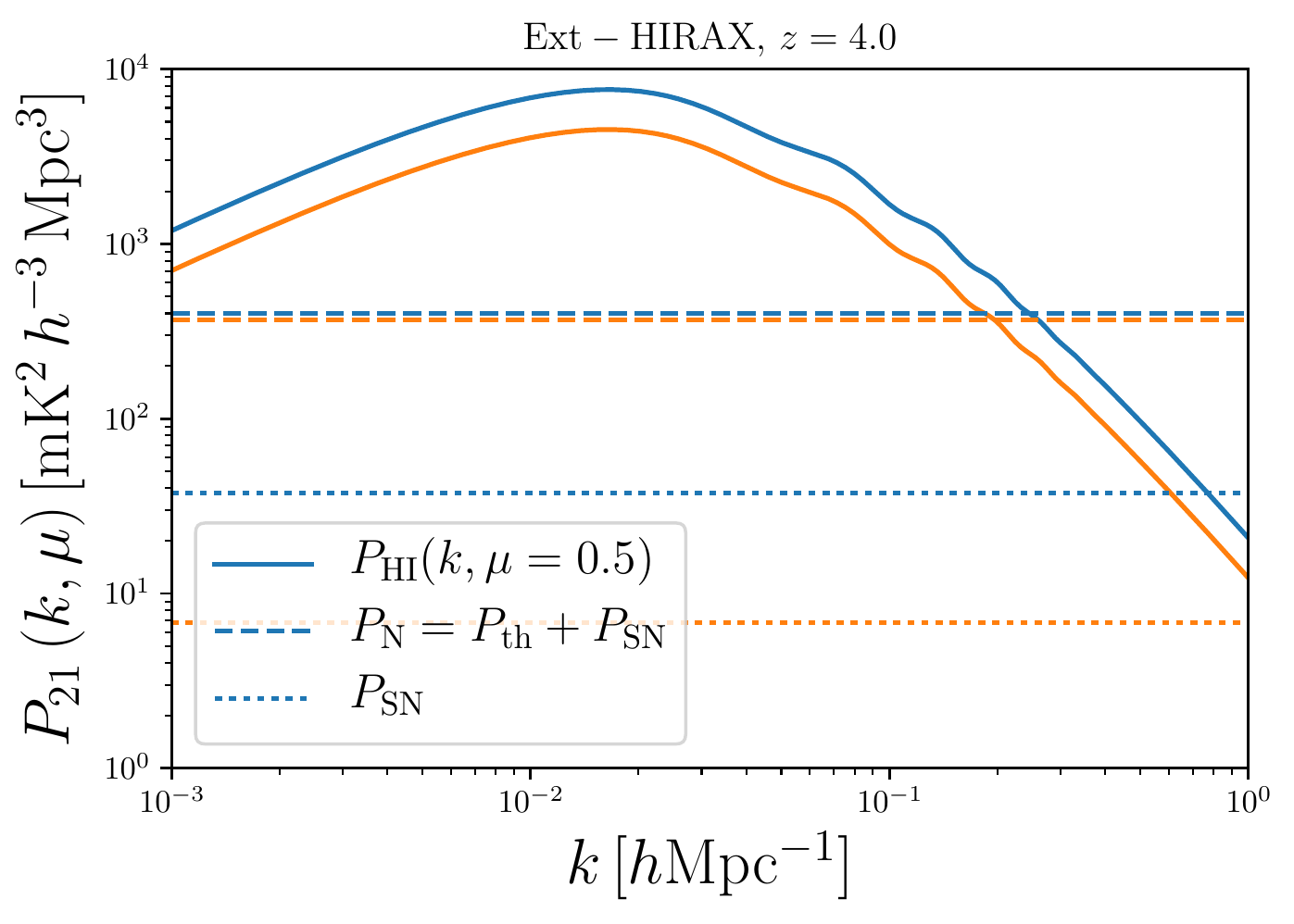}}
\subfloat{\includegraphics[scale = 0.5]{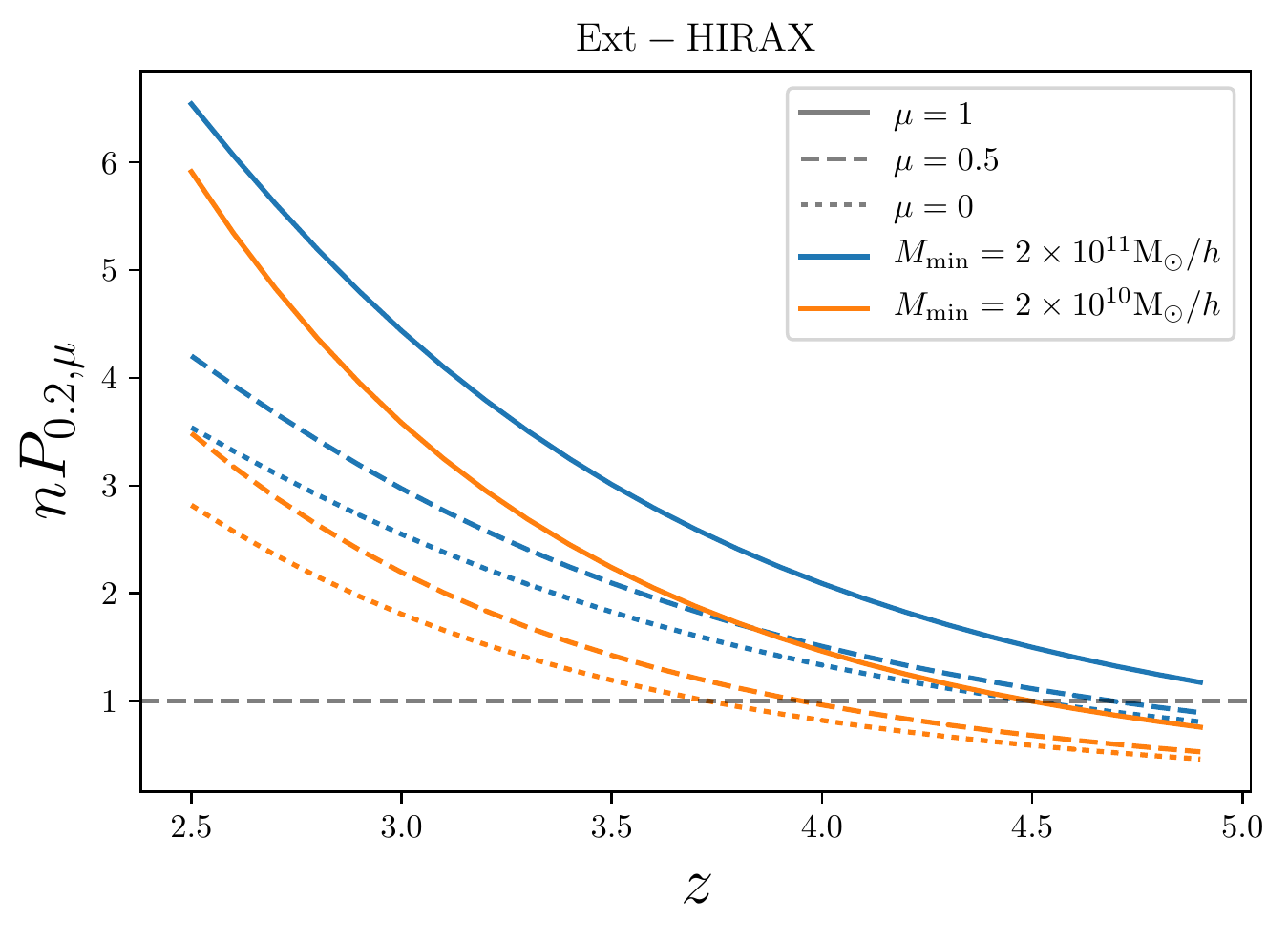}}\\
\subfloat{\includegraphics[scale = 0.5]{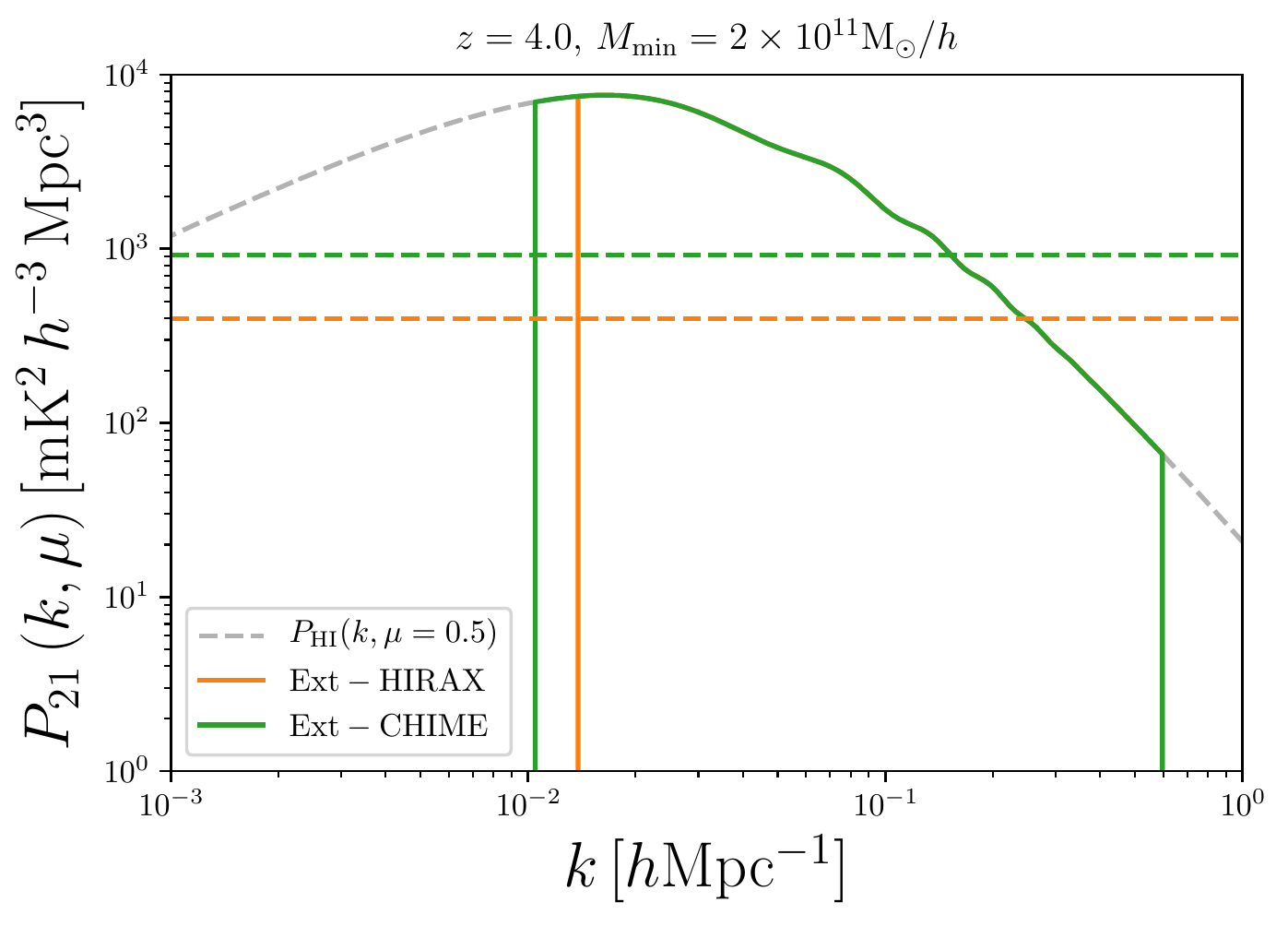}}
\subfloat{\includegraphics[scale = 0.5]{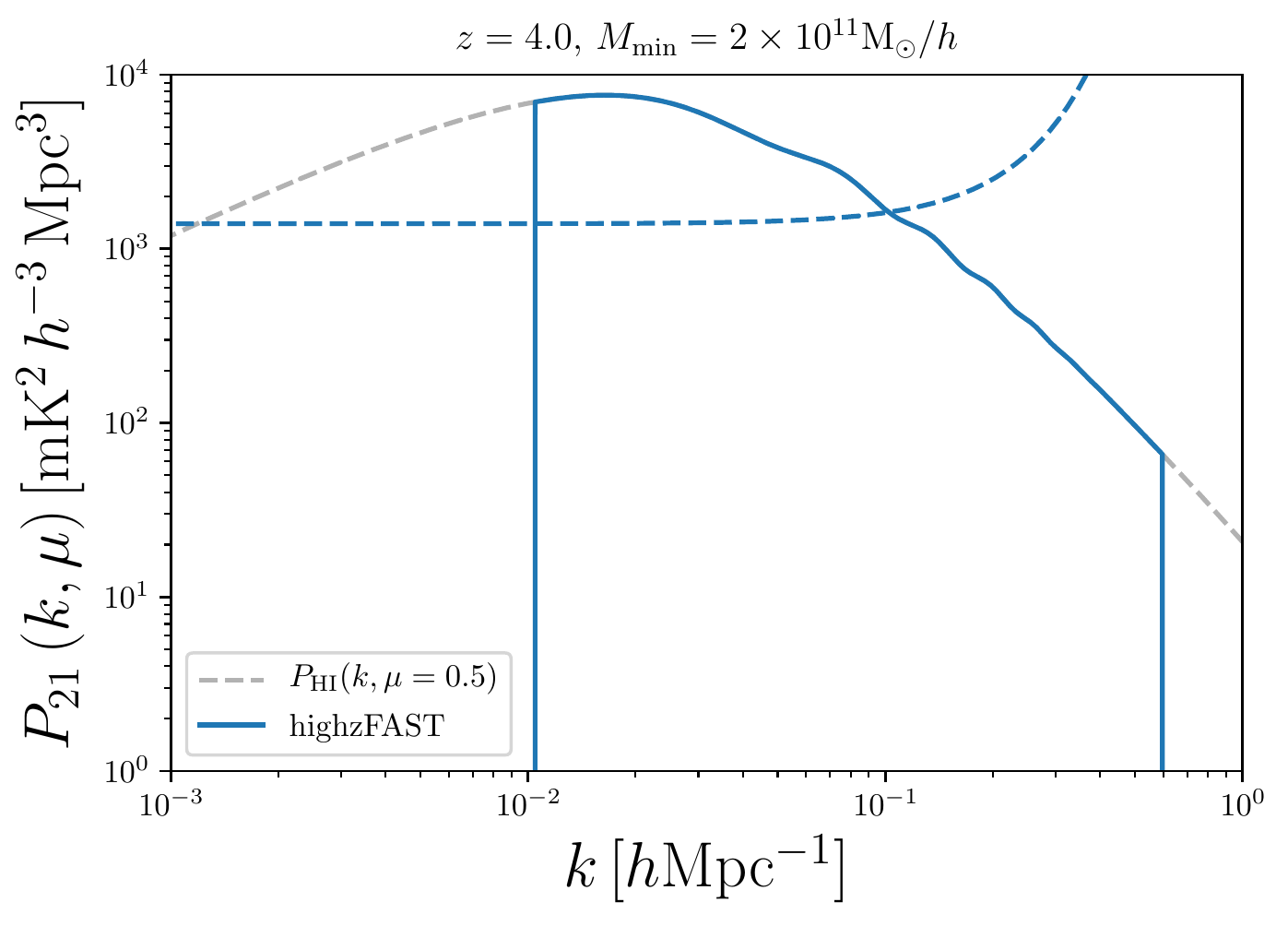}}
\caption{\textit{Top-left:} HI power spectrum at $\mu=0.5$ (solid lines) together with the total noise (dashed) and shot-noise (dotted) for Ext-HIRAX at $z=4$. The different colors show the results for different values of $M_{\rm min}=2\times10^{10}~h^{-1}M_\odot$ (orange) and $2\times10^{11}~h^{-1}M_\odot$ (blue). \textit{Top-right:} Signal-to-noise ratio at $k=0.2 h\mathrm{Mpc^{-1}}$ for different values of $\mu$, 1 (solid), 0.5 (dashed) and 0 (dotted), for the two models with different values of $M_{\rm min}$. \textit{Bottom-left:} Effects of the window function on the HI power spectrum  for $\mu=0.5$ for Ext-HIRAX (orange) and Ext-CHIME (green) at $z=4$. The dashed lines show the amplitude of the noise power spectrum. \textit{Bottom-right:} Same as before but for highzFAST.}
\label{fig:PHI_noises}
\end{figure}

The total noise power spectrum is therefore equal to
\be
\label{eq:Pnoise}
P_\mathrm{N}^\mathrm{tot}=P^\mathrm{th}_\mathrm{N}(z)+P_\mathrm{SN}(z)\overline{T}_b^2(z).
\ee
We plot the different noise contributions for a survey like Ext-HIRAX at $z=4$ in the top-left panel of Figure \ref{fig:PHI_noises}. In the top-right panel we show the expected signal-to-noise ratio for different $\mu$ values and different values of $M_\mathrm{min}$ in the case of Ext-HIRAX. Bottom panels show the total noise power spectra for both interferometers (left panel) and single-dish (right panel) in dashed lines at $z=4$ and fixed value of $M_\mathrm{min}$. As expected the thermal noise is the major source of uncertainties, especially going to high-redshift. We see that on cosmological scales smaller than $k\simeq 0.2 \kMpc$ all surveys we consider become noise dominated. In the upper left panel we plot the signal-to-noise at $k\simeq 0.2 \kMpc$ for different values of $\mu$
\be
nP_{0.2}(z) \equiv \frac{P_{21}(k,\mu;\,z)}{P_\mathrm{N}^\mathrm{tot}}.
\ee
Since the shot-noise is always much smaller than the thermal noise in the antennas, a higher $M_{\rm min}$, i.e.\ a higher HI bias, results in a higher signal to noise ratio.

\subsection{Fisher Matrix formalism}

Given the signal and noise models discussed in the previous section we can use the Fisher matrix formalism to forecast the constraints on a set of parameters of interest $\{p\}$ from a 21cm IM survey. The Fisher matrix for a single redshift bin is given by \cite{Tegmark}
\be \label{eq:Fisher}
F_{ij}=\frac{1}{8\pi^2}\int_{-1}^1d\mu\int k^2 dk \frac{\partial \ln P_{\rm 21}(k,\mu)}{\partial p_i}\frac{\partial \ln P_{\rm 21}(k,\mu)}{\partial p_j}V_\mathrm{eff}(k,\mu).
\ee
The effective volume is related to the comoving survey volume in a redshift bin $V_\mathrm{sur}$ by
\be
V_\mathrm{eff}(k,\mu)=V_\mathrm{sur}\left(\frac{P_\mathrm{21}(k,\mu)W(k,\mu)}{P_\mathrm{21}(k,\mu)W(k,\mu)+P_\mathrm{N}^\mathrm{tot}(k,\mu)}\right)^2.
\ee
The instrument response/window function $W(k,\mu)$ is included to account for the modes that are unavailable to a given instrument. The total noise $P_\mathrm{N}^\mathrm{tot}$ accounts for the instrument thermal noise and the shot-noise of the HI sources as defined in eq. \eqref{eq:Pnoise}. 

The total redshift window of the 21cm IM surveys considered here, $2.5<z<5$, is divided into smaller redshift bins, in each of them we calculate the Fisher matrix for a set of parameters. Under the assumption that each redshift bin is independent, the sum of the Fisher matrices in each redshift bin gives the total Fisher matrix. 

Having obtained the Fisher matrix for a given redshift bin or the total redshift window, the forecasted constraint on a given parameter $p_i$ marginalised over all the other parameters is related to the Fisher matrix by $\sigma(p_i)=\sqrt{[F^{-1}]_{ii}}$. We quote these numbers as the 1$\sigma$ constraints throughout the paper.

In observations one needs to convert angular coordinates and redshifts to wave vectors in Fourier space. That is done by assuming a fiducial cosmology, which could be different from the true underlying one. By using a cosmology different from the fiducial one, we introduce an additional artificial anisotropy in the clustering measurements. This effect, known as Alcock-Paczynski(AP) effect \cite{AP}, must be taken into account in the forecast. We include all the geometrical distortions arising from the assumption of a fiducial cosmology \cite{Ballinger, Beutler2013}. The fiducial wavenumber vector $\vec{k}$ is determined by two numbers, namely, the amplitude $k_f$ and by the cosine of the angle between the $\vec{k}$ and the line-of-sight, $\mu_f=k_{\parallel,f}/k_f$. The component along and transverse to the line-of-sight are related to the true values by $k_\parallel=k_{\parallel,f}(H/H_f)$ and $k_\perp=k_{\perp,f}(D_{A,f}/D_A)$. Furthermore the amplitudes of the wavenumber vectors $\vec{k}$ and $\vec{k}_f$ are related
\be
k^2=\left((1-\mu^2_f)\frac{D^2_{A,f}}{D^2_A} + \mu^2_f\frac{H^2}{H^2_f}\right)k^2_f.
\ee

We write our full signal power spectrum in a redshift bin centred at redshift $z$ as \cite{SeoEisenstein}:
\be
P_\mathrm{21}(k_f,\mu_f,z) = \overline{T}_b^2(z) \frac{D_A(z)_f^2 H(z)}{D_A(z)^2 H(z)_f}
\left(b_\mathrm{HI}(z) + f(k,z)\mu^2)^2 P(k,z\right).
\ee

We will also use another way of writing the full power spectrum, particularly in the case we are interested in the redshift-space distortions where an important parameter is $f\sigma_8$:
\be \label{eq:PHIkmu}
P_\mathrm{21}(k_f,\mu_f,z) = \overline{T}_b^2(z) \frac{D_A(z)_f^2 H(z)}{D_A(z)^2 H(z)_f}
(b_\mathrm{HI}\sigma_8(z) + f\sigma_8(z)\mu^2)^2 \frac{P(k,z)}{\sigma_\mathrm{8,f}^2}.
\ee

Within our model, the derivative of the HI power spectrum with respect to a generic parameter $X$ looks like

\begin{multline}
\frac{\partial \ln P_\mathrm{21}}{\partial X} = \frac{\partial \ln \overline{T}_b^2}{\partial X} 
+ \frac{\partial \ln D_A^{-2}}{\partial X} 
+ \frac{\partial \ln H}{\partial X}
+ \frac{\partial \ln P(k)}{\partial X} 
+ \frac{\partial \ln P(k)}{\partial k} \frac{\partial k}{\partial X} \\
+ \frac{2\mu^2}{b_\mathrm{HI}+f(k)\mu^2}
\left[
\frac{\partial f(k)}{\partial X} 
+ \frac{\partial f(k)}{\partial k} \frac{\partial k}{\partial X} + 2f(k)(1-\mu^2)
\left(\frac{\partial \ln H}{\partial X} +  \frac{\partial \ln D_A}{\partial X}\right)
\right],
\end{multline}
where 
\be
\frac{\partial k}{\partial X} = k\left[\mu^2\frac{\partial \ln H}{\partial X} - (1-\mu^2)\frac{\partial \ln D_A}{\partial X}\right].
\ee

The Fisher matrix formalism can also accommodate theoretical errors in the signal model, for instance due to poor knowledge of non-linearities, along the way described  in \cite{Audren,Baldauf}. We do not include such terms in our analysis, but in Section \ref{sec:1-loop} we will comment on the effect of marginalising over uncertainties in non-linear models using templates.

%%%%%%%%%%%%%%%%%%%%%%%%%%%%%%%%%%%%%%%%%%%%%%%%%%%%%%%%%%%%%%%%%
%%%%%%%%%%%%%%%%%%%%%%%%%%%%%%%%%%%%%%%%%%%%%%%%%%%%%%%%%%%%%%%%%
%%%%%%%%%%%%%%%%%%%%%%%%%%%%%%%%%%%%%%%%%%%%%%%%%%%%%%%%%%%%%%%%%
%%%%%%%%%%%%%%%%%%%%%%%%%%%%%%%%%%%%%%%%%%%%%%%%%%%%%%%%%%%%%%%%%

\section{Results from IM alone}

In this section we present the constraints on the value of the cosmological parameters that can be achieved by using 21cm data alone. We focus our analysis on the growth rate and on the BAO distance scale parameters.

%%%%%%%%%%%%%%%%%% GROWTH %%%%%%%%%%%%%%

\subsection{Growth of structures}

\label{sec:growth}
We will first consider the case in which we assume that the geometry is fixed, e.g.\ by CMB measurements, and we can fit for the amplitude of the power spectrum using two parameters only: $b\sigma_8(z)$ and $f\sigma_8(z)$. This choice is motivated by a comparison with the forecasts presented in \cite{DESInu,DESI2016,Amendola}, where BAO distance scale parameters have been held fixed when quoiting constraints on RSD. 
Following the discussion in Section \ref{sec:21cmsignal}, such a measurement in an IM survey is possible only with external information on $\Omega_{\rm HI}$ and $b_{\rm HI}$, that we therefore include as priors in our Fisher forecasts. We present results for 2\%  5\% and 10\% priors in each redshift bin, but it should be kept in mind that, in reality, these priors will be a function of redshift, worsening at higher $z$.
We show the outcome of the Fisher calculation for both a conservative choice of $k_{\rm max} = 0.2 \kMpc$ and for $k_{\rm max} = k_{\rm nl}$.

\begin{figure}[t!]
\centering
\subfloat{\includegraphics[scale = 0.5]{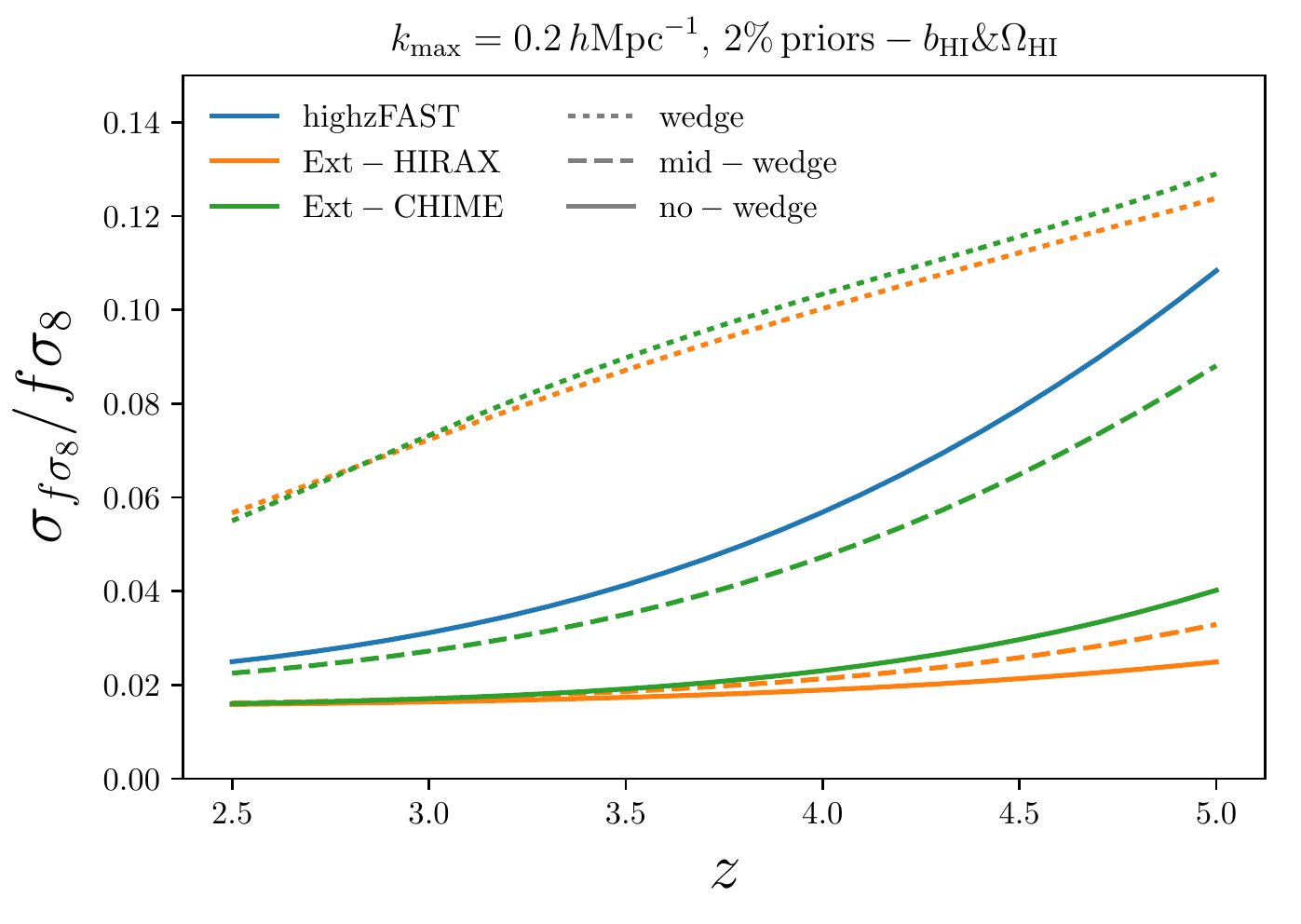}}
\subfloat{\includegraphics[scale = 0.5]{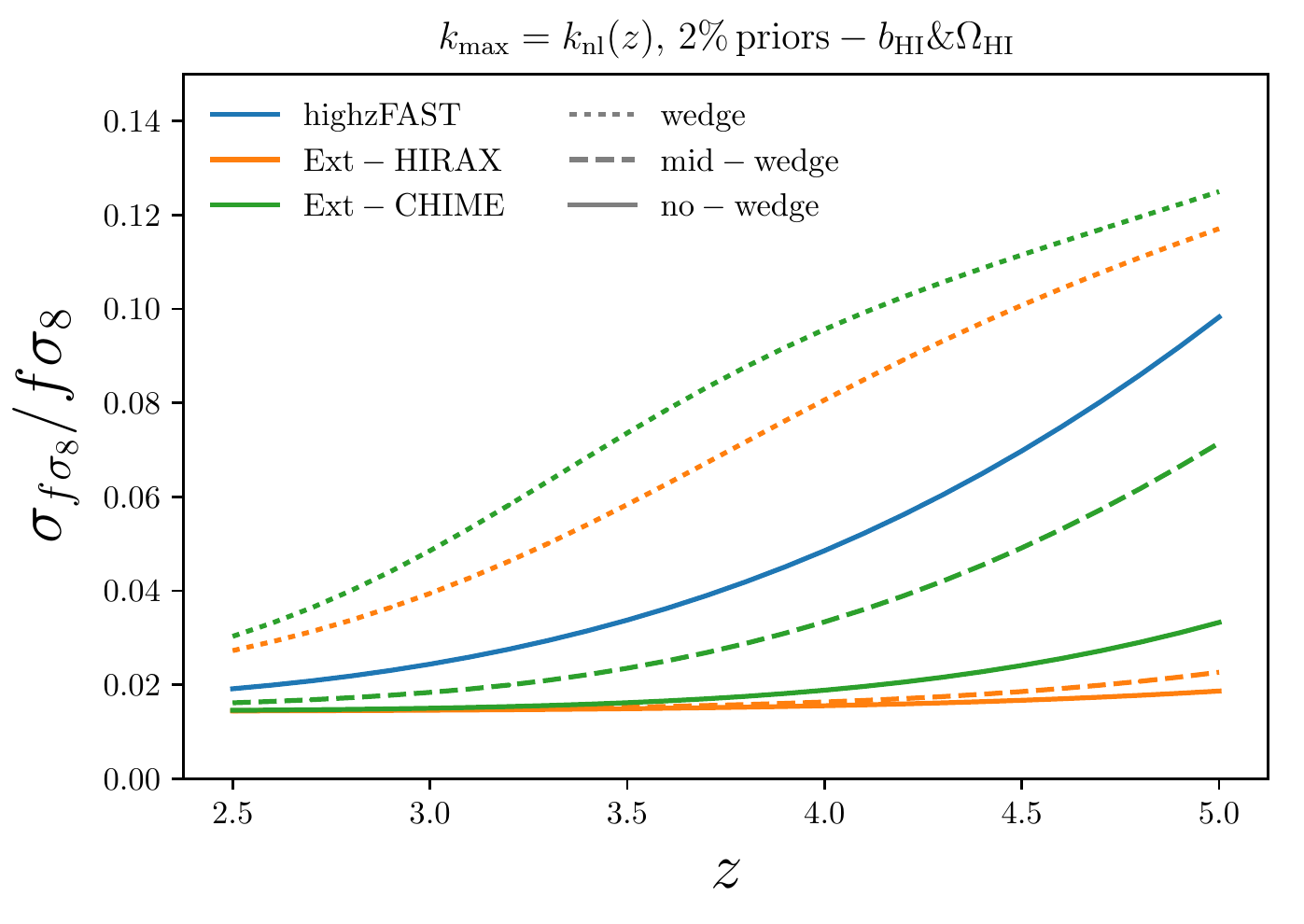}}\\
\subfloat{\includegraphics[scale = 0.5]{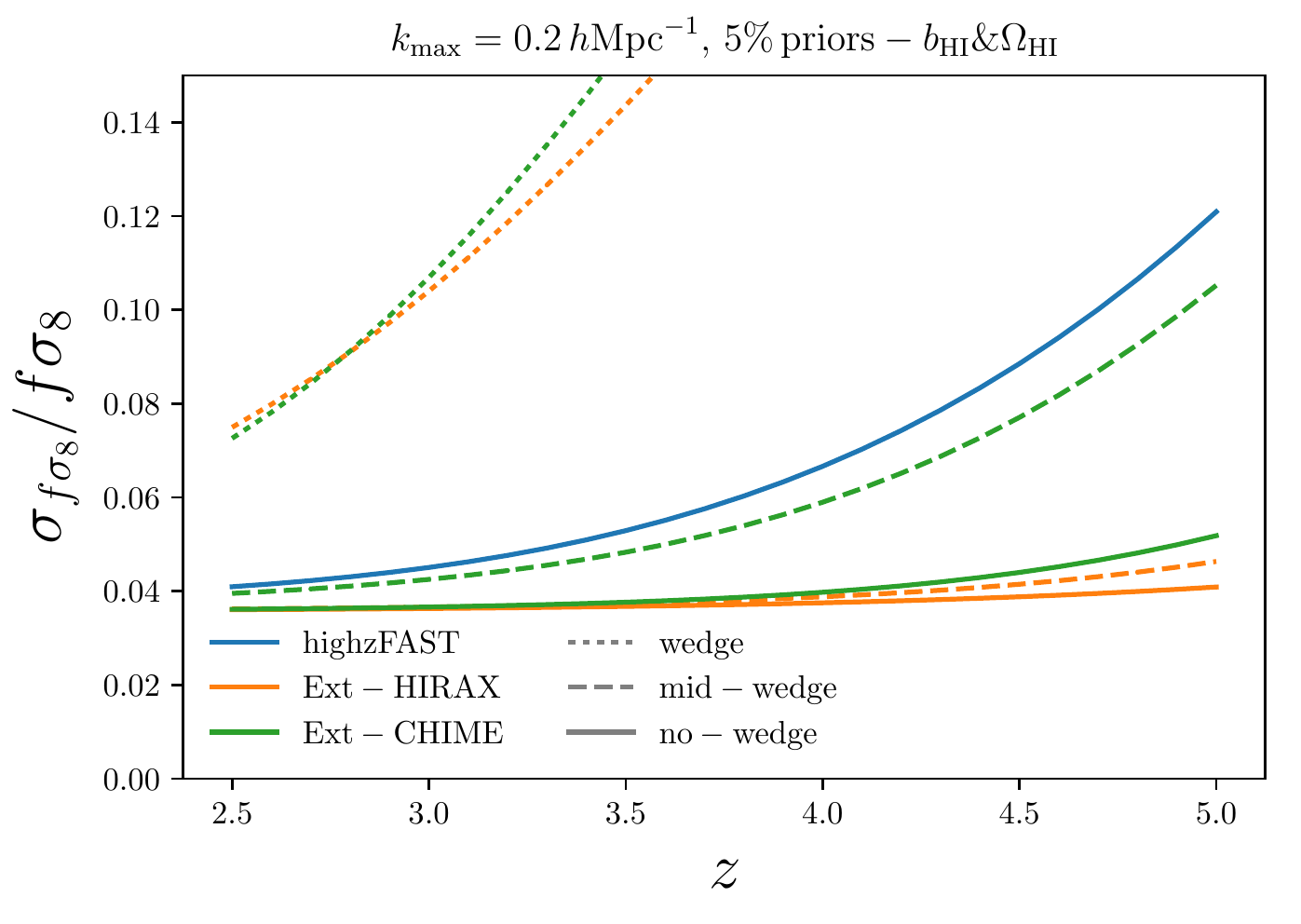}}
\subfloat{\includegraphics[scale = 0.5]{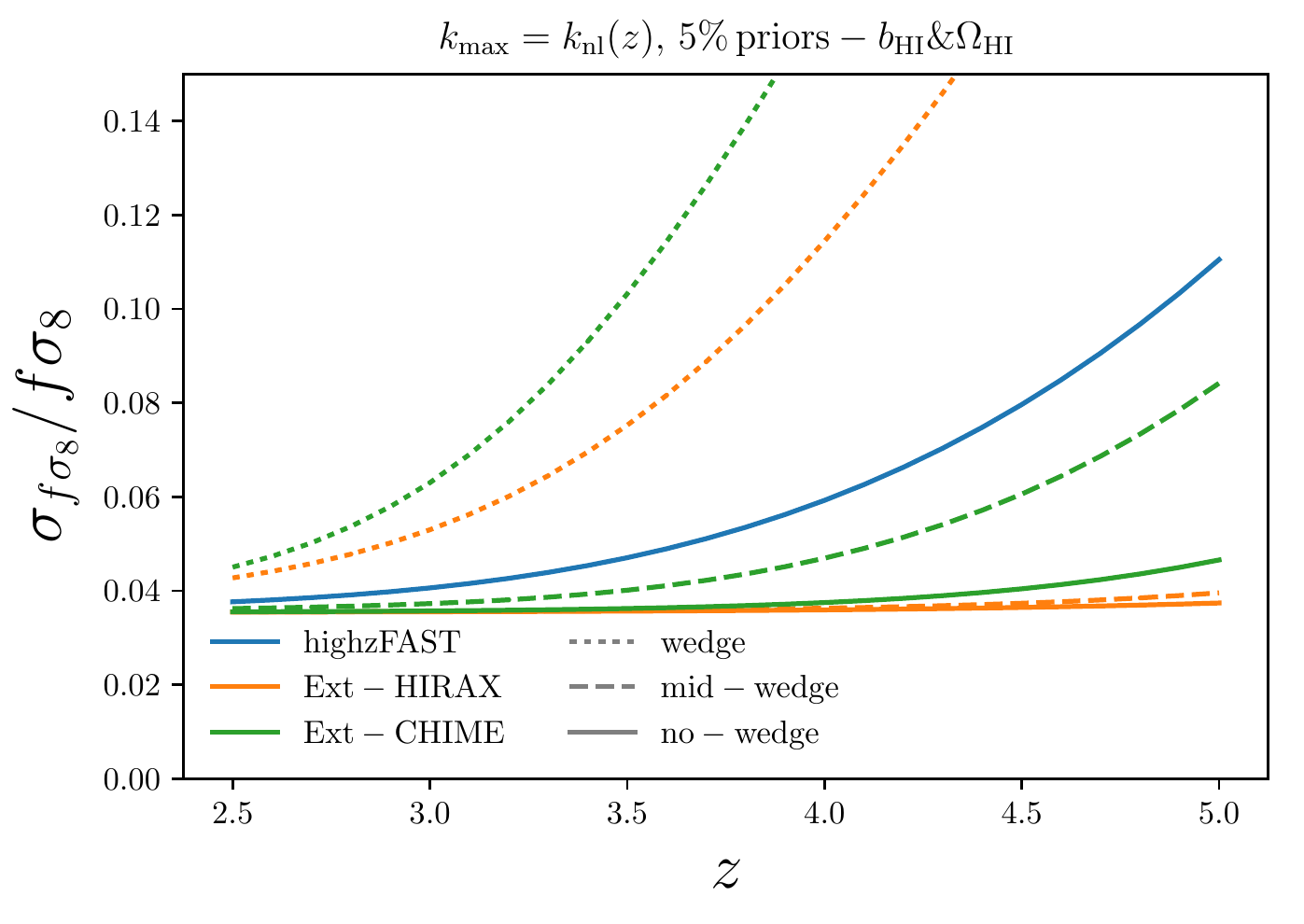}}\\
\subfloat{\includegraphics[scale = 0.5]{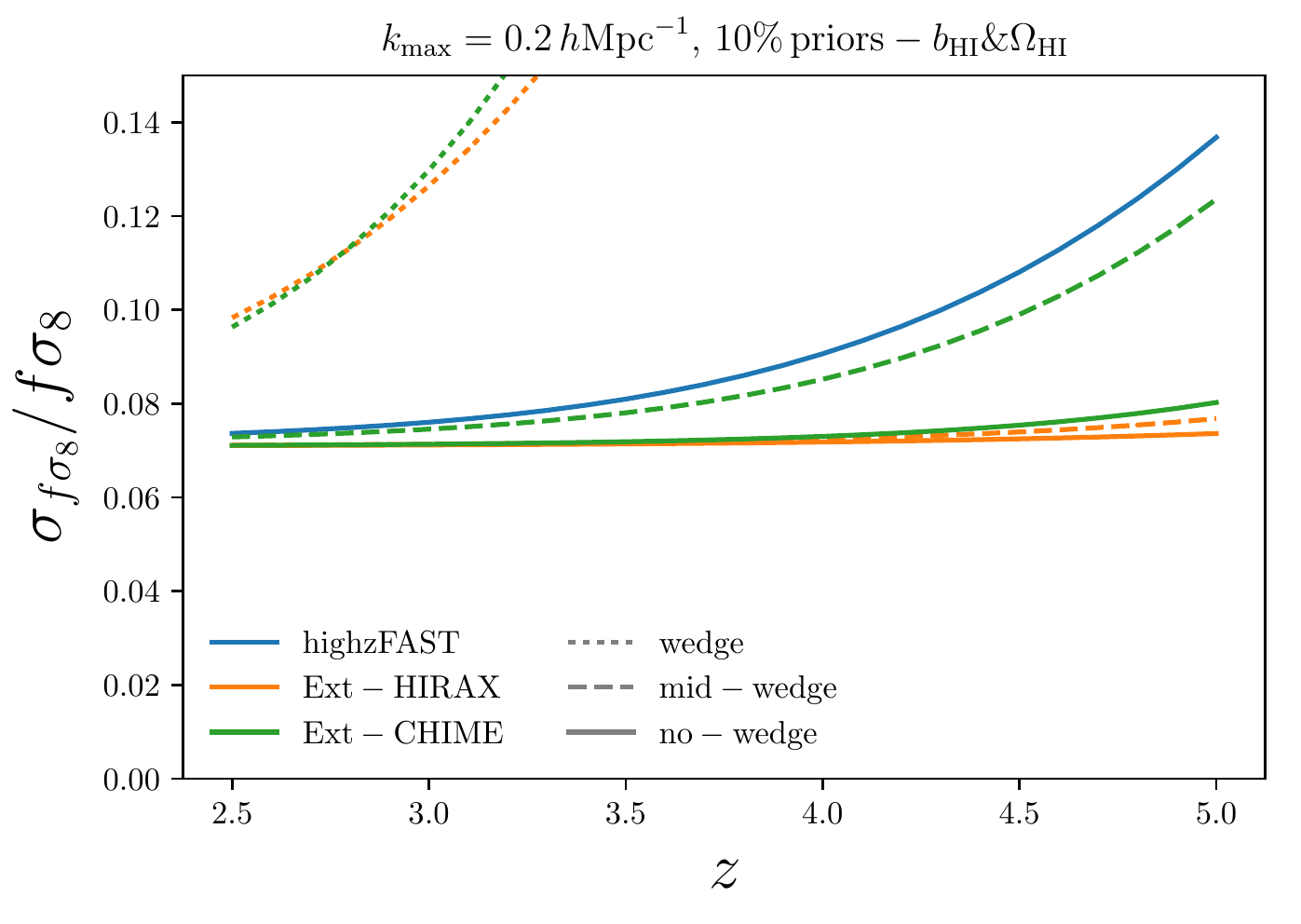}}
\subfloat{\includegraphics[scale = 0.5]{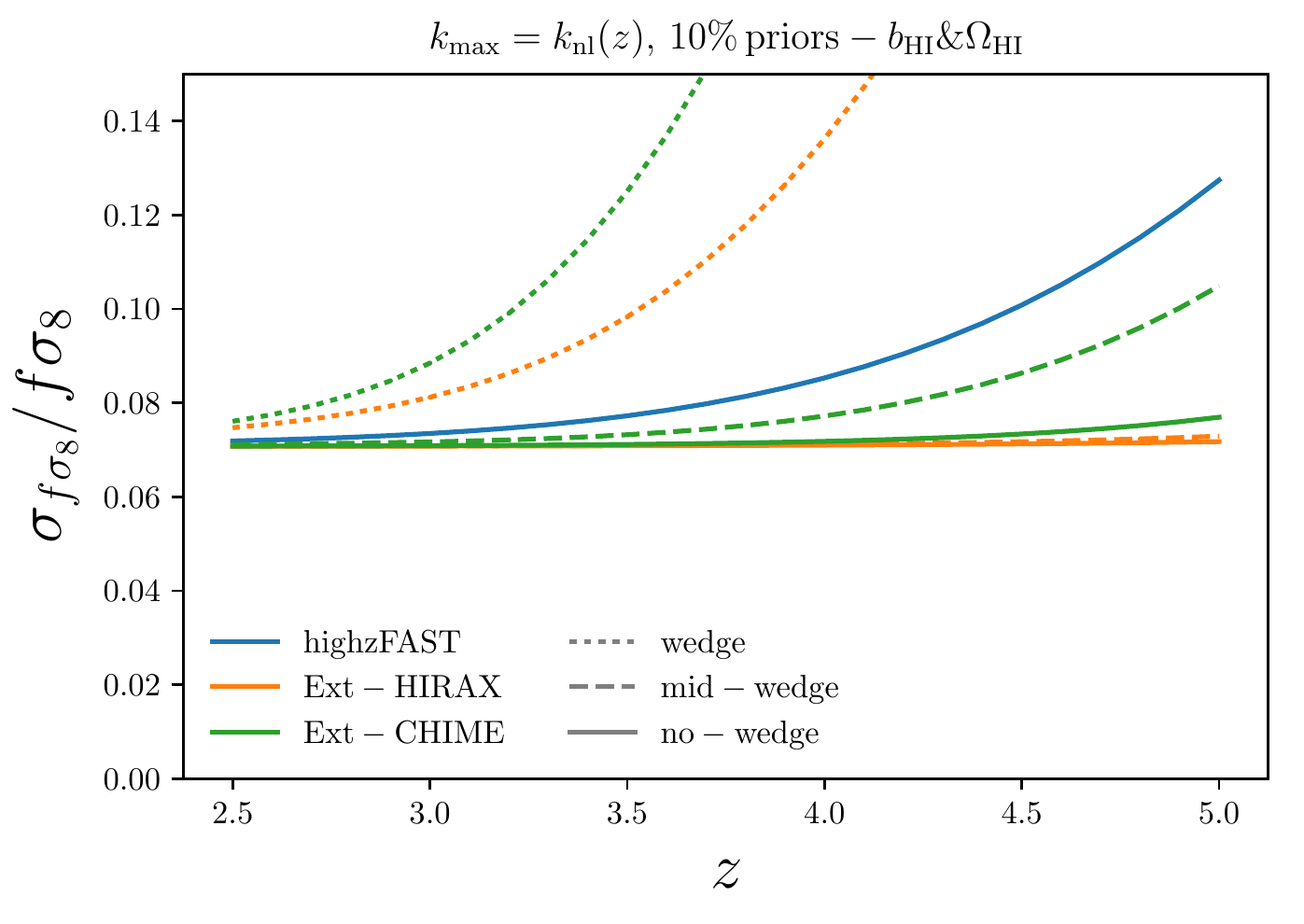}}\\
\caption{$1\sigma$ constraints on $f\sigma_8$ for highFAST (blue), Ext-HIRAX (orange) and Ext-CHIME (green) for different wedge configurations: no-wedge (solid), mid-wedge (dashed) and full wedge (dotted). Notice that the wedge only apply to interferometers. The different panels show the results for different assumptions of $k_{\rm max}$, $0.2~h{\rm Mpc}^{-1}$ (left column) and $k_{\rm nl}$ (right column), and for different priors on both $b_{\rm HI}$ and $\Omega_{\rm HI}$, $2\%$ (top row), $5\%$ (middle row) and $10\%$ (bottom row).}
\label{fig:fs8_comparison}
\end{figure}

Results are summarized in Figure \ref{fig:fs8_comparison}. Starting from the continuous lines, which show the no-wedge case for Ext-CHIME, Ext-HIRAX, and highzFAST, we see that in the absence of foreground contamination 21cm surveys can measure RSD parameters to exquisite precision. Interferometric experiments saturate the priors up to $z=3.5$, and are still able to measure $f \sigma_8$ with 4\% precision at $z=5$ in the case of aggresive priors.. In our  analysis Ext-HIRAX perfoms better that the other two hypothetical facilities, mostly because of a better angular resolution. Single dish experiments like highzFAST pay the price of low angular resolution, but they are unaffected by the wedge, so the blue line in Figure \ref{fig:fs8_comparison} can be considered the final result for auto-correlation experiments. 
If we now turn our attention to the case with partial wedge contamination, the mid-wedge case described in Section \ref{sec:wedge}, we see a degradation in the constraining power, with surveys with larger FOV yielding larger error bars as expected. For instance we find similar results comparing Ext-CHIME and highzFast.
If the wedge extends up to the horizon the picture changes dramatically and interferometer performs much worse than single dishes. 
As it can be noticed in the right and lower panels, the inclusion of all the modes down to the non-linear scale $k_{\rm nl}$ does not substantially change the results. This is a consequence of the very low S/N ratio on scales smaller than $k \simeq0.2 \kMpc$ (see Figure \ref{fig:PHI_noises}). Only in the case where we consider the full wedge, we find marginal improvement from the inclusion of smaller scales.

\begin{figure}[t!]
\centering
\subfloat{\includegraphics[scale = 0.5]{./figures/fs8/fs8_z_k02_5priors}}
\subfloat{\includegraphics[scale = 0.5]{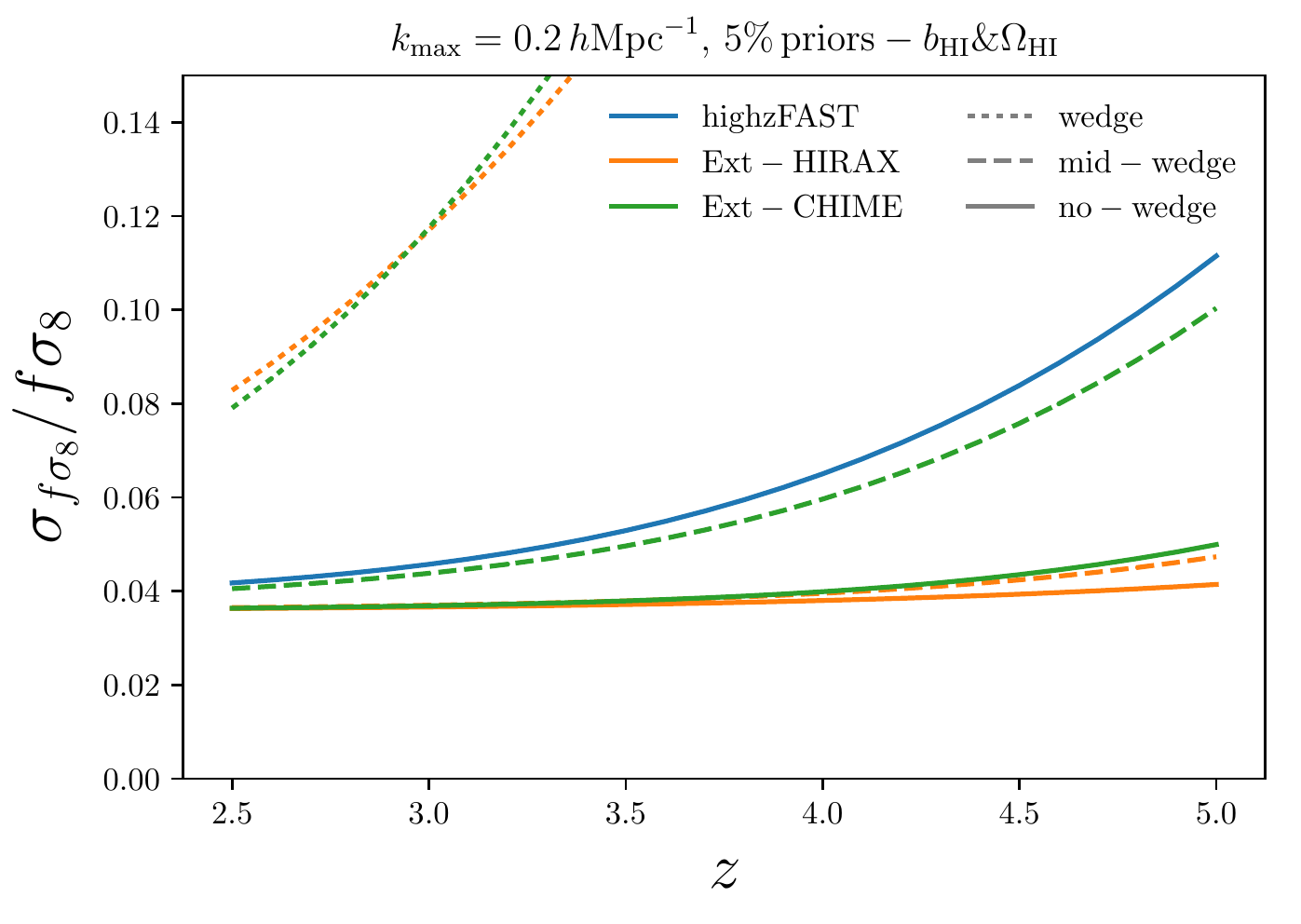}}\\
\caption{Dependence of the constraints on $f\sigma_8$ on the fiducial value of the HI bias. The left and right panels show the results when the value of $M_{\rm min}$, a parameter that controls the cut-off mass in the $M_{\rm HI}(M,z)$ function and therefore the amplitude of the HI bias, is set to $2\times10^{10} M_\odot/h$ and $2\times10^{11} M_\odot/h$, respectively. In both panels we consider a $5\%$ prior on the value of both $\Omega_{\rm HI}$ and $b_{\rm HI}$. Different colors and line types represent the different instruments and assumptions about the wedge (see legend), correspondingly.}
\label{fig:fs8_comparison_diff_Mmin}
\end{figure}

As discussed in Sec. \ref{sec:21cmsignal}, the value of the HI bias as a function of redshift is quite uncertain, reflecting our poor knowledge of what halos host neutral hydrogen. Figure \ref{fig:fs8_comparison_diff_Mmin} shows the dependence of our results on the HI bias. We do that by assuming two different values for $M_{\rm min}$: $M_{\rm min} =2\times 10^{10} \Msun/h$ (left panel) and $M_{\rm min} =2\times 10^{11} \Msun/h$ (right panel). The numbers we obtain are very similar in most cases, with the exception of the full wedge scenario, indicating our forecasts should be robust to different choices of fiducial astrophysical parameters.
For completeness, the numbers with the errors on $f\sigma_8$ for Ext-HIRAX and various considerations using $5\%$ priors on the values of both $\Omega_{\rm HI}$ and $b_{\rm HI}$ are shown in Table \ref{table:Table_fs8}.

\begin{figure}[t!]
\centering
\includegraphics[width=0.8\textwidth]{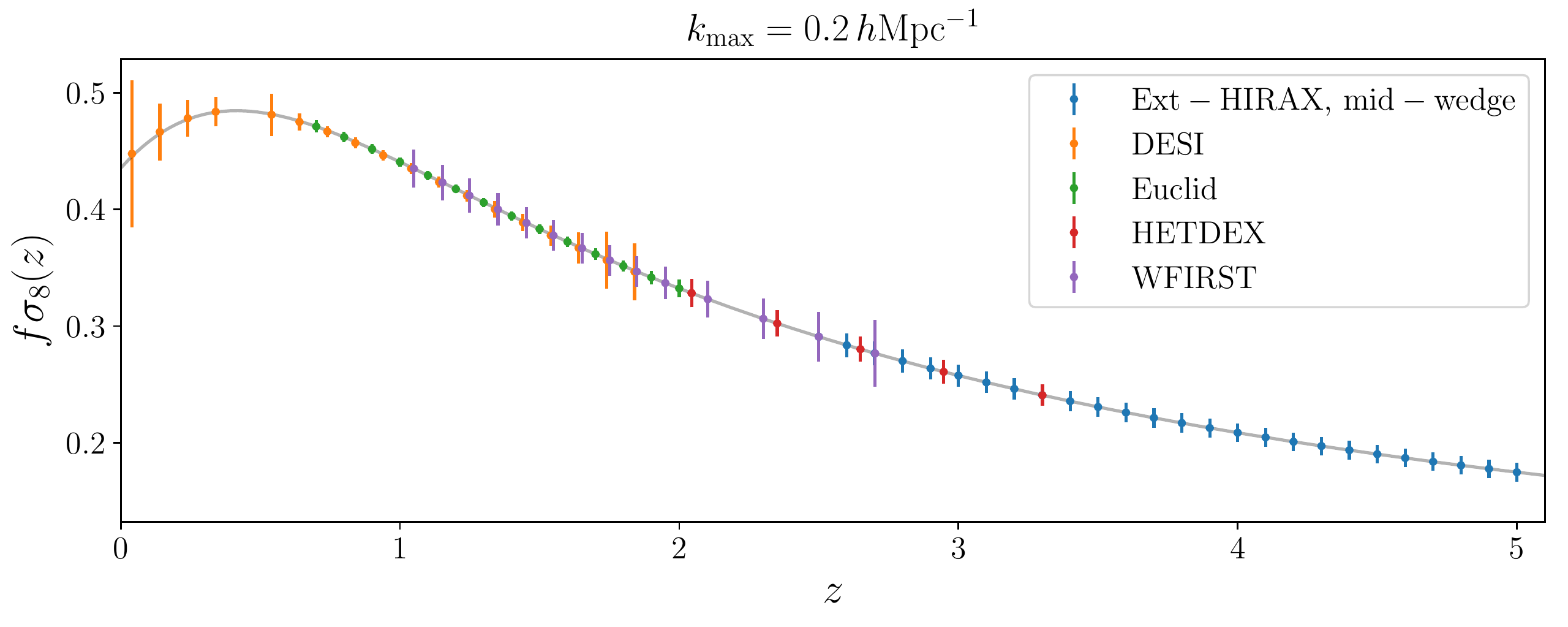}
\caption{The evolution of $f\sigma_8$ and the predicted constraints as a function of redshift using Ext-HIRAX considering mid wedge configuration with $5\%$ priors on both $\Omega_{\rm HI}$ and $b_{\rm HI}$ (blue). For comparison we also show predicted measurements (see text for the references) from upcoming galaxy surveys: DESI (orange, data points shifted by -0.01 in redshift), Euclid (green), HETDEX (red) and WFIRST (purple).}
\label{fig:fs8_money_plot}
\end{figure}

To better illustrate the synergies of a 21cm IM probe with galaxy surveys, in Figure \ref{fig:fs8_money_plot} we plot, as a function of redshift, the forecasted constraints on $f \sigma_8$ from: DESI (Tables 2.3 and 2.5 from \cite{DESI2016}), Euclid (Table 1.4 from \cite{Amendola}, reference case), HETDEX (Figure 6. from \cite{BullSKAGR}), WFIRST (Figure C.6 from \cite{WFIRST}) and Ext-HIRAX for the mid-wedge case (Figure \ref{fig:fs8_comparison}). We emphasise that the one should not directly compare the precision a survey could possibly achieve in the Figure \ref{fig:fs8_money_plot}, since different analysis have different underlying assumptions, but the redshift coverage. 21 cm surveys could be able to fill the gap almost to the end of reionization, in a redshift range difficult to reach for conventional surveys. As a reference we used the case of 5\% priors and partial cleaning of the wedge.

%%%%%%%%%%%%%%%%%%%
%%%%%%%%%%%%%%%%%%%
\subsection{BAO distance scale parameters}
We now turn our attention to the BAO distance scale parameters. This case is similar to a real analysis if the modelling of the broadband power spectrum has introduced enough nuisance parameters that only an amplitude can be measured with enough precision. In this case any constraint on $f \sigma_8$ makes sense only in combination with CMB data.

\begin{figure}[t!]
\centering
%\subfloat{\includegraphics[scale = 0.5]{./figures/BAO_fs8/Ext-HIRAX_fs8_z_2priors_rsdonly}}
%\subfloat{\includegraphics[scale = 0.5]{./figures/BAO_fs8/Ext-HIRAX_fs8_z_5priors}}\\
\subfloat{\includegraphics[width = 0.475\textwidth]{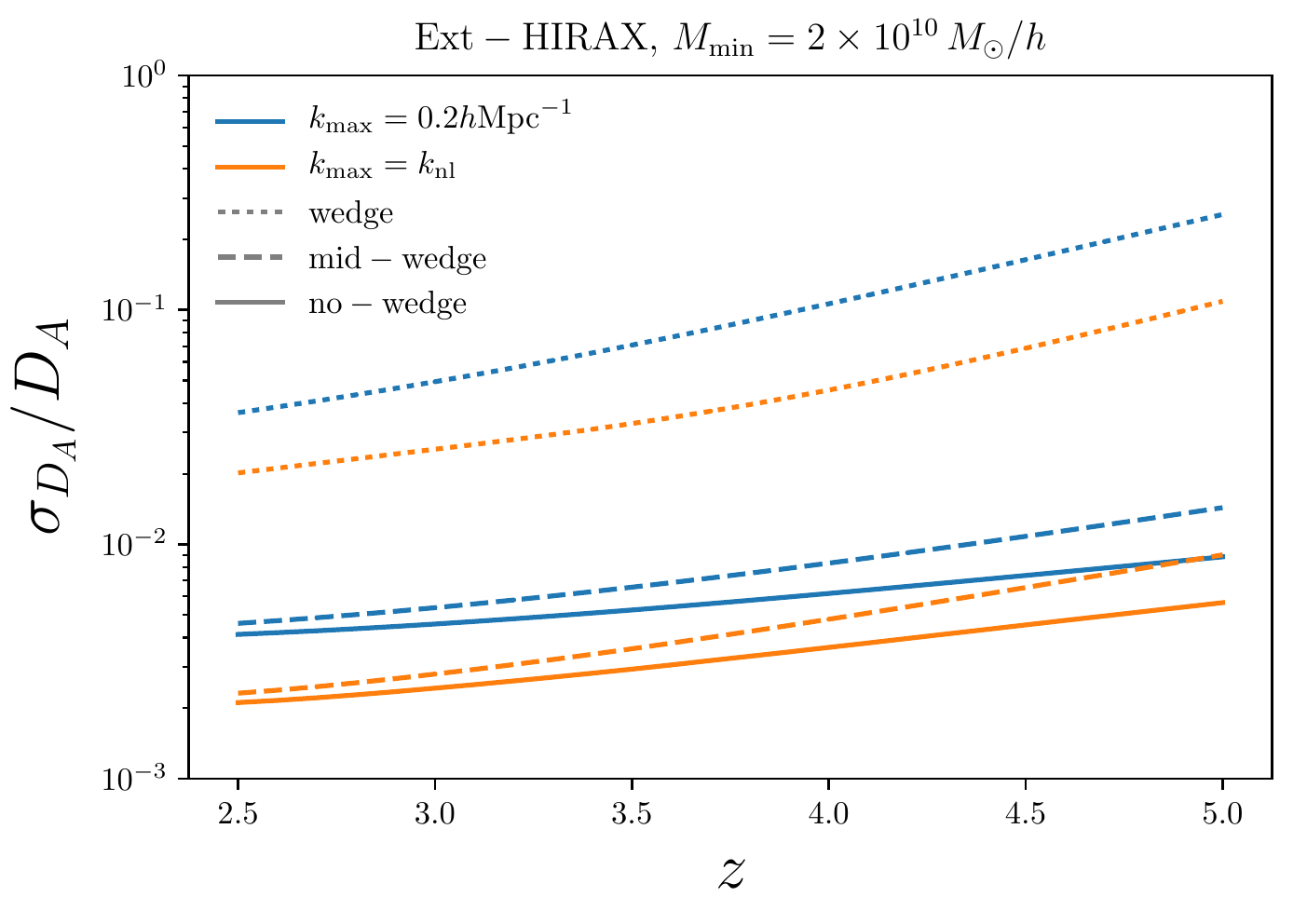}}
\subfloat{\includegraphics[width = 0.475\textwidth]{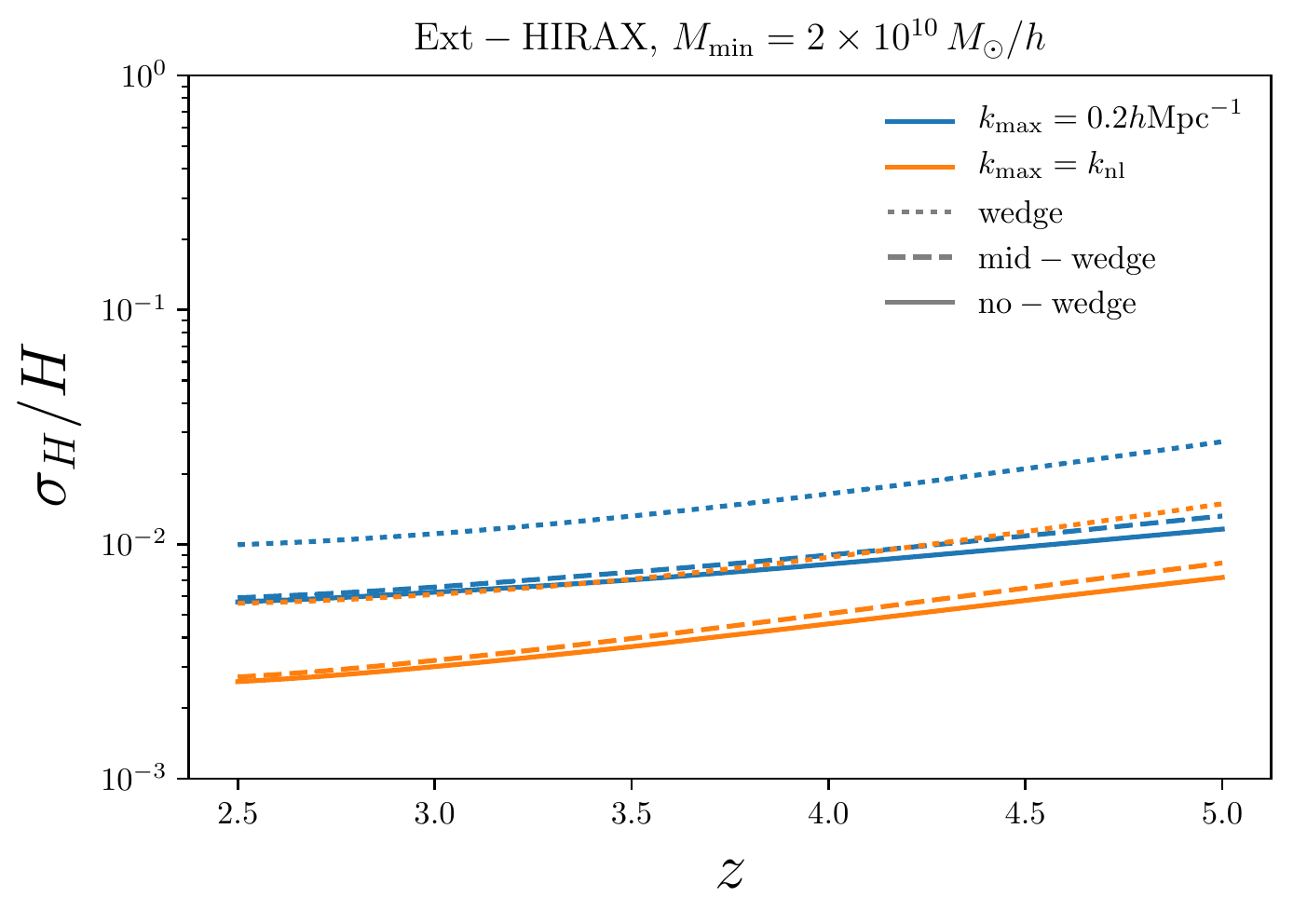}}\\
\caption{$1\sigma$ constraints on the angular diameter distance (left) and Hubble function (right) from the Ext-HIRAX setup. The forecasts have been derived for two different values of $k_{\rm max} = 0.2~h{\rm Mpc}^{-1}$ (blue) and $k_{\rm nl}$ (orange). The solid, dashed and dotted lines represent different assumptions over the wedge (see legend). The fiducial value of the HI bias is computed assuming $M_{\rm min}=2\times10^{10} M_\odot/h$.}
\label{fig:BAO_fs8_Ext-HIRAX}
\end{figure}

\begin{figure}[t!]
\centering
\subfloat{\includegraphics[width = 0.475\textwidth]{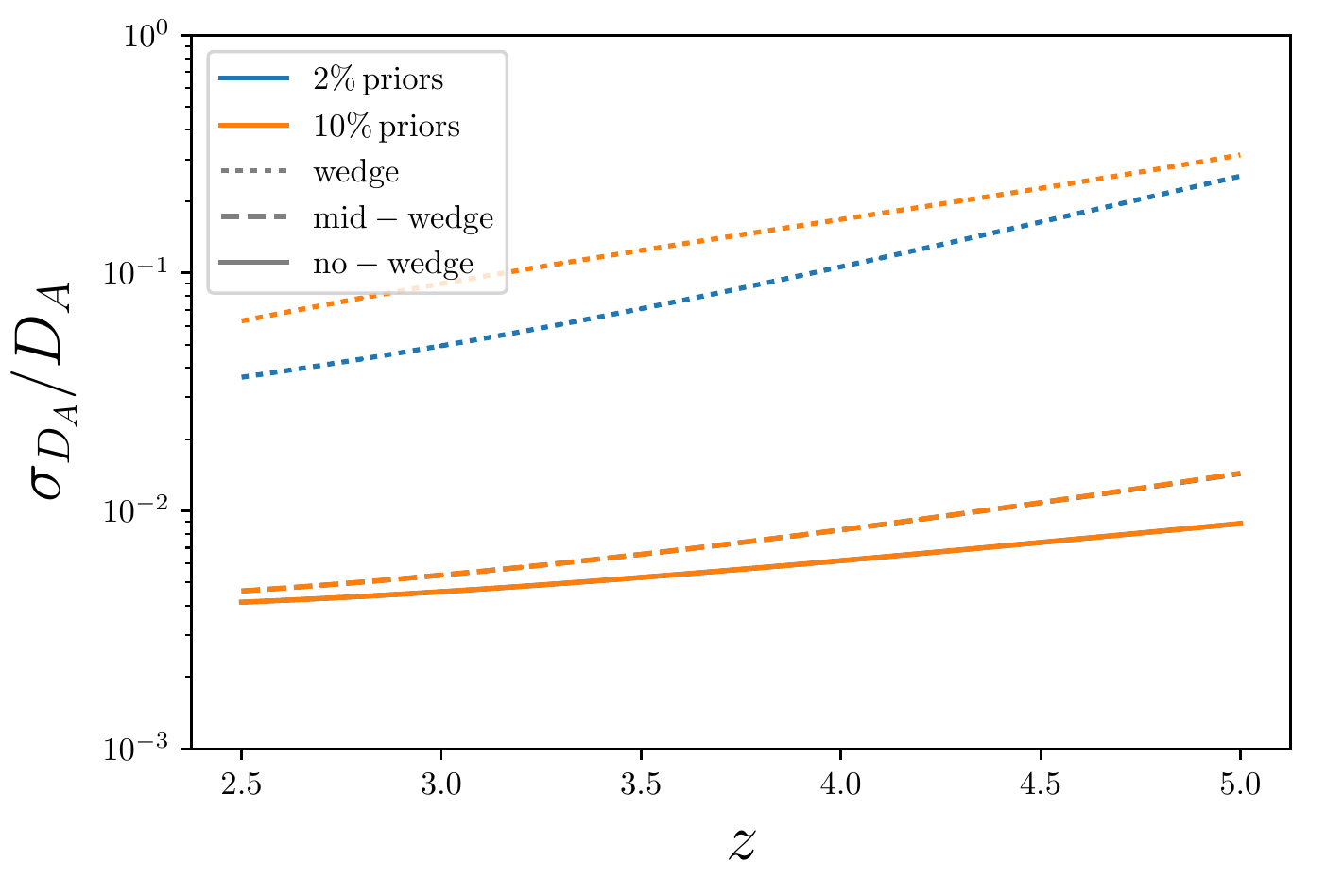}}
\subfloat{\includegraphics[width = 0.475\textwidth]{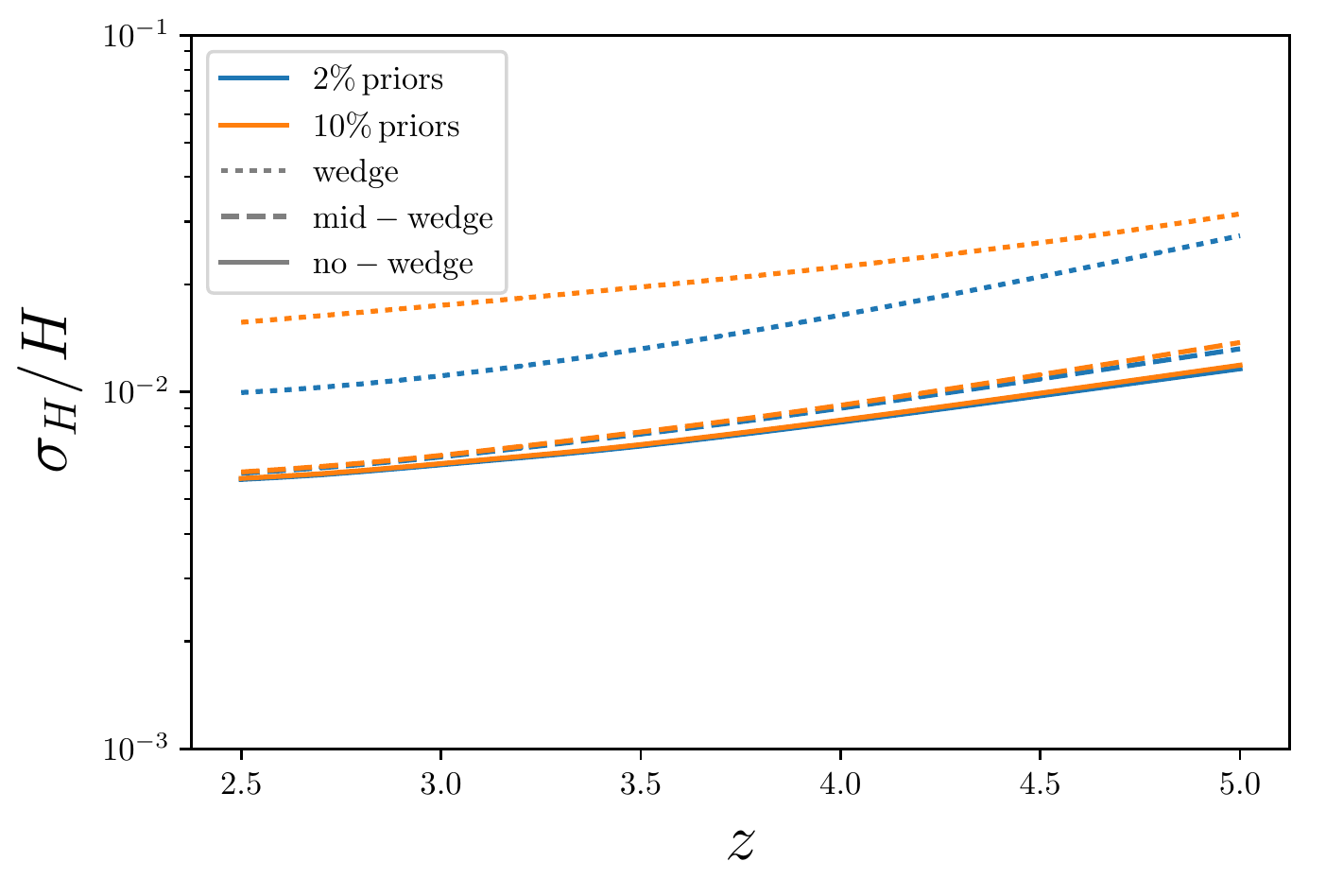}}\\
\caption{$1\sigma$ constraints on the angular diameter distance (left) and Hubble function (right) from the Ext-HIRAX setup assuming different prior knowledge on both $b_{\rm HI}$ and $\Omega_{\rm HI}$: $2\%$ (blue) and $10\%$ (orange). The forecasts have been derived for $k_{\rm max}$=$0.2~h{\rm Mpc}^{-1}$. The solid, dashed and dotted lines represent different assumptions over the wedge (see legend). The fiducial value of the HI bias is computed assuming $M_{\rm min}=2\times10^{10} M_\odot/h$. }
\label{fig:BAO_fs8_Ext-HIRAX_priors}
\end{figure}

Results are shown in Figure \ref{fig:BAO_fs8_Ext-HIRAX}. We find that 21cm surveys can provide very accurate measurements of both the Hubble parameter and the angular diameter distance. For most cases, the geometry of the Universe can be constrained at the sub \% precision. Similar to the analysis of the growth of structure in the previous section, the presence of the wedge suppresses the amount of information that can be extracted. The Hubble parameter however can still be measured quite accurately. This is expected since purely radial modes are the only ones surviving the wedge.

The precision at which we can measure the BAO distance scale parameters depends weakly on the prior on astrophysical parameters in the cases of low wedge coverage, while the effect is more pronounced in the full wedge case. We show the 1$\sigma$ constraints on $H(z)$ and $D_A(z)$ dependance on the prior on astrophysical parameters in Figure \ref{fig:BAO_fs8_Ext-HIRAX_priors} in the case of Ext-HIRAX.

\begin{figure}[t!]
\centering
\subfloat{\includegraphics[scale = 0.5]{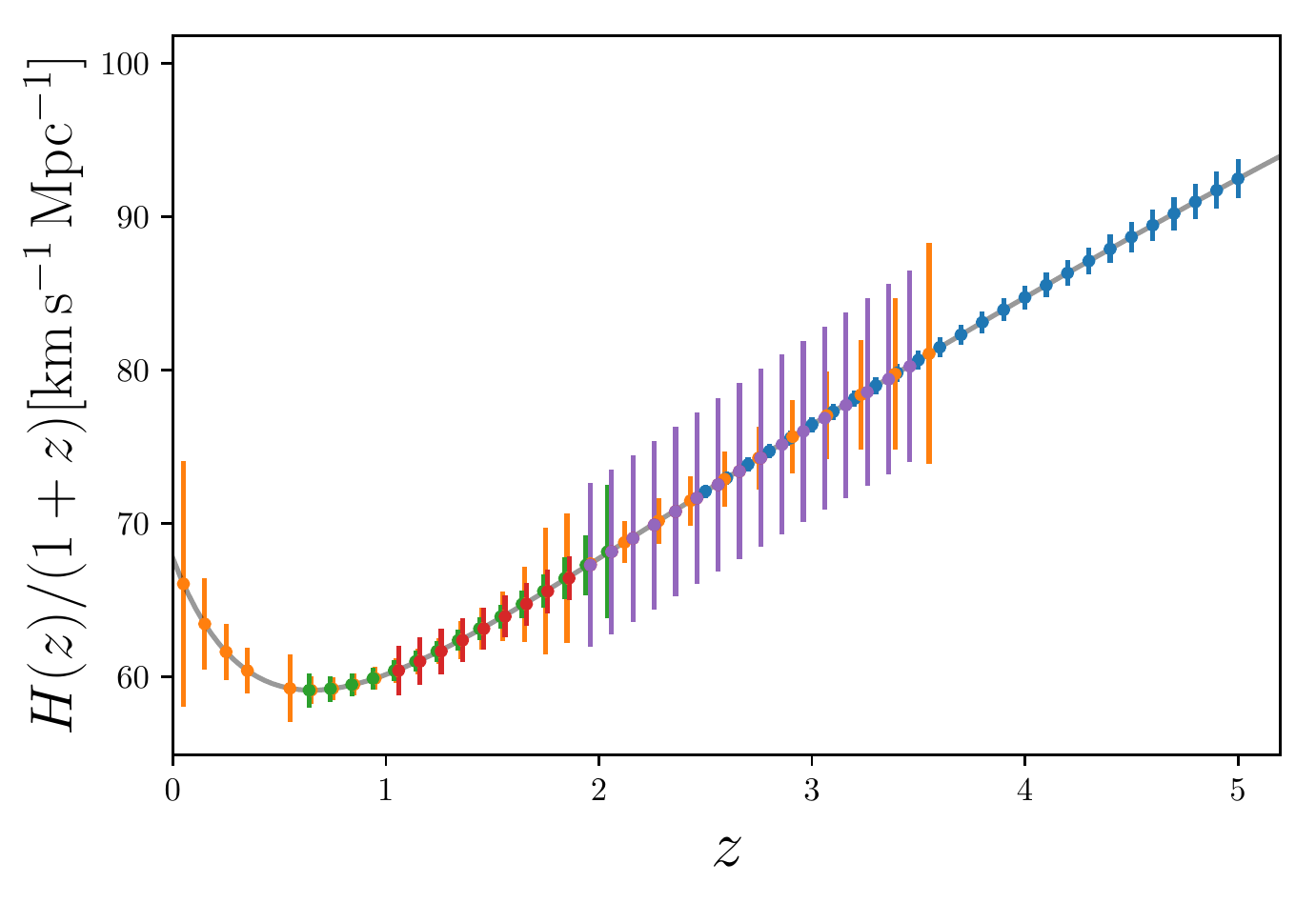}}
\subfloat{\includegraphics[scale = 0.5]{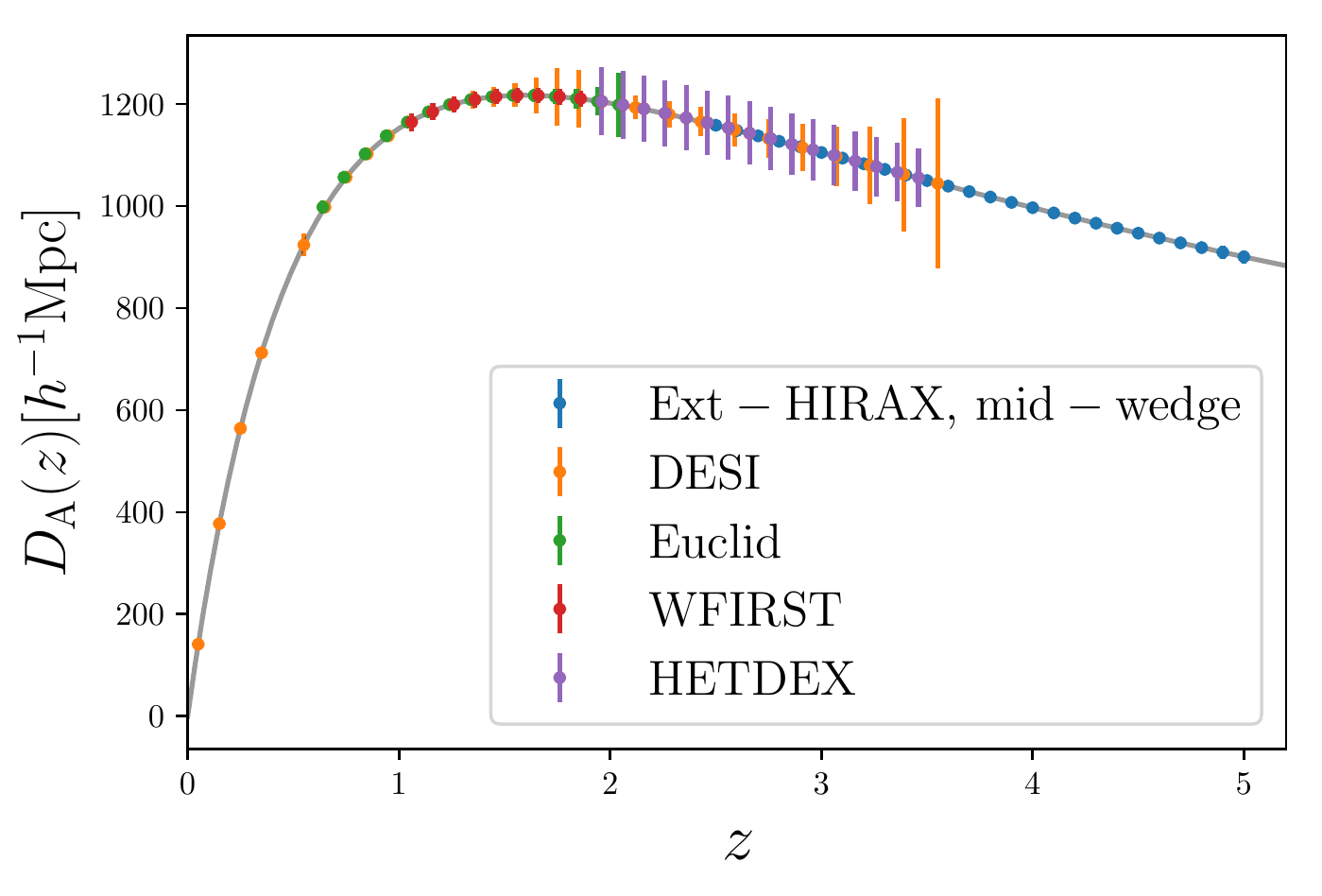}}
\caption{The evolution of the Hubble function $H(z)$ (left panel) and the angular diameter distance $D_A(z)$ (right panel) with the predicted constraints as a function of redshift using Ext-HIRAX considering mid wedge configuration with $5\%$ priors on both $\Omega_{\rm HI}$ and $b_{\rm HI}$ (blue). For comparison we also show predicted measurements (see text for the references) from upcoming galaxy surveys: DESI (orange, Euclid (green), HETDEX (red) and WFIRST (purple). Quoted data points have been shifted in redshift by $\pm$0.01 in order to avoid overlapping.}
\label{fig:H_DA_money_plot}
\end{figure}

Similar to figure \ref{fig:fs8_money_plot}, in Figure \ref{fig:H_DA_money_plot} we plot the Hubble function $H(z)$ and the angular diameter distance $D_A(z)$ projected measurements from low to high-redshift combining different probes. The forecasted constraints for DESI are taken from \cite{DESI2016}, while the numbers for other galaxy surveys -- Euclid, HETDEX and WFIRST, have all been taken from the corresponding tables in \cite{DESInu}. We also show the forecasted constraints from Ext-HIRAX for the mid-wedge case and assuming $5\%$ priors on $b_{\rm HI}$ and $\Omega_{\rm HI}$. Once again it is clear the contribution of a 21cm survey in filling up the whole redshift range. In Table \ref{table:Table_H_DA} we list the constraints for both $H(z)$ and $D_A$ as a function of redshift for different foreground configurations of Ext-HIRAX, fixing the $5\%$ priors on $b_{\rm HI}$ and $\Omega_{\rm HI}$.

\section{Extension to $\Lambda$CDM: Results from probe combination}
\label{sec:probecombination}
Cosmological constraints from a single survey usually are limited by the presence of degeneracies between parameters, which can be partially broken by combining different datasets. The purpose of this section is to study the possible benefits of combining data from CMB, galaxy surveys data at low-redshift and 21cm data at high-redshift.
Since the three probes we consider do not overlap in the redshift range, the Fisher forecasts become very simple as the covariance is diagonal and the different Fisher matrices can be added up.

We include the current constraints on the cosmological parameters measured from Planck 2015 \cite{Planck2015} in combination with BAO measurements from BOSS at redshift $z \simeq 0.3$ and $z\simeq 0.5$ \cite{Anderson2014}. In order to obtain the corresponding Fisher matrix we use the publicly available MCMC chains\footnote{We use {\tt base\_mnu\_plikHM\_TT\_lowTEB\_BAO} chains for the study on the sum of neutrino masses $\Sigma m_\nu$ and {\tt base\_nnu\_plikHM\_TT\_lowTEB\_BAO} chains for the study of the number of relativistic species $N_\mathrm{eff}$.} which are based on Planck temperature (TT) $+$ low TEB polarisation $+$ external BAO constraints. We then compute the covariance matrix for the parameters we consider and invert it.

Additionally, we use the Fisher matrix calculated for a generic CMB Stage 4 experiment\footnote{We thank An\v ze Slosar for providing us the Fisher matrix of CMB S-4 made by Joel Meyers.}. We refer the reader to the CMB S4 Science Book \cite{CMBS4} for more details about next generation of CMB experiments. For consistency we remove Planck information if CMB S-4 information is used. 

We also separately perform Fisher forecasts for future spectroscopic galaxy redshift surveys such as Euclid \cite{Euclid} or DESI \cite{DESInu}, rather than relying on existing results. This choice has been made in order to avoid the final constraints to be affected by different modeling assumptions in different analysis. It is also motivated by the need for the full Fisher matrix rather than the marginalized error on each individual parameter.

To model the galaxy power spectrum, we employ a similar expression to eq. \eqref{eq:PHIkmu}, but without the brightness temperature factor and using the galaxy bias instead of the HI bias. The effect of non-linearities on the galaxy power spectrum is larger at lower redshifts and effectively damps the BAO oscillations in the power spectrum. We use the following description to model this effect in redshift-space \cite{SeoEisenstein2007}:
\be
\label{eq:damping}
P^\mathrm{RSD}_\mathrm{nl}(k,\mu)\propto P^\mathrm{RSD}_\mathrm{lin}(k,\mu)\exp{\left(-\frac{k^2_\perp\Sigma^2_\perp}{2} - \frac{k^2_\parallel\Sigma^2_\parallel}{2}\right)},
\ee
where the term $P^\mathrm{RSD}_\mathrm{lin}(k,\mu)$ contains the bias and linear redshift-space distortions terms, while $\Sigma_\perp$ and $\Sigma_\parallel$ are the non-linear damping scales in the transverse and the radial direction, respectively. We compute the damping scales using $\Sigma_\perp=9.4(\sigma_{8}(z)/0.9)\,h^{-1}\mathrm{Mpc}$ and $\Sigma_\parallel=(1+f(z))\Sigma_\perp$\cite{DESInu}. This effect is particularly important for the BAO measurements at low-redshifts since it deteriorates the constraints on $D_A(z)$ and $H(z)$. In our forecasts for the massive neutrinos we will not assume the BAO reconstruction (see Section \ref{sec:21cmsignal}) has been applied and we will use the full values of $\Sigma_\perp$ and $\Sigma_\parallel$, while in our forecasts on the $N_\mathrm{eff}$ we do assume BAO reconstruction and half the damping scales accordingly. The exponential factor in eq. \eqref{eq:damping} is taken outside the logarithmic derivatives in the Fisher matrix since we do not want to infer any distance information from these factors. In practice, the exponential only degrades the effective volume.

For Euclid, we assume an area equal to $15000\,\deg^2$ covering the redshift range $0.65<z<2.05$. We take the fiducial galaxy bias $b(z)=\sqrt{1+z}$ and treat it as a free parameter that we marginalise over in each redshift bin, $\Delta z=0.1$. The only noise term in the power spectrum we consider is the shot-noise coming from the discreetness of galaxy sample and equal to the inverse number density of galaxies: $P_\mathrm{N}^\mathrm{shot}=\bar{n}^{-1}\,[h^{-3} \mathrm{Mpc}^3]$. We compute the $\bar{n}$ using the data in Table 1.3 (reference case) of \cite{Amendola}. We will show the results for two different values of the maximum wavenumber: Euclid02 for which we take $k_\mathrm{max}=0.2\,h \mathrm{Mpc}^{-1}$  and Euclidnl for which we take $k_\mathrm{max}=k_\mathrm{nl}(z)$.

Of all possible extension of the standard cosmological model we focus on the sum of the neutrino masses, $\sum {m_\nu}$, and on the effective number of relativistic degrees of freedom, $N_{\rm eff}$, since, as we will see, these two have a clear target to be reached.

%%%%%%%%%%%%%%% neutrinos %%%%%%%%%%%%%%%%%%
\subsection{Massive neutrinos}
\label{sec:BROADBAND_mnu}
Constraining neutrino masses is one of the main goals of current and future cosmological surveys. Massive neutrinos suppress power on intermediate and small scales and can therefore be detected by combining probes of large scales, like CMB anisotropies, with data from intermediate/small scales \cite{Lesgourgues}. Currently the best constraints are $\sum m_{\nu} < 0.11$ eV \cite{Lyalpha}, but we expect the next generation of galaxy surveys and CMB probes to significantly improve those bounds.
The benchmark to achieve for the error on neutrino masses is set by the laboratory measurements of neutrino oscillations. These indicate two possible scenarios for the sum of neutrino masses, the normal hierarchy where $\sum m_{\nu} \ge 0.06$ eV and the inverted one where $\sum m_{\nu} \ge 0.1$ eV \cite{PDG}.

The goal of this section is to study the improvements on the constraints on the neutrino masses that can be achieved by adding information from 21cm intensity mapping surveys in the redshift range $2.5<z<5$. Improvements on the constraints coming from the epoch of reionization have been presented in \cite{Pritchard2008,Kohri2013,Kohri2016}.

In the case of massive neutrinos we have to slightly modify our signal model. In fact, as shown \cite{Paco2014,Ema2014,Ema2015,FVN2017}, halos and galaxies are biased tracers of the CDM+baryons field only, as oposed to the total matter field -- CDM+baryons+neutrinos. This means that we have to change the $P(k)$ and $f(k)$ in eq. \eqref{eq:P21} to include only the CDM+baryons component. This important physical effect has the consequence of reducing the constraining power of a broadband analysis of $P_{21}(k,z)$, since the total matter power spectrum is more suppressed by the presence of massive neutrinos than the CDM+baryons only.
In our case, we find the difference being around 30\% in terms of the forecasted $\sigma_{m_\nu}$, an absolutely non-negligible effect. 
To the best of our knowledge, existing forecasts on neutrino masses do not include this effect in their analysis (with some notable exceptions as in \cite{Villaescusa-Navarro_2015a}), suggesting some level of bias in their final answers.
Our forecast is therefore more realistic, but it yields worse constraints.

\begin{figure}[t!]
\centering
\includegraphics[scale = 0.6]{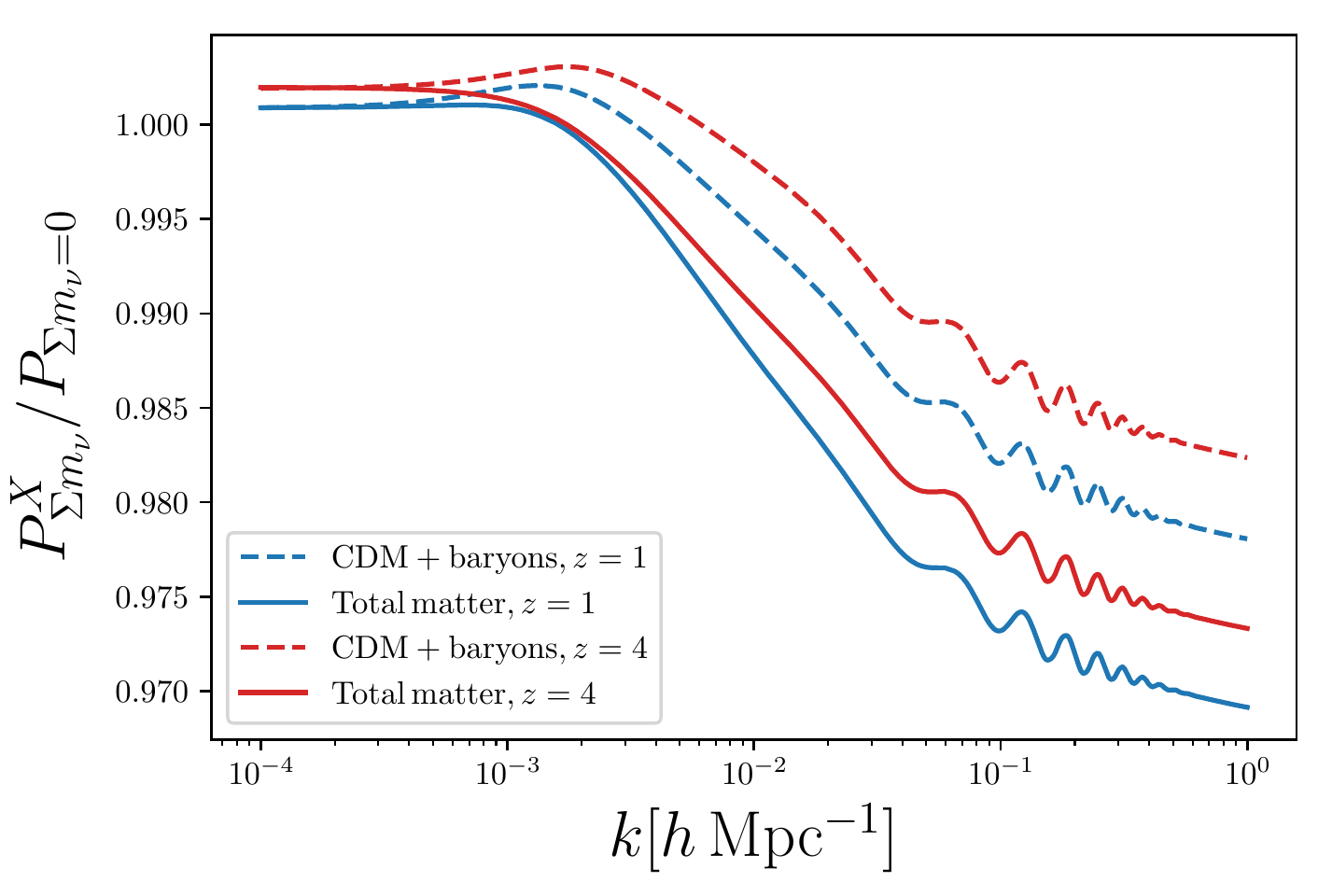}
\caption{Ratio between the linear CDM+baryons (dashed) and the total matter (solid) power spectrum in a model with massive neutrinos ($\Sigma m_\nu=0.06$ eV) to the same quantities in the standard  $\Lambda$CDM model ($\Sigma m_\nu=0$ eV). Different colors represent different redshift, in blue $z=1$ and in red $z=4$.}
\label{fig:Pcm_Pnonu}
\end{figure}

Another important thing to notice is that the effect of neutrinos on the power spectrum become smaller with redshift. This is due to the fact that neutrinos have less time to delay the growth of CDM+baryons perturbations on small scales. This means that the constraint on $\Sigma m_\nu$ from two surveys, one at low-redshift and another at high-redshift, covering the same volume and using the same amount of information ($k_{\rm min}<k<k_{\rm max}$) will be different,  the high-redshift one being worse. In contrast, going to higher redshifts has two advantages: larger survey volume and larger $k_\mathrm{nl}$ up to which one can hope to use perturbation theory. Figure \ref{fig:Pcm_Pnonu} shows the ratio between the linear CDM+baryons (dashed) and the total matter (solid) power spectrum in a model with massive neutrinos ($\Sigma m_\nu=0.06$ eV) to the same quantities in the standard  $\Lambda$CDM model ($\Sigma m_\nu=0$ eV). Different colors represent different redshift, in blue $z=1$ and in red $z=4$.  We notice that the difference between CDM+baryons and total matter remains constant with redshift, but the overall suppression compared to the vanilla models decrease by almost a factor of two between low and high $z$. 

\begin{figure}[t!]
\centering
\includegraphics[scale = 0.7]{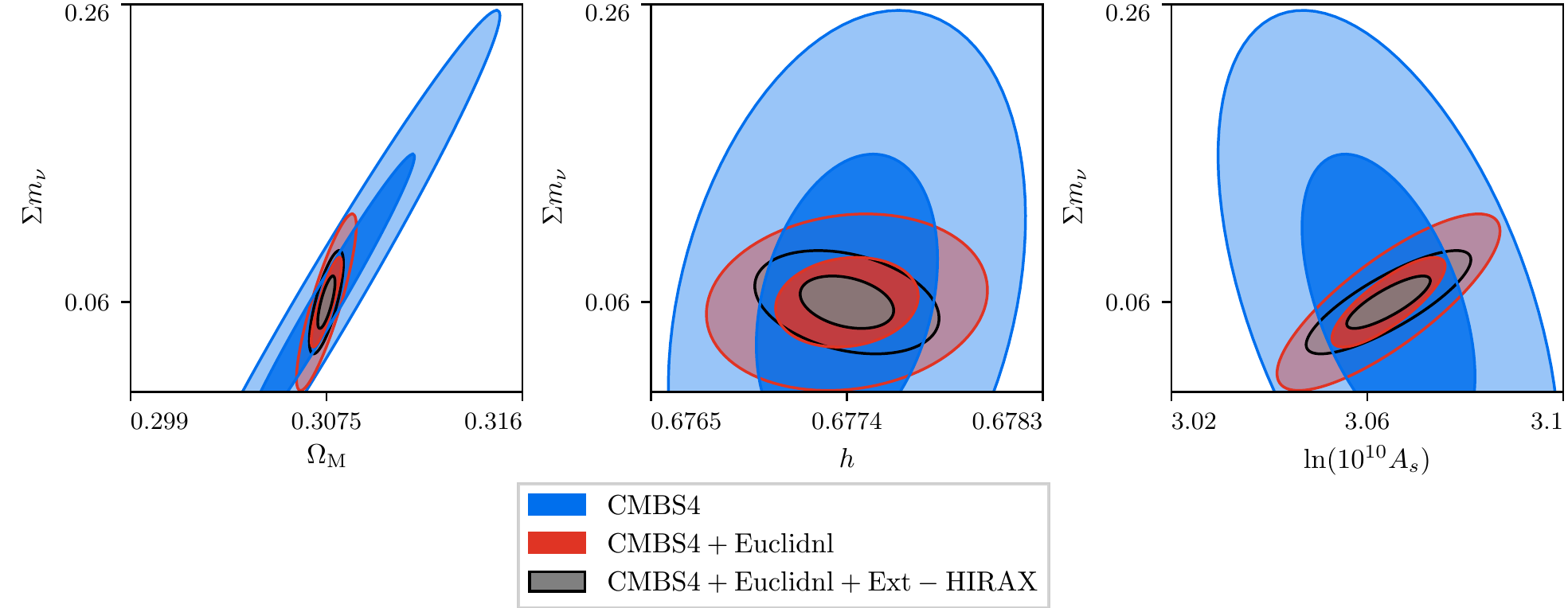}
\caption{Constraints on the neutrino masses and their degeneracies with $\Omega_{\rm m}$ (left), $h$ (middle) and $A_s$ (right) from CMBS4 (blue), CMBS4+Euclid (red) and CMBS4+Euclid+Ext-HIRAX (black). For Ext-HIRAX we have assumed $k_\mathrm{max}=k_\mathrm{nl}$, no wedge and $2\%$ priors on both $b_\mathrm{HI}$ and $\Omega_\mathrm{HI}$.}
\label{fig:Mnu_degeneracy}
\end{figure}

We show the forecasted constraints on the neutrino masses and other cosmological parameters coming from external datasets alone in table \ref{table:Priors}. Constraints on the neutrino masses for various combinations of probes and different foreground assumptions are shown in Table \ref{table:Table_mnu_Ext-HIRAX_21cm_gain} and \ref{table:Table_mnu_ExtCHIME_21cm_gain} for Ext-CHIME and Ext-HIRAX, respectively. 
We see that in the most pessimistic case, i.e.\ full wedge and $k_{\rm max} = 0.2\,\kMpc$, Ext-HIRAX plus Euclid and CMB-S4 is able to pin down neutrino masses to $\sigma_{m_\nu}=0.028$ eV, with intensity mapping bringing a 10\% improvement on the errorbars. In the most optimistic case this combination of probes yields $\sigma_{m_\nu}=0.018$ eV in which case the gain of adding IM can reach 40\%. It is important to stress again that the use of CDM+baryons to describe the clustering of galaxies and 21cm is worsening the error bars by roughly 30\%. In Figure \ref{fig:Mnu_degeneracy} we show the constraints and the degeneracies of the neutrino masses with other cosmological parameters when Ext-HIRAX is combined with CMB-S4 and CMB-S4+Euclid in the most optimistic case.
Single-dish experiments do not help in constraining neutrinos masses as well as interferometers, as one can see from results presented in Tables \ref{table:Table_mnu_FAST} for highzFAST alone and Table \ref{table:Table_mnu_FAST_21cm_gain} in which we show the results when combining different probes together.

We notice that the effect of the wedge does not worsen the constraints too much. In Section \ref{sec:growth} we have shown that the presence of the wedge deteriorate significantly the constraints on $f\sigma_8$. Thus, we conclude that most of the improvement brought in by the 21cm data in the combined analysis is due to the very accurate distance measurements, which are less affected by foregrounds. A similar argument applies to external priors on the density and the bias of the HI field, which do not change much the final result on neutrino masses.
On the other hand, an even more aggressive, in terms of signal-to-noise, instrument could potentially improve on neutrino masses constraints.

\begin{figure}[t!]
\centering
\subfloat{\includegraphics[scale = 0.5]{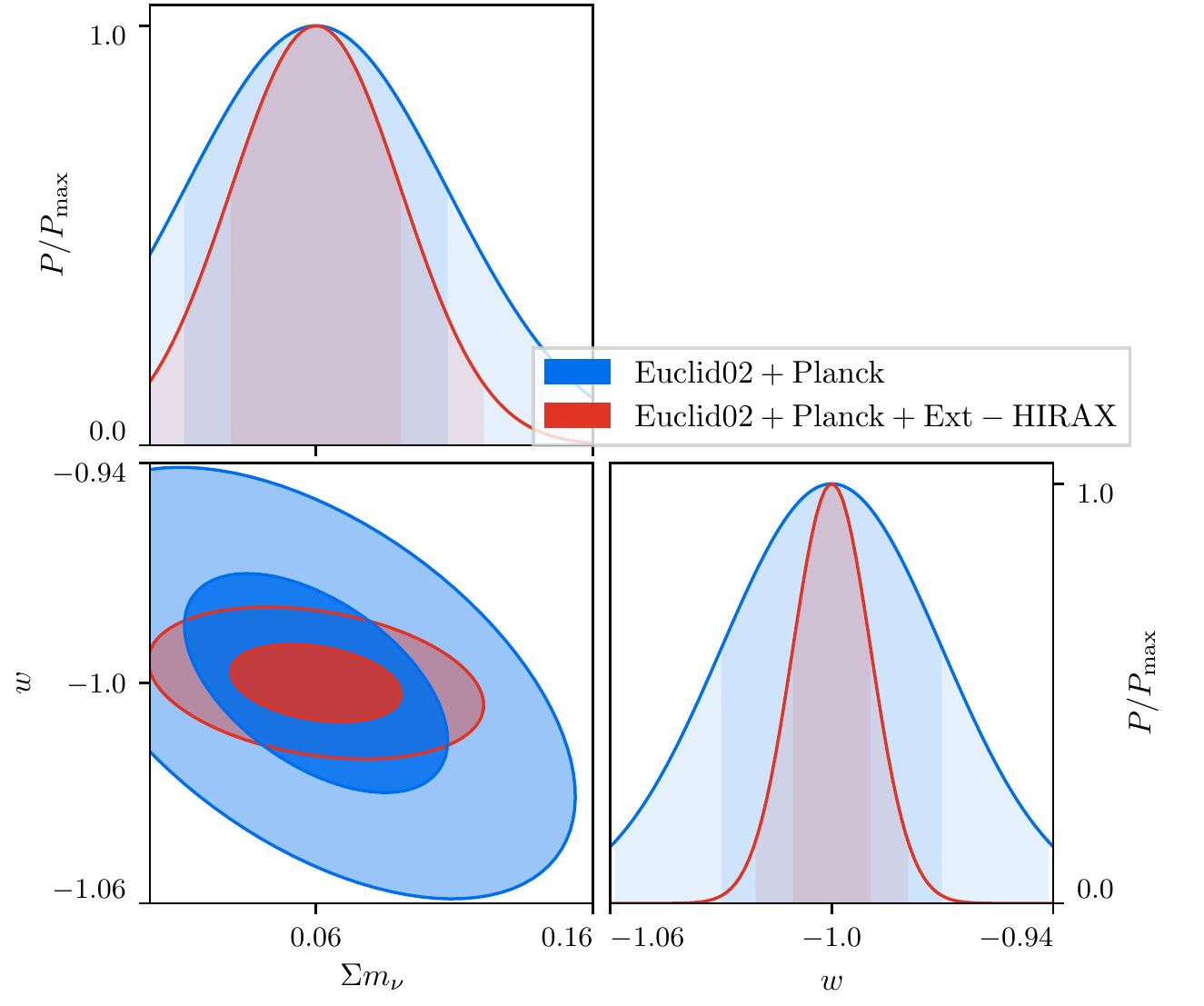}}
\subfloat{\includegraphics[scale = 0.5]{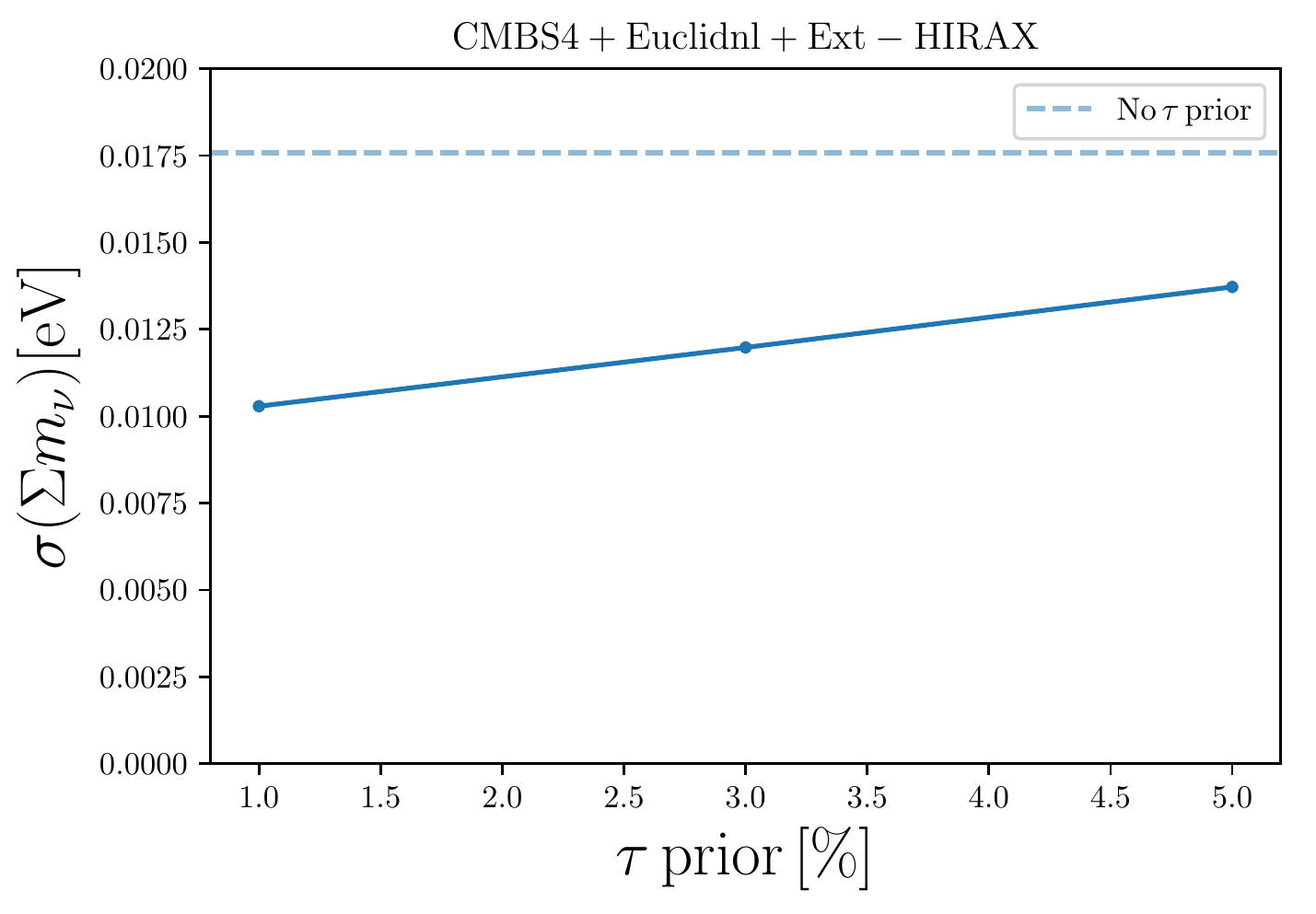}}
\caption{(Left panel) Degeneracy between $\Sigma{m_\nu}$ and $w$ when considering only Euclid02+Planck (blue) and when adding information from Ext-HIRAX (red). (Right panel) The effect of an additional $\tau$ prior on the neutrino masses constraints coming from the combination of CMBS4 + Euclidnl + Ext-HIRAX. The blue dashed line shows the case on no additional prior assumed.}
\label{fig:Mnu_w_tau_prior}
\end{figure}

Another advantage of observing the high-redshift Universe is that dark energy is a sub dominant fraction of the energy density. It is in fact well known that in the extended parameter spaces the equation of state of dark energy $w$ is very degenerate with other parameters, e.g. massive neutrinos and curvature \citep{Leonard16}. In Figure \ref{fig:Mnu_w_tau_prior} we show the degeneracy $\Sigma {m_\nu}-w$ degeneracy in the case where only low redshift measurements in combination with CMB are available (Euclid02+Planck) and when the information from above $z>2.5$ (Ext-HIRAX) is added on top. We find a factor of 3 improvement on the error on $w$ when adding Ext-HIRAX, constraining $w$ to the 1\% level even if neutrinos are allowed to vary. For Ext-HIRAX in this case we have used $k_{\rm max}=0.2\,\kMpc$, 2\% priors on both $b_{\rm HI}$ and $\Omega_{\rm HI}$, and mid-wedge case.

To conclude this Section we would like to emphasize that even adding high-redshift information the constraint on neutrino masses is primarily limited by the uncertainty in the amplitude of the primordial fluctuation spectra $A_{\rm s}$ and the optical depth of reionization -- $\tau$ -- in CMB data. The parameter $\tau$ is weakly constrained by the current Planck results \cite{Planck2015}, with an error bar almost a factor of two worse than the forecasted value \citep{PlanckBlue}. Improvement in the understanding of systematics in the Planck polarization data will likely improve the constraint on $\tau$, but to be more conservative we decided to adopt the current estimate. This is not the case for pre-Planck forecasts, e.g.\ DESI \cite{DESInu} or more recent ones \citep{Sprenger18,DESI2016}, who assume in their forecasts the blue book value for $\sigma_\tau$. This fact, together with our correct choice of the modeling at the beginning of this Section, explains why our constraint on neutrino masses is in general worse than others in the literature. Concerning the future improvements on constraining $\tau$, aside from including all polarization measurements from Planck, another hope is that future reionization epoch IM surveys, e.g.\ HERA, could bring tight constraints on $\tau$ \cite{Liutau}, thereby bringing substantial improvements on other cosmological parameters. Figure \ref{fig:Mnu_w_tau_prior} shows how much a better measurement of $\tau$ in the CMB could help in constraining the neutrino masses in the best case of Ext-HIRAX with $k_{\rm max}=k_{\rm nl}$, 2\% prior on both $b_{\rm HI}$ and $\Omega_{\rm HI}$, and no wedge.

%Figure \ref{fig:Mnu_w_tau_prior} shows how much a cosmic variance limited measurement of $\tau$ in the CMB could help constraining neutrino masses much better.

\subsubsection{Beyond linear theory}
\label{sec:1-loop}
So far we have considered a simple model for the 21cm power spectrum consisting of linear theory and linear bias term. Going beyond linear theory requires new nuisance parameters that need to be marginalised over {\cite{Baldauf,Audren,Sprenger18,Obinna2017}. This can degrade the constraints on the neutrino masses. In order to estimate the effect of these extra terms we will consider a simple order-of-magnitude extension to our power spectrum model and write the 21cm power spectrum as:
\be
P_\mathrm{HI}(k,z)=\bar{T}_b^2(b_\mathrm{HI}^2+f\mu^2)^2 P_\mathrm{lin}(k,z=0)D(z)^2\left(1+2k^2R^2+\alpha P_\mathrm{lin}(k,z=0)D(z)^2\frac{k^3}{2\pi^2}\right).
\ee
The second term in the parentheses has the form of the most relevant counter-term in the effective field theory approach to perturbation theory \cite{Baumann,Carrasco}. The third term is included in order to mimic the effects of 1-loop corrections to the power spectrum and can be understood as the ``theoretical error'' \cite{Baldauf} on the linear theory model. This approximation for one-loop contributions is a good order-of-magnitude estimate on scales relevant for constraining the sum of the neutrino masses. The coefficients $R$ and $\alpha$ are free parameters in each redshift bin and we take the fiducial values $R=\alpha=0$. We marginalise over $R$ and $\alpha$ in each redshift bin. We show the resulting constraints on the cosmological parameters from Ext-HIRAX in Table \ref{table:Table_mnu_Ext-HIRAX}. Compared to the standard case of linear theory and bias, the constraints on the neutrino masses become weaker by a factor of $\sim25\%$ and $31\%$ in the no wedge case of $k_\mathrm{max} = 0.2\,\kMpc$ and $k_\mathrm{max}=k_\mathrm{nl}(z)$, respectively. Adding other probes (Euclid, CMB S-4) makes this difference smaller and the constraints are weaker by a factor of $5-10\%$ (see Table \ref{table:Table_mnu_Ext-HIRAX_21cm_gain}). This reinforces our previous argument that most of the constraining power is coming from geometrical measurements of the BAO distances.

%%%%%%%%%%%%%%%%%%%
%%%%%%%%%%%%%%%%%%%

\subsection{The effective numbers of relativistic degrees of freedom}
\label{sec:BROADBAND_Neff}

\begin{figure}[t!]
\centering
\includegraphics[scale = 0.7]{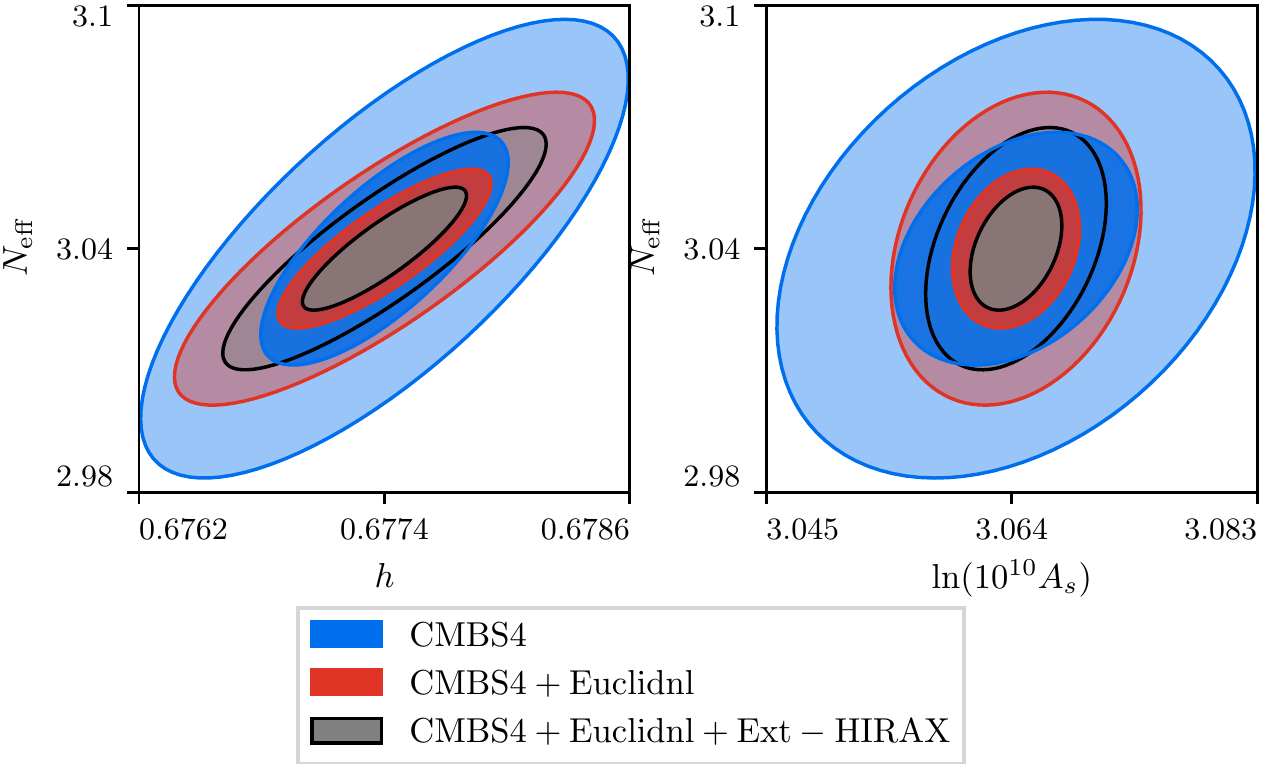}
\caption{Constraints on $N_{\rm eff}$ and its degeneracies with $h$ (left) and $A_s$ (right) from CMB-S4 (blue), CMB-S4+Euclid (red) and CMB-S4+Euclid+Ext-HIRAX (black). For Ext-HIRAX we have assumed $k_\mathrm{max}=k_\mathrm{nl}$, no wedge and $2\%$ priors on both $b_\mathrm{HI}$ and $\Omega_\mathrm{HI}$. We used \texttt{cosmicfish} software \cite{Cosmicfish} to make this Figure and also Figure \ref{fig:Mnu_degeneracy}.}
\label{fig:Neff_degeneracy}
\end{figure}

Another extension of the standard $\Lambda$CDM model is represented by the presence of extra radiation species. In the standard model, with 3 massive neutrinos, one has $N_{\rm eff} = 3.046$ \citep{Mangano_2005}, but any light species in thermal equilibrium during the history of the Universe will increase this value. Theoretical estimates in \cite{Brust2013,Chacko2015} set the minimum $\Delta N_{\rm eff}$ due to any light particles between $z=0$ and the QCD phase transition to $\Delta N_{\rm eff}=0.027$. This number is therefore the natural target of any analysis of $N_{\rm eff}$.

In CMB data, $N_{\rm eff}$ is very degenerate with other parameters, and the 1$\sigma$ forecasted error of CMB S-4 is slightly above the theory benchmark. 
However, it has been recently realized that a clean signature of $N_{\rm eff}$ exists and can be detected in both CMB and large scale structure \cite{Bashinsky03,Baumann2015,Baumann17}, as the presence of extra relativistic degrees of freedom introduce a phase shift in the acoustic peaks of the baryon-photon fluid \cite{Bashinsky03}. This effect has already been detected in Planck data \cite{Follin15,Baumann2015}.
In \cite{Baumann17}, the authors have shown that this phase shift survives non-linear evolution and therefore could in principle be detected in BAO analysis of galaxy clustering. 

Building and testing a template for the phase shift goes beyond the scope of this paper, but motivated by the above discussion we study the constraints on $N_{\rm eff}$ from various combination of probes. We fit the full broadband power spectrum, which by definition contains also the phase shift, to constraint the number of relativistic degrees of freedom. In a realistic data analysis one would use reconstructed data, which we mimic by dividing by a factor of 2 the damping factor in eq. \eqref{eq:damping} \citep{DESInu}.

We show the forecasted constraints on $N_{\rm eff}$ and other cosmological parameters coming from external datasets alone in table \ref{table:Priors_Neff}. Our results for various combinations of probes and different foreground assumptions are presented in tables \ref{table:Table_Neff_Ext-HIRAX},  \ref{table:Table_Neff_FAST}, \ref{table:Table_Neff_Ext-HIRAX_21cm_gain}, \ref{table:Table_Neff_Ext-CHIME_21cm_gain}, \ref{table:Table_Neff_FAST_21cm_gain} and figure \ref{fig:Neff_degeneracy}. Our forecasted error for Ext-HIRAX ranges from $\sigma_{N_{\rm eff}}=0.015$, for the very optimistic case of no wedge and $k_{\rm max} = k_{\rm nl}$, to  $\sigma_{N_{\rm eff}}=0.021$ for the full wedge and $k_{\rm max} = 0.2 \, \kMpc$ case.
These numbers are interesting, but on the other hand any deviation from the minimal value will be detected at less than 1$\sigma$. It is also important to point out that 21cm surveys will add very little to the combination of CMB data and galaxy surveys. Our results are in broad agreement with the analysis of \cite{Baumann17}.

%%%%%%%%%%%%%%%%%%%
%%%%%%%%%%%%%%%%%%%

%%%%%%%%%%%%%%%%%%%
%%%%%%%%%%%%%%%%%%%

%%%%%%%%%%%%%%%%%%%%%%%%%%%
\section{Summary and conclusions}
\label{sec:conclusions}

In this work we have explored the possibility of studying cosmology through radio-telescopes that operate in the redshift range $2.5<z<5$. The reason behind this is that, while this is a redshift-range not considered in current and upcoming setups, the volume it encloses is much larger then the one probed by current and future spectroscopic surveys. The question we try to answer is: how much cosmological information is contained in this redshift window?

We focus our analysis on four key cosmological quantities: 1) the growth rate, $f\sigma_8$, the BAO distance scale parameters, $D_A$ and $H$, the sum of the neutrino masses, $\Sigma m_\nu$, and the number of relativistic degrees of freedom, $N_{\rm eff}$. We consider four extensions of current or upcoming radio-telescopes like HIRAX, CHIME and FAST, and two different observational strategies: interferometry (for Ext-HIRAX and Ext-CHIME) and single-dish (highzFast).

We carry out our analysis using the Fisher matrix formalism. We model the amplitude and shape of the 21cm signal using the model proposed in \cite{EmaPaco}. We also account for cosmological and instrumental effects such as the presence of the wedge, the window function, the instrument thermal noise, the angular resolution, the presence of shot-noise...etc. 

We point out that measurements that are sensitive to the overall amplitude of the 21cm power spectrum, like $f\sigma_8$, will be completely degenerate with astrophysical parameters like $\Omega_{\rm HI}$ and $b_{\rm HI}$. In order to break that degeneracy it is necessary that we use independent datasets that constrain those quantities. We show how the value of these parameters can be determined through either the Ly$\alpha$-forest alone or via cross-correlations between 21cm and the Ly$\alpha$-forest or DLAs.

Under the assumption of the primary beam foreground wedge contamination (mid-wedge case in the text), $5\%$ priors on $b_{\rm HI}$ and $\Omega_{\rm HI}$ and $k_{\rm max}=0.2~h{\rm Mpc}^{-1}$, that we term the fiducial setup, we find that Ext-HIRAX can constrain the value of $f\sigma_8$ within bins of $\Delta z=0.1$ at $\simeq4\%$ in the redshift range $2.5<z<5$. A modest improvement is achieved by changing $k_{\rm max}$ from 0.2 $h{\rm Mpc}^{-1}$ to $k_{\rm nl}$. If data from the whole wedge need to be discarded, these constraints degrade between a factor $2$ (at $z=2.5$) and $7$ (at $z=5$). 

Under the fiducial setup, we find that Ext-HIRAX will place $\simeq1\%$ constraints on $D_A$ and $H$. As with the growth rate, our results point out that going to smaller scales has only a very modest impact on the results. Being able to use a fraction of the modes in the wedge has a huge impact on our results, as removing the information in the whole wedge degrades the constraints between a factor of 10 (at $z=2.5$) and 20 (at $z=5$). 

We have also studied the impact that the theory model has on the results. By using a theory template that accounts for 1-loop corrections and incorporates 2 free parameters that we marginalise over, we find that the constraints on the cosmological parameters worsen between $10\%$ and $500\%$ when $2\%$ priors on $\Omega_{\rm HI}$ and $b_{\rm HI}$ are used. In the case of the neutrino masses, the constraints worsen between $10\%$ and $30\%$.

We find that data from Ext-HIRAX in the fiducial setup can constrain the neutrino masses with an error of $0.10$ eV. In combination with data from CMB S4 and galaxy clustering from Euclid the errors shrink to $\simeq20$ meV. Our results are not very sensitive to the wedge coverage, the minimum scale employed and the priors on $b_{\rm HI}$ and $\Omega_{\rm HI}$.

Finally, we find that data from Ext-HIRAX, in the fiducial setup, plus CMB S4 plus Euclid can constrain $N_{\rm eff}$ with a very competitive error of 0.02. As with the neutrino masses, our constraints do not depend much on the $\Omega_{\rm HI}$ and $b_{\rm HI}$ priors, $k_{\rm max}$ and the wedge coverage.

Results for the Ext-CHIME and highzFAST instruments are similar to those of Ext-HIRAX, with the exception of neutrino masses, where highzFAST performs worse than Ext-CHIME or Ext-HIRAX.

We conclude that there is a large amount of cosmological information embedded in the, poorly constrained, redshift range $2.5<z<5$. Suitable extensions of existing and upcoming radio-telescopes targeting at this redshift window can provide very tight constraint on key cosmological parameters.

%
%
%\section{Conclusions} \label{sec:conclusions}
%

\section{Tables} \label{Ap:Tables}

In this section we show the tables with the constraints on RSD, {BAO distance scale parameters} and cosmological parameters coming from various configurations of the proposed 21cm IM surveys or in combination with current and foreseen galaxy surveys and CMB experiements.

%% RSD Table

\begin{table}[ht!]
\centering
\begin{tabular}{ccccccc}
\hline \hline
\multicolumn{7}{c}{Ext-HIRAX $\sigma_{f\sigma_8}/(f\sigma_8)$, 5\% $b_\mathrm{HI}\,\&\,\Omega_\mathrm{HI}$}\\
\hline
& \multicolumn{	3}{c}{$k_\mathrm{max}=0.2 h\mathrm{Mpc^{-1}}$} & \multicolumn{3}{c}{$k_\mathrm{max}=k_\mathrm{nl}(z)$}\\
$z$ & No wedge & Mid wedge & Wedge & No wedge & Mid wedge & Wedge\\\hline
2.5 & 0.036 & 0.036 & 0.075 & 0.035 & 0.035 & 0.043\\ 
2.6 & 0.036 & 0.036 & 0.080 & 0.035 & 0.036 & 0.044\\ 
2.7 & 0.036 & 0.036 & 0.085 & 0.035 & 0.036 & 0.046\\ 
2.8 & 0.036 & 0.036 & 0.091 & 0.035 & 0.036 & 0.048\\ 
2.9 & 0.036 & 0.037 & 0.097 & 0.036 & 0.036 & 0.050\\ 
3.0 & 0.036 & 0.037 & 0.104 & 0.036 & 0.036 & 0.053\\ 
3.1 & 0.036 & 0.037 & 0.111 & 0.036 & 0.036 & 0.056\\ 
3.2 & 0.036 & 0.037 & 0.119 & 0.036 & 0.036 & 0.060\\ 
3.3 & 0.037 & 0.037 & 0.127 & 0.036 & 0.036 & 0.065\\ 
3.4 & 0.037 & 0.037 & 0.135 & 0.036 & 0.036 & 0.070\\ 
3.5 & 0.037 & 0.037 & 0.144 & 0.036 & 0.036 & 0.075\\ 
3.6 & 0.037 & 0.038 & 0.153 & 0.036 & 0.036 & 0.082\\ 
3.7 & 0.037 & 0.038 & 0.162 & 0.036 & 0.036 & 0.089\\ 
3.8 & 0.037 & 0.038 & 0.171 & 0.036 & 0.036 & 0.097\\ 
3.9 & 0.037 & 0.038 & 0.181 & 0.036 & 0.036 & 0.105\\ 
4.0 & 0.038 & 0.039 & 0.190 & 0.036 & 0.036 & 0.114\\ 
4.1 & 0.038 & 0.039 & 0.200 & 0.036 & 0.036 & 0.124\\ 
4.2 & 0.038 & 0.040 & 0.209 & 0.036 & 0.037 & 0.135\\ 
4.3 & 0.038 & 0.040 & 0.218 & 0.036 & 0.037 & 0.146\\ 
4.4 & 0.039 & 0.041 & 0.228 & 0.036 & 0.037 & 0.158\\ 
4.5 & 0.039 & 0.041 & 0.237 & 0.036 & 0.037 & 0.170\\ 
4.6 & 0.039 & 0.042 & 0.246 & 0.037 & 0.038 & 0.182\\ 
4.7 & 0.040 & 0.043 & 0.254 & 0.037 & 0.038 & 0.194\\ 
4.8 & 0.040 & 0.044 & 0.263 & 0.037 & 0.038 & 0.206\\ 
4.9 & 0.040 & 0.045 & 0.271 & 0.037 & 0.039 & 0.219\\ 
5.0 & 0.041 & 0.046 & 0.279 & 0.037 & 0.040 & 0.231\\
\hline
\hline
\end{tabular}
\caption{Forecasted 1$\sigma$ constraints on $f\sigma_8$ for Ext-HIRAX as a function of redshift for different wedge configurations and different $k_{\rm max}$. The numbers correspond to $5\%$ priors on the values of both $\Omega_{\rm HI}$ and $b_{\rm HI}$.}
\label{table:Table_fs8}
\end{table}

%% AP table

\begin{table}[ht!]
\centering
\begin{tabular}{ccccccc}
\hline \hline
\multicolumn{7}{c}{Ext-HIRAX, $k_\mathrm{max}=0.2 h\mathrm{Mpc^{-1}}$, 5\% $b_\mathrm{HI}\,\&\,\Omega_\mathrm{HI}$,}\\
\hline
& \multicolumn{	3}{c}{$\sigma_{D_A}/D_A$} & \multicolumn{3}{c}{$\sigma_H/H$}\\
$z$ & No wedge & Mid wedge & Wedge & No wedge & Mid wedge & Wedge\\\hline
2.5 & 0.004 & 0.005 & 0.053 & 0.006 & 0.006 & 0.014\\ 
2.6 & 0.004 & 0.005 & 0.057 & 0.006 & 0.006 & 0.014\\ 
2.7 & 0.004 & 0.005 & 0.060 & 0.006 & 0.006 & 0.014\\ 
2.8 & 0.004 & 0.005 & 0.064 & 0.006 & 0.006 & 0.014\\ 
2.9 & 0.004 & 0.005 & 0.067 & 0.006 & 0.006 & 0.014\\ 
3.0 & 0.005 & 0.005 & 0.071 & 0.006 & 0.007 & 0.014\\ 
3.1 & 0.005 & 0.006 & 0.075 & 0.006 & 0.007 & 0.015\\ 
3.2 & 0.005 & 0.006 & 0.080 & 0.007 & 0.007 & 0.015\\ 
3.3 & 0.005 & 0.006 & 0.084 & 0.007 & 0.007 & 0.015\\ 
3.4 & 0.005 & 0.006 & 0.089 & 0.007 & 0.007 & 0.016\\ 
3.5 & 0.005 & 0.007 & 0.095 & 0.007 & 0.008 & 0.016\\ 
3.6 & 0.005 & 0.007 & 0.10 & 0.007 & 0.008 & 0.016\\ 
3.7 & 0.006 & 0.007 & 0.11 & 0.008 & 0.008 & 0.017\\ 
3.8 & 0.006 & 0.008 & 0.11 & 0.008 & 0.009 & 0.017\\ 
3.9 & 0.006 & 0.008 & 0.12 & 0.008 & 0.009 & 0.018\\ 
4.0 & 0.006 & 0.008 & 0.13 & 0.008 & 0.009 & 0.019\\ 
4.1 & 0.006 & 0.009 & 0.14 & 0.009 & 0.010 & 0.019\\ 
4.2 & 0.007 & 0.009 & 0.15 & 0.009 & 0.010 & 0.020\\ 
4.3 & 0.007 & 0.010 & 0.16 & 0.009 & 0.010 & 0.021\\ 
4.4 & 0.007 & 0.010 & 0.17 & 0.010 & 0.011 & 0.022\\ 
4.5 & 0.007 & 0.011 & 0.18 & 0.010 & 0.011 & 0.023\\ 
4.6 & 0.008 & 0.011 & 0.20 & 0.010 & 0.012 & 0.024\\ 
4.7 & 0.008 & 0.012 & 0.21 & 0.011 & 0.012 & 0.025\\ 
4.8 & 0.008 & 0.013 & 0.23 & 0.011 & 0.013 & 0.026\\ 
4.9 & 0.009 & 0.014 & 0.25 & 0.011 & 0.013 & 0.027\\ 
5.0 & 0.009 & 0.014 & 0.27 & 0.012 & 0.014 & 0.029\\ 
\hline
\hline
\end{tabular}
\caption{Forecasted 1$\sigma$ constraints on the expansion rate $H(z)$ and the angular diameter distance $D_A(z)$ for Ext-HIRAX as a function of redshift for different wedge configurations. The numbers correspond to $5\%$ priors on the values of both $\Omega_{\rm HI}$ and $b_{\rm HI}$ and $k_\mathrm{max}=0.2 h\mathrm{Mpc^{-1}}$.}
\label{table:Table_H_DA}
\end{table}

%%%% Mnu tables

\begin{table}[ht!]
\centering
\begin{tabular}{lccccc}
\hline \hline
\multicolumn{6}{c}{External datasets -- $\Sigma m_\nu$}\\
\hline	
 & $\Omega_\mathrm{M}$ & $h$ & $\Sigma m_\nu$[eV] & $\ln(10^{10} A_s)$ & $n_s$ \\
\hline
\hline
Fiducial values & 0.3075 & 0.6774 & 0.060 & 3.064 & 0.9667\\
\hline
\hline
PlanckBAO  & 0.0082 & 0.0065 & 0.067 & 0.037 & 0.0048\\
\hline
Euclid02 No Rec. & 0.0055 & 0.0025 & 0.20 & 0.091 & 0.0086\\
\hline
Euclidnl No Rec. & 0.0050 & 0.0021 & 0.17 & 0.077 & 0.0065\\
\hline
CMB-S4 & 0.0038 & 0.0004 & 0.10 & 0.018 & 0.0019\\
\hline
\end{tabular}
\caption{Forecasted 1$\sigma$ constraints on cosmological parameters and the $\Sigma m_\nu$ considering each external dataset alone. These constraints are obtained using the Fisher matrices explained in Section \ref{sec:probecombination}. PlanckBAO represents Planck CMB combined with BOSS BAO measurements. We show the projected constraints for Euclid for two different $k_{\rm max}$ used and the constraints coming from CMB-S4 experiment.}
\label{table:Priors}
\end{table}

%%%%

%%%%

\begin{table}[ht!]
\centering
\begin{tabular}{lccccc}
\hline \hline
\multicolumn{6}{c}{Ext-HIRAX $\Sigma m_\nu$}\\
\hline	
 & $\Omega_\mathrm{M}$ & $h$ & $\Sigma m_\nu$[eV] & $\ln(10^{10} A_s)$ & $n_s$\\
\hline
Fiducial values & 0.3075 & 0.6774 & 0.060 & 3.064 & 0.9667\\
\hline
\hline
\multicolumn{6}{c}{No wedge $k_\mathrm{max}=0.2 h\mathrm{Mpc^{-1}}$}\\
\hline
2\% $b_\mathrm{HI}\,\&\,\Omega_\mathrm{HI}$ & 0.0016 & 0.0010 & 0.059 & 0.026 & 0.0036\\
2\% $b_\mathrm{HI}\,\&\,\Omega_\mathrm{HI}$, diff $M_{\rm min}$ & 0.0015 & 0.0009 & 0.056 & 0.025 & 0.0033\\
2\% $b_\mathrm{HI}\,\&\,\Omega_\mathrm{HI}$, 1-loop & 0.0020 & 0.0010 & 0.081 & 0.038 & 0.014\\
5\% $b_\mathrm{HI}\,\&\,\Omega_\mathrm{HI}$ & 0.0024 & 0.0015 & 0.093 & 0.047 & 0.0038\\
10\% $b_\mathrm{HI}\,\&\,\Omega_\mathrm{HI}$ & 0.0029 & 0.0018 & 0.11 & 0.065 & 0.0040\\
\hline
\multicolumn{6}{c}{No wedge $k_\mathrm{max}=k_\mathrm{nl}(z)$}\\
\hline
2\% $b_\mathrm{HI}\,\&\,\Omega_\mathrm{HI}$ & 0.0010 & 0.0007 & 0.040 & 0.020 & 0.0016\\
2\% $b_\mathrm{HI}\,\&\,\Omega_\mathrm{HI}$, diff $M_{\rm min}$ & 0.0009 & 0.0006 & 0.037 & 0.019 & 0.0014\\
2\% $b_\mathrm{HI}\,\&\,\Omega_\mathrm{HI}$, 1-loop & 0.0011 & 0.0007 & 0.058 & 0.027 & 0.0060\\
5\% $b_\mathrm{HI}\,\&\,\Omega_\mathrm{HI}$ & 0.0014 & 0.0009 & 0.054 & 0.032 & 0.0016\\
10\% $b_\mathrm{HI}\,\&\,\Omega_\mathrm{HI}$ & 0.0014 & 0.0009 & 0.054 & 0.032 & 0.0016\\
\hline
\hline
\multicolumn{6}{c}{Mid wedge $k_\mathrm{max}=0.2 h\mathrm{Mpc^{-1}}$}\\ \hline
2\% $b_\mathrm{HI}\,\&\,\Omega_\mathrm{HI}$ & 0.0017 & 0.0010 & 0.062 & 0.027 & 0.0039\\
2\% $b_\mathrm{HI}\,\&\,\Omega_\mathrm{HI}$, diff $M_{\rm min}$ & 0.0016 & 0.0010 & 0.060 & 0.027 & 0.0037\\
2\% $b_\mathrm{HI}\,\&\,\Omega_\mathrm{HI}$, 1-loop & 0.0021 & 0.0011 & 0.088 & 0.041 & 0.016\\
5\% $b_\mathrm{HI}\,\&\,\Omega_\mathrm{HI}$ & 0.0025 & 0.0016 & 0.098 & 0.049 & 0.0042\\
10\% $b_\mathrm{HI}\,\&\,\Omega_\mathrm{HI}$ & 0.0032 & 0.0020 & 0.12 & 0.071 & 0.0044\\
\hline
\multicolumn{6}{c}{Mid wedge $k_\mathrm{max}=k_\mathrm{nl}(z)$}\\
\hline
2\% $b_\mathrm{HI}\,\&\,\Omega_\mathrm{HI}$  & 0.0011 & 0.0007 & 0.042 & 0.021 & 0.0017\\
2\% $b_\mathrm{HI}\,\&\,\Omega_\mathrm{HI}$, diff $M_{\rm min}$ & 0.0010 & 0.0006 & 0.039 & 0.020 & 0.0015\\
2\% $b_\mathrm{HI}\,\&\,\Omega_\mathrm{HI}$, 1-loop & 0.0011 & 0.0008 & 0.061 & 0.028 & 0.0066\\
5\% $b_\mathrm{HI}\,\&\,\Omega_\mathrm{HI}$ & 0.0015 & 0.0010 & 0.058 & 0.035 & 0.0017\\
10\% $b_\mathrm{HI}\,\&\,\Omega_\mathrm{HI}$ & 0.0017 & 0.0011 & 0.065 & 0.051 & 0.0018\\
\hline
\hline
\multicolumn{6}{c}{Wedge $k_\mathrm{max}=0.2 h\mathrm{Mpc^{-1}}$}\\ \hline
2\% $b_\mathrm{HI}\,\&\,\Omega_\mathrm{HI}$ & 0.0066 & 0.0048 & 0.24 & 0.091 & 0.010\\
2\% $b_\mathrm{HI}\,\&\,\Omega_\mathrm{HI}$, diff $M_{\rm min}$ & 0.0068 & 0.0050 & 0.25 & 0.092 & 0.010\\
2\% $b_\mathrm{HI}\,\&\,\Omega_\mathrm{HI}$, 1-loop & 0.0076 & 0.0050 & 0.34 & 0.138 & 0.054\\ 
5\% $b_\mathrm{HI}\,\&\,\Omega_\mathrm{HI}$ & 0.0072 & 0.0053 & 0.27 & 0.107 & 0.011\\
10\% $b_\mathrm{HI}\,\&\,\Omega_\mathrm{HI}$ & 0.0084 & 0.0062 & 0.31 & 0.136 & 0.011\\
\hline
\multicolumn{6}{c}{Wedge $k_\mathrm{max}=k_\mathrm{nl}(z)$}\\
\hline
2\% $b_\mathrm{HI}\,\&\,\Omega_\mathrm{HI}$ & 0.0027 & 0.0019 & 0.11 & 0.042 & 0.0044\\
2\% $b_\mathrm{HI}\,\&\,\Omega_\mathrm{HI}$, diff $M_{\rm min}$ & 0.0026 & 0.0019 & 0.10 & 0.041 & 0.0040\\
2\% $b_\mathrm{HI}\,\&\,\Omega_\mathrm{HI}$, 1-loop & 0.0029 & 0.0021 & 0.16 & 0.060 & 0.018\\
5\% $b_\mathrm{HI}\,\&\,\Omega_\mathrm{HI}$ & 0.0034 & 0.0025 & 0.13 & 0.059 & 0.0044\\
10\% $b_\mathrm{HI}\,\&\,\Omega_\mathrm{HI}$ & 0.0048 & 0.0034 & 0.17 & 0.092 & 0.0045\\
\hline
\hline
\end{tabular}
\caption{Fiducial values and 68\% confidence intervals on cosmological parameters and the $\Sigma m_\nu$ using Ext-HIRAX alone for various considerations. Going from top to bottom, we show the constraints considering different priors on both $b_\mathrm{HI}\,\&\,\Omega_\mathrm{HI}$, different $k_{\rm max}$ we use and different level of foreground wedge contamination. Alongside, focusing only on the case of 2\% priors on $b_\mathrm{HI}\,\&\,\Omega_\mathrm{HI}$, we show the results obtained when we use a different value of $M_\mathrm{min}=2\times10^{11}\,{\rm M}_\odot/h$ (labeled diff $M_{\rm min}$.) and when we go beyond linear theory and marginalise over 1-loop and counter-terms parameters (labeled 1-loop, see Section \ref{sec:1-loop}).}
\label{table:Table_mnu_Ext-HIRAX}
\end{table}

%%%%%

\begin{table}[ht!]
\centering
\begin{tabular}{lccccc}
\hline \hline
\multicolumn{6}{c}{highzFAST $\Sigma m_\nu$}\\
\hline	
 & $\Omega_\mathrm{M}$ & $h$ & $\Sigma m_\nu$[eV] & $\ln(10^{10} A_s)$ & $n_s$ \\
\hline
Fiducial values & 0.3075 & 0.6774 & 0.060 & 3.064 & 0.9667\\
\hline\hline
\multicolumn{6}{c}{$k_\mathrm{max}=0.2 h\mathrm{Mpc^{-1}}$}\\
\hline
2\% $b_\mathrm{HI}\,\&\,\Omega_\mathrm{HI}$  & 0.0045 & 0.0026 & 0.14 & 0.054 & 0.013\\
5\% $b_\mathrm{HI}\,\&\,\Omega_\mathrm{HI}$ & 0.0054 & 0.0032 & 0.18 & 0.076 & 0.013\\
10\% $b_\mathrm{HI}\,\&\,\Omega_\mathrm{HI}$ & 0.0069 & 0.0041 & 0.24 & 0.112 & 0.013\\
\hline\hline
\multicolumn{6}{c}{$k_\mathrm{max}=k_\mathrm{nl}(z)$}\\
\hline
2\% $b_\mathrm{HI}\,\&\,\Omega_\mathrm{HI}$  & 0.0031 & 0.0021 & 0.11 & 0.044 & 0.0076\\
5\% $b_\mathrm{HI}\,\&\,\Omega_\mathrm{HI}$ & 0.0041 & 0.0026 & 0.15 & 0.064 & 0.0078\\
10\% $b_\mathrm{HI}\,\&\,\Omega_\mathrm{HI}$ & 0.0054 & 0.0034 & 0.19 & 0.095 & 0.0079\\ 
\hline
\hline
\end{tabular}
\caption{Fiducial values and 68\% confidence intervals on cosmological parameters and the $\Sigma m_\nu$ using highzFAST alone for various considerations. We show the constraints considering different astrophysical priors on both $b_\mathrm{HI}\,\&\,\Omega_\mathrm{HI}$, as well as the dependence of the constraints on the $k_{\rm max}$ used.}
\label{table:Table_mnu_FAST}
\end{table}

%%%%

\begin{table}[ht!]
\centering
\begin{tabular}{l|ccc|ccc}
\hline \hline
\multicolumn{7}{c}{Ext-HIRAX $\Sigma m_\nu$[eV]}\\
\hline
& \multicolumn{3}{c|}{ $k_\mathrm{max}=0.2 h\mathrm{Mpc^{-1}}$} & \multicolumn{3}{c}{$k_\mathrm{max}=k_\mathrm{nl}(z)$}\\ \hline\hline
Euclid No Rec.+PlanckBAO & \multicolumn{3}{c|}{0.050} & \multicolumn{3}{c}{0.049}\\ \hline
$+$21cm & No wedge & Mid wedge & Wedge & No wedge & Mid wedge & Wedge\\ \hline
2\% $b_\mathrm{HI}\,\&\,\Omega_\mathrm{HI}$ & 0.037 & 0.038 & 0.045 & 0.030 & 0.031 & 0.040\\
2\% $b_\mathrm{HI}\,\&\,\Omega_\mathrm{HI}$, diff $M_\mathrm{min}$ & 0.037 & 0.038 & 0.044 & 0.028 & 0.029 & 0.039\\ 
2\% $b_\mathrm{HI}\,\&\,\Omega_\mathrm{HI}$, 1-loop & 0.038 & 0.039 & 0.046 & 0.035 & 0.035 & 0.042\\
5\% $b_\mathrm{HI}\,\&\,\Omega_\mathrm{HI}$ & 0.042 & 0.043 & 0.048 & 0.034 & 0.035 & 0.043\\
10\% $b_\mathrm{HI}\,\&\,\Omega_\mathrm{HI}$ & 0.044 & 0.045 & 0.049 & 0.035 & 0.036 & 0.044\\\hline
\hline
Euclid No Rec.+CMB-S4 & \multicolumn{3}{c|}{0.031} & \multicolumn{3}{c}{0.030}\\ \hline
$+$21cm & No wedge & Mid wedge & Wedge & No wedge & Mid wedge & Wedge\\ \hline
2\% $b_\mathrm{HI}\,\&\,\Omega_\mathrm{HI}$ & 0.022 & 0.022 & 0.028 & 0.018 & 0.018 & 0.025\\
2\% $b_\mathrm{HI}\,\&\,\Omega_\mathrm{HI}$, diff $M_\mathrm{min}$ & 0.021 & 0.021 & 0.028 & 0.017 & 0.017 & 0.024\\
2\% $b_\mathrm{HI}\,\&\,\Omega_\mathrm{HI}$, 1-loop & 0.023 & 0.023 & 0.028 & 0.020 & 0.020 & 0.025\\
5\% $b_\mathrm{HI}\,\&\,\Omega_\mathrm{HI}$ & 0.023 & 0.023 & 0.028 & 0.018 & 0.019 & 0.025\\
10\% $b_\mathrm{HI}\,\&\,\Omega_\mathrm{HI}$ & 0.023 & 0.023 & 0.028 & 0.019 & 0.019 & 0.025\\\hline
\end{tabular}
\caption{Forecasted 1$\sigma$ constraints on $\Sigma m_\nu$ considering Ext-HIRAX instrument combined with CMB+galaxy probes. We show the constraints considering different priors on both $b_\mathrm{HI}\,\&\,\Omega_\mathrm{HI}$, different level of wedge contamination and different $k_{\rm max}$ used. In the top half of the table we show what are the gains of adding Ext-HIRAX measurements on top of combined Fisher matrix for Euclid galaxy survey  and PlanckBAO. In the bottom half we do the same only using CMBS-4 Fisher matrix instead of the PlanckBAO one. Additionally, focusing only on the case of 2\% priors on $b_\mathrm{HI}\,\&\,\Omega_\mathrm{HI}$, we also show the results obtained when we use a different value of $M_\mathrm{min}=2\times10^{11}\,{\rm M}_\odot/h$ (labeled diff $M_{\rm min}$.) and when we go beyond linear theory and marginalise over 1-loop and counter-terms parameters (labeled 1-loop, see Section \ref{sec:1-loop}).}
\label{table:Table_mnu_Ext-HIRAX_21cm_gain}
\end{table}

%%%%

\begin{table}[ht!]
\centering
\begin{tabular}{l|ccc|ccc}
\hline \hline
\multicolumn{7}{c}{Ext-CHIME $\Sigma m_\nu$[eV]}\\
\hline
& \multicolumn{3}{c|}{ $k_\mathrm{max}=0.2 h\mathrm{Mpc^{-1}}$} & \multicolumn{3}{c}{$k_\mathrm{max}=k_\mathrm{nl}(z)$}\\ \hline\hline
Euclid No Rec.+PlanckBAO & \multicolumn{3}{c|}{0.050} & \multicolumn{3}{c}{0.049}\\ \hline
$+$21cm & No wedge & Mid wedge & Wedge & No wedge & Mid wedge & Wedge\\ \hline
2\% $b_\mathrm{HI}\,\&\,\Omega_\mathrm{HI}$ & 0.038 & 0.041 & 0.044 & 0.033 & 0.035 & 0.040\\
5\% $b_\mathrm{HI}\,\&\,\Omega_\mathrm{HI}$ & 0.043 & 0.045 & 0.048 & 0.037 & 0.040 & 0.043\\
10\% $b_\mathrm{HI}\,\&\,\Omega_\mathrm{HI}$ & 0.044 & 0.047 & 0.049 & 0.037 & 0.041 & 0.044\\ 
\hline
\hline
Euclid No Rec.+CMB-S4 & \multicolumn{3}{c|}{0.031} & \multicolumn{3}{c}{0.030}\\ \hline
$+$21cm & No wedge & Mid wedge & Wedge & No wedge & Mid wedge & Wedge\\ \hline
2\% $b_\mathrm{HI}\,\&\,\Omega_\mathrm{HI}$ & 0.023 & 0.026 & 0.029 & 0.020 & 0.022 & 0.026\\
5\% $b_\mathrm{HI}\,\&\,\Omega_\mathrm{HI}$ & 0.024 & 0.026 & 0.029 & 0.021 & 0.023 & 0.026\\
10\% $b_\mathrm{HI}\,\&\,\Omega_\mathrm{HI}$ & 0.025 & 0.026 & 0.029 & 0.021 & 0.023 & 0.027\\
\hline
\end{tabular}
\caption{Forecasted 1$\sigma$ constraints on $\Sigma m_\nu$ considering Ext-CHIME instrument combined with CMB+galaxy probes. We show the constraints considering different priors on both $b_\mathrm{HI}\,\&\,\Omega_\mathrm{HI}$, different level of wedge contamination and different $k_{\rm max}$ used. In the top half of the table we show what are the gains of adding Ext-CHIME measurements on top of combined Fisher matrix for Euclid galaxy survey and PlanckBAO. In the bottom half we do the same only using CMBS-4 Fisher matrix instead of the PlanckBAO one.}
\label{table:Table_mnu_ExtCHIME_21cm_gain}
\end{table}

%%%%

\begin{table}[ht!]
\centering
\begin{tabular}{lcccccc}
\hline \hline
\multicolumn{3}{c}{highzFAST $\Sigma m_\nu$[eV]}\\
\hline
& $k_\mathrm{max}=0.2 h\mathrm{Mpc^{-1}}$ & $k_\mathrm{max}=k_\mathrm{nl}(z)$ \\ \hline\hline
Euclid No Rec.+PlanckBAO & 0.050 & 0.049\\ \hline
$+$21cm & & \\ \hline
2\% $b_\mathrm{HI}\,\&\,\Omega_\mathrm{HI}$ & 0.043 & 0.041\\
5\% $b_\mathrm{HI}\,\&\,\Omega_\mathrm{HI}$ & 0.048 & 0.045\\
10\% $b_\mathrm{HI}\,\&\,\Omega_\mathrm{HI}$ & 0.049 & 0.047\\ 
\hline
\hline
Euclid No Rec.+CMB-S4 & 0.031 & 0.030\\ \hline
$+$21cm & & \\ \hline
2\% $b_\mathrm{HI}\,\&\,\Omega_\mathrm{HI}$ & 0.029 & 0.028\\
5\% $b_\mathrm{HI}\,\&\,\Omega_\mathrm{HI}$ & 0.030 & 0.028\\ 
10\% $b_\mathrm{HI}\,\&\,\Omega_\mathrm{HI}$ & 0.030 & 0.029\\
\hline
\end{tabular}
\caption{Forecasted 1$\sigma$ constraints on $\Sigma m_\nu$ considering highzFAST instrument combined with CMB+galaxy probes. We show the constraints considering different priors on both $b_\mathrm{HI}\,\&\,\Omega_\mathrm{HI}$ and different $k_{\rm max}$ used. In the top half of the table we show what are the gains of adding highzFAST measurements on top of combined Fisher matrix for Euclid galaxy survey and PlanckBAO. In the bottom half we do the same only using CMBS-4 Fisher matrix instead of the PlanckBAO one.}
\label{table:Table_mnu_FAST_21cm_gain}
\end{table}

%%%%%
%%%%%
%%%%% Neff tables
%%%%%

\begin{table}[ht!]
\centering
\begin{tabular}{lccccc}
\hline \hline
\multicolumn{6}{c}{External datasets -- $N_\mathrm{eff}$}\\
\hline	
& $\Omega_\mathrm{M}$ & $h$ & $\ln(10^{10} A_s)$ & $n_s$ & $N_\mathrm{eff}$\\
\hline
\hline
Fiducial values & 0.3075 & 0.6774 & 3.064 & 0.9667 & 3.04\\
\hline
\hline
PlanckBAO  &  0.0088 & 0.015 & 0.038 & 0.0088 & 0.23\\
\hline
Euclid02 & 0.0031 & 0.0064 & 0.0136 & 0.0106 & 0.37\\
\hline
Euclidnl  & 0.0027 & 0.0041 & 0.0096 & 0.0056 & 0.23\\
\hline
CMB-S4 & 0.0012 & 0.00061 & 0.0094 & 0.0028 & 0.029\\
\hline
\end{tabular}
\caption{Forecasted 1$\sigma$ constraints on cosmological parameters and the $N_\mathrm{eff}$ considering each external dataset alone. These constraints are obtained using the Fisher matrices explained in Section \ref{sec:probecombination}. PlanckBAO represents Planck CMB combined with BOSS BAO measurements. We show the projected constraints for Euclid  for two different $k_{\rm max}$ used and the constraints coming from CMB-S4 experiment.}
\label{table:Priors_Neff}
\end{table}

%%%%%

\begin{table}[ht!]
\centering
\begin{tabular}{lccccc}
\hline \hline
\multicolumn{6}{c}{Ext-HIRAX $N_\mathrm{eff}$}\\
\hline	
& $\Omega_\mathrm{M}$ & $h$ & $\ln(10^{10} A_s)$ & $n_s$ & $N_\mathrm{eff}$\\
\hline
Fiducial values & 0.3075 & 0.6774 & 3.064 & 0.9667 & 3.04\\
\hline\hline 
\multicolumn{6}{c}{No wedge $k_\mathrm{max}=0.2 h\mathrm{Mpc^{-1}}$}\\
\hline
2\% $b_\mathrm{HI}\,\&\,\Omega_\mathrm{HI}$ & 0.0017 & 0.0036 & 0.013 & 0.0061 & 0.22\\ \hline
5\% $b_\mathrm{HI}\,\&\,\Omega_\mathrm{HI}$ & 0.0019 & 0.0037 & 0.023 & 0.0061 & 0.23\\ \hline
10\% $b_\mathrm{HI}\,\&\,\Omega_\mathrm{HI}$ & 0.0020 & 0.0037 & 0.041 & 0.0061 & 0.23\\ 
\hline\hline 
\multicolumn{6}{c}{No wedge $k_\mathrm{max}=k_\mathrm{nl}(z)$}\\
\hline
2\% $b_\mathrm{HI}\,\&\,\Omega_\mathrm{HI}$ & 0.0010 & 0.0013 & 0.009 & 0.0014 & 0.075\\ \hline
5\% $b_\mathrm{HI}\,\&\,\Omega_\mathrm{HI}$ & 0.0011 & 0.0013 & 0.020 & 0.0015 & 0.077\\ \hline
10\% $b_\mathrm{HI}\,\&\,\Omega_\mathrm{HI}$ & 0.0011 & 0.0013 & 0.040 & 0.0015 & 0.077\\ 
\hline\hline 
\multicolumn{6}{c}{Mid wedge $k_\mathrm{max}=0.2 h\mathrm{Mpc^{-1}}$}\\
\hline
2\% $b_\mathrm{HI}\,\&\,\Omega_\mathrm{HI}$ & 0.0018 & 0.0039 & 0.013 & 0.0067 & 0.24\\ \hline
5\% $b_\mathrm{HI}\,\&\,\Omega_\mathrm{HI}$ & 0.0021 & 0.0041 & 0.023 & 0.0067 & 0.25\\ \hline
10\% $b_\mathrm{HI}\,\&\,\Omega_\mathrm{HI}$ & 0.0022 & 0.0041 & 0.042 & 0.0067 & 0.25\\ 
\hline\hline 
\multicolumn{6}{c}{Mid wedge $k_\mathrm{max}=k_\mathrm{nl}(z)$}\\
\hline
2\% $b_\mathrm{HI}\,\&\,\Omega_\mathrm{HI}$ & 0.0011 & 0.0014 & 0.010 & 0.0016 & 0.081\\ \hline
5\% $b_\mathrm{HI}\,\&\,\Omega_\mathrm{HI}$ & 0.0012 & 0.0014 & 0.021 & 0.0016 & 0.084\\ \hline
10\% $b_\mathrm{HI}\,\&\,\Omega_\mathrm{HI}$ & 0.0013 & 0.0014 & 0.040 & 0.0016 & 0.085\\ 
\hline\hline 
\multicolumn{6}{c}{Wedge $k_\mathrm{max}=0.2 h\mathrm{Mpc^{-1}}$}\\
\hline
2\% $b_\mathrm{HI}\,\&\,\Omega_\mathrm{HI}$ & 0.0057 & 0.0117 & 0.03 & 0.0192 & 0.73\\ \hline
5\% $b_\mathrm{HI}\,\&\,\Omega_\mathrm{HI}$ & 0.0058 & 0.0118 & 0.04 & 0.0192 & 0.74\\ \hline
10\% $b_\mathrm{HI}\,\&\,\Omega_\mathrm{HI}$ & 0.0060 & 0.0119 & 0.06 & 0.0193 & 0.75\\ 
\hline\hline 
\multicolumn{6}{c}{Wedge $k_\mathrm{max}=k_\mathrm{nl}(z)$}\\
\hline
2\% $b_\mathrm{HI}\,\&\,\Omega_\mathrm{HI}$ & 0.0027 & 0.0035 & 0.01 & 0.0040 & 0.21\\ \hline
5\% $b_\mathrm{HI}\,\&\,\Omega_\mathrm{HI}$ & 0.0031 & 0.0035 & 0.03 & 0.0041 & 0.21\\ \hline
10\% $b_\mathrm{HI}\,\&\,\Omega_\mathrm{HI}$ & 0.0035 & 0.0035 & 0.05 & 0.0042 & 0.22\\ 
\hline
\hline
\end{tabular}
\caption{Fiducial values and 68\% confidence intervals on cosmological parameters and the $N_{\rm eff}$ using Ext-HIRAX alone for various considerations. Going from top to bottom, we show the constraints considering different priors on both $b_\mathrm{HI}\,\&\,\Omega_\mathrm{HI}$, different $k_{\rm max}$ we use and different level of foreground wedge contamination.}
\label{table:Table_Neff_Ext-HIRAX}
\end{table}

%%%%

\begin{table}[h]
\centering
\begin{tabular}{lccccc}
\hline \hline
\multicolumn{6}{c}{highzFAST $N_\mathrm{eff}$}\\
\hline	
& $\Omega_\mathrm{M}$ & $h$ & $\ln(10^{10} A_s)$ & $n_s$ & $N_\mathrm{eff}$\\
\hline
Fiducial values & 0.3075 & 0.6774 & 3.064 & 0.9667 & 3.04\\
\hline\hline 
\multicolumn{6}{c}{$k_\mathrm{max}=0.2 h\mathrm{Mpc^{-1}}$}\\ 
\hline
2\% $b_\mathrm{HI}\,\&\,\Omega_\mathrm{HI}$& 0.0053 & 0.011 & 0.027 & 0.021 & 0.67\\
5\% $b_\mathrm{HI}\,\&\,\Omega_\mathrm{HI}$& 0.0061 & 0.012 & 0.036 & 0.021 & 0.70\\
10\% $b_\mathrm{HI}\,\&\,\Omega_\mathrm{HI}$& 0.0067 & 0.012 & 0.053 & 0.021 & 0.72\\
\hline\hline 
\multicolumn{6}{c}{$k_\mathrm{max}=k_\mathrm{nl}(z)$}\\
\hline
2\% $b_\mathrm{HI}\,\&\,\Omega_\mathrm{HI}$ & 0.0038 & 0.0058 & 0.018 & 0.0078 & 0.34\\
5\% $b_\mathrm{HI}\,\&\,\Omega_\mathrm{HI}$& 0.0046 & 0.0059 & 0.028 & 0.0079 & 0.36\\
10\% $b_\mathrm{HI}\,\&\,\Omega_\mathrm{HI}$& 0.0052 & 0.0060 & 0.047 & 0.0079 & 0.37\\
\hline
\hline
\end{tabular}
\caption{Fiducial values and 68\% confidence intervals on cosmological parameters and the $N_{\rm eff}$ using highzFAST alone for various considerations. We show the constraints considering different astrophysical priors on both $b_\mathrm{HI}\,\&\,\Omega_\mathrm{HI}$, as well as the dependence of the constraints on the $k_{\rm max}$ used.}
\label{table:Table_Neff_FAST}
\end{table}

%%%%

\begin{table}[ht!]
\centering
\begin{tabular}{l|ccc|ccc}
\hline \hline
\multicolumn{7}{c}{Ext-HIRAX $N_\mathrm{eff}$}\\
\hline
& \multicolumn{3}{c|}{ $k_\mathrm{max}=0.2 h\mathrm{Mpc^{-1}}$} & \multicolumn{3}{c}{$k_\mathrm{max}=k_\mathrm{nl}(z)$}\\ \hline\hline
Euclid Rec.+PlanckBAO & \multicolumn{3}{c|}{0.067} & \multicolumn{3}{c}{0.064}\\ \hline
$+$21cm & No wedge & Mid wedge & Wedge & No wedge & Mid wedge & Wedge\\ \hline
2\% $b_\mathrm{HI}\,\&\,\Omega_\mathrm{HI}$ & 0.046 & 0.047 & 0.057 & 0.038 & 0.039 & 0.049\\
5\% $b_\mathrm{HI}\,\&\,\Omega_\mathrm{HI}$ & 0.047 & 0.048 & 0.057 & 0.039 & 0.040 & 0.049\\
10\% $b_\mathrm{HI}\,\&\,\Omega_\mathrm{HI}$ & 0.047 & 0.048 & 0.057 & 0.039 & 0.041 & 0.050\\\hline
\hline
Euclid Rec.+CMB-S4 & \multicolumn{3}{c|}{0.022} & \multicolumn{3}{c}{0.020}\\ \hline
$+$21cm & No wedge & Mid wedge & Wedge & No wedge & Mid wedge & Wedge\\ \hline
2\% $b_\mathrm{HI}\,\&\,\Omega_\mathrm{HI}$ & 0.019 & 0.019 & 0.021 & 0.015 & 0.015 & 0.017\\
5\% $b_\mathrm{HI}\,\&\,\Omega_\mathrm{HI}$ & 0.020 & 0.020 & 0.022 & 0.015 & 0.016 & 0.018\\
10\% $b_\mathrm{HI}\,\&\,\Omega_\mathrm{HI}$ & 0.020 & 0.020 & 0.022 & 0.015 & 0.016 & 0.018\\\hline
\end{tabular}
\caption{Forecasted 1$\sigma$ constraints on $N_{\rm eff}$ considering Ext-HIRAX instrument combined with CMB+galaxy probes. We show the constraints considering different priors on both $b_\mathrm{HI}\,\&\,\Omega_\mathrm{HI}$, different level of wedge contamination and different $k_{\rm max}$ used. In the top half of the table we show what are the gains of adding Ext-HIRAX measurements on top of combined Fisher matrix for Euclid galaxy survey  and PlanckBAO. In the bottom half we do the same only using CMBS-4 Fisher matrix instead of the PlanckBAO one.}\label{table:Table_Neff_Ext-HIRAX_21cm_gain}
\end{table}

%%%%

\begin{table}[ht!]
\centering
\begin{tabular}{l|ccc|ccc}
\hline \hline
\multicolumn{7}{c}{Ext-CHIME $N_\mathrm{eff}$}\\
\hline
& \multicolumn{3}{c|}{ $k_\mathrm{max}=0.2 h\mathrm{Mpc^{-1}}$} & \multicolumn{3}{c}{$k_\mathrm{max}=k_\mathrm{nl}(z)$}\\ \hline\hline
Euclid Rec.+PlanckBAO & \multicolumn{3}{c|}{0.067} & \multicolumn{3}{c}{0.064}\\ \hline
$+$21cm & No wedge & Mid wedge & Wedge & No wedge & Mid wedge & Wedge\\ \hline
2\% $b_\mathrm{HI}\,\&\,\Omega_\mathrm{HI}$ & 0.049 & 0.052 & 0.058 & 0.043 & 0.045 & 0.052\\
5\% $b_\mathrm{HI}\,\&\,\Omega_\mathrm{HI}$ & 0.050 & 0.053 & 0.058 & 0.044 & 0.047 & 0.052\\
10\% $b_\mathrm{HI}\,\&\,\Omega_\mathrm{HI}$ & 0.050 & 0.053 & 0.058 & 0.044 & 0.047 & 0.052\\\hline
\hline
Euclid Rec.+CMB-S4 & \multicolumn{3}{c|}{0.022} & \multicolumn{3}{c}{0.020}\\ \hline
$+$21cm & No wedge & Mid wedge & Wedge & No wedge & Mid wedge & Wedge\\ \hline
2\% $b_\mathrm{HI}\,\&\,\Omega_\mathrm{HI}$ & 0.020 & 0.021 & 0.021 & 0.016 & 0.017 & 0.018\\
5\% $b_\mathrm{HI}\,\&\,\Omega_\mathrm{HI}$ & 0.020 & 0.021 & 0.022 & 0.016 & 0.017 & 0.018\\
10\% $b_\mathrm{HI}\,\&\,\Omega_\mathrm{HI}$ & 0.020 & 0.021 & 0.022 & 0.016 & 0.017 & 0.018\\\hline
\end{tabular}
\caption{Forecasted 1$\sigma$ constraints on $N_{\rm eff}$ considering Ext-CHIME instrument combined with CMB+galaxy probes. We show the constraints considering different priors on both $b_\mathrm{HI}\,\&\,\Omega_\mathrm{HI}$, different level of wedge contamination and different $k_{\rm max}$ used. In the top half of the table we show what are the gains of adding Ext-CHIME measurements on top of combined Fisher matrix for Euclid galaxy survey and PlanckBAO. In the bottom half we do the same only using CMBS-4 Fisher matrix instead of the PlanckBAO one.}
\label{table:Table_Neff_Ext-CHIME_21cm_gain}
\end{table}

%%%%%

\clearpage

%%%%

\begin{table}[ht!]
\centering
\begin{tabular}{lcccccc}
\hline \hline
\multicolumn{3}{c}{highzFAST $N_\mathrm{eff}$}\\
\hline
& $k_\mathrm{max}=0.2 h\mathrm{Mpc^{-1}}$ & $k_\mathrm{max}=k_\mathrm{nl}(z)$ \\ \hline\hline
Euclid Rec.+PlanckBAO & 0.067 & 0.064\\ \hline
$+$21cm & & \\ \hline
2\% $b_\mathrm{HI}\,\&\,\Omega_\mathrm{HI}$ & 0.062 & 0.057\\
5\% $b_\mathrm{HI}\,\&\,\Omega_\mathrm{HI}$ & 0.062 & 0.058\\
10\% $b_\mathrm{HI}\,\&\,\Omega_\mathrm{HI}$ & 0.062 & 0.058\\
\hline
\hline
Euclid Rec.+CMB-S4 & 0.022 & 0.020\\ \hline
$+$21cm & & \\ \hline
2\% $b_\mathrm{HI}\,\&\,\Omega_\mathrm{HI}$ & 0.022 & 0.019\\
5\% $b_\mathrm{HI}\,\&\,\Omega_\mathrm{HI}$ & 0.022 & 0.019\\
10\% $b_\mathrm{HI}\,\&\,\Omega_\mathrm{HI}$ & 0.022 & 0.019\\
\hline
\end{tabular}
\caption{Forecasted 1$\sigma$ constraints on $N_\mathrm{eff}$ considering highzFAST instrument combined with CMB+galaxy probes. We show the constraints considering different priors on both $b_\mathrm{HI}\,\&\,\Omega_\mathrm{HI}$, different level of wedge contamination and different $k_{\rm max}$ used. In the top half of the table we show what are the gains of adding Ext-CHIME measurements on top of combined Fisher matrix for Euclid galaxy survey  and PlanckBAO. In the bottom half we do the same only using CMBS-4 Fisher matrix instead of the PlanckBAO one.}
\label{table:Table_Neff_FAST_21cm_gain}
\end{table}

%%\bibliographystyle{apsrev}
%%\bibliography{runnings}
\section*{Acknowledgments}
We thank An\v ze Slosar for providing us with the CMB S4 Fisher matrix made by Joel Meyers and for insightful conversations. We also thank Marko Simonovi\' c and Alkistis Pourtsidou for useful conversations. EC thanks Dan Green for useful discussion about $N_{\rm eff}$. MV and AO are supported by the INFN grant PD 51 INDARK. The work of FVN is supported by the Simons Foundation. MV is also supported by the ERC StG cosmoIGM. AO acknowledges the hospitality of the Center for Computational Astrophysics of the Simons Foundation in NY.

\bibliographystyle{JHEP}
\bibliography{References}

\end{document}